\documentclass[prb,twocolumn,superscriptaddress,showpacs]{revtex4}
\pdfoutput=1
\usepackage{graphicx}
\usepackage{subfigure}
\usepackage{amsmath}
\usepackage{xspace}
\usepackage{color}
\usepackage{times}

\usepackage{listings}

\newfont{\tensy}{cmsy10}

\renewcommand{\Im}[0]{\text{Im}\,}

\newcommand{\ie}[0]{i.e.\@\xspace}
\newcommand{\eg}[0]{e.g.\@\xspace}
\newcommand{\etal}[0]{et al.\@\xspace}

\newcommand{\las}[0]{\langle}
\newcommand{\ras}[0]{\rangle}
\newcommand{\la}[0]{\left\las}
\newcommand{\ra}[0]{\right\ras}
\newcommand{\ket}[1]{\left|#1\ra}
\newcommand{\bra}[1]{\la#1\right|}
\newcommand{\braket}[1]{\la #1\ra}

\newcommand{\rmi}{\text{i}}
\newcommand{\UP}[0]{\uparrow}
\newcommand{\DO}[0]{\downarrow}

\newcommand{\on}{\hat{n}}

\newcommand{\om}[0]{\omega}
\newcommand{\en}[0]{\epsilon}
\newcommand{\nag}{{\phantom{\dag}}}

\renewcommand{\tilde}[1]{\widetilde{#1}}

\begin{document}
%\lstset{frame=trbl,language=C++,basicstyle=\tiny}

%%%%%%%%%%%%%%%%%%%%%%%%%%%%%%%%%%%%%%%%%%%%%%%%%%%%%%%%%%%%%%%%%%%%%%%%%%%%%
%%%%%%%%%%%%%%%%%%%%%       TITLE & ABSTRACT          %%%%%%%%%%%%%%%%%%%%%%%
%%%%%%%%%%%%%%%%%%%%%%%%%%%%%%%%%%%%%%%%%%%%%%%%%%%%%%%%%%%%%%%%%%%%%%%%%%%%%
\title{Excitation spectra of strongly correlated lattice bosons and
  polaritons}

\author{Peter Pippan}
\author{Hans Gerd Evertz}
\affiliation{%
Institute for Theoretical and Computational Physics, TU Graz,
8010 Graz, Austria}
\author{Martin Hohenadler}
\altaffiliation{Present address: OSRAM Opto Semiconductors, 93055 Regensburg, Germany}
\affiliation{Cavendish Laboratory, University of Cambridge, Cambridge, CB3
  0HE, United Kingdom}

\begin{abstract}
  Spectral properties of the Bose-Hubbard model and a recently proposed
  coupled-cavity model are studied by means of quantum Monte Carlo
  simulations in one dimension. Both models exhibit a quantum phase
  transition from a Mott insulator to a superfluid phase. The dynamic
  structure factor $S(k,\omega)$ and the single-particle spectrum
  $A(k,\omega)$ are calculated, focusing on the parameter region around the
  phase transition from the Mott insulator with density one to the superfluid
  phase, where correlations are important. The strongly interacting nature of
  the superfluid phase manifests itself in terms of additional gapped modes
  in the spectra. Comparison is made to recent analytical work on the
  Bose-Hubbard model. Despite some subtle differences due to the polaritonic
  particles in the cavity model, the gross features are found to be very
  similar to the Bose-Hubbard case. For the polariton model, emergent
  particle-hole symmetry near the Mott lobe tip is demonstrated, and
  temperature and detuning effects are analyzed. A scaling analysis for the
  generic transition suggests mean field exponents, in accordance with field
  theory results.
\end{abstract} 

\date{\today}

% Dynamic properties of condensates; collective and hydrodynamic excitations, superfluid flow

\maketitle

\section{Introduction}\label{sec:introduction}

The Mott insulator (MI) to superfluid (SF) quantum phase transition in the
generic Bose-Hubbard model\cite{PhysRevB.40.546} has attracted a lot of
attention in recent years due to the progress in experiments on cold atomic
gases in optical lattices.\cite{Gr.Ma.Es.Ha.Bl.02} More recently, there have
also been significant advances in the coherent coupling of single atoms and
cold atomic gases to cavity radiation (cavity quantum
electrodynamics).\cite{BiBoMiBoNoKi05,BrDoRiBoKoEs07} A clean realization of
the Jaynes-Cummings Hamiltonian has been achieved by coupling a
superconducting qubit to a microwave cavity.\cite{Fi.Go.Ba.Bi.Le.Bl.Wa.08} On
the theory side, multi-component Bose gases coupled to light have \eg been
shown to support a superradiant Mott insulator phase with polariton
condensation.\cite{BhHoSiSi08} 

In parallel, several theoretical proposals have shown the possibility of
having a state of strongly correlated photons or polaritons in solid-state
systems of coupled cavity arrays (also referred to as polariton models or
Jaynes-Cummings-Hubbard models),\cite{GrTaCoHo06,HaBrPl06} and a review of
work along these lines has been given.\cite{Ha.Br.Pl.08} The possibility of
preparing a system of photons in a Mott state with one photon per site is a
promising starting point for quantum information processing. An
important feature shared with cold atomic gases coupled to light is the composite nature of the
polaritons.  Particularly attractive properties of cavity arrays would include
accessibility of local properties in measurements and scalability in terms of
size. Perhaps the most likely candidate for setting up such a model
experimentally is based on extending the work on superconducting qubits to
arrays.\cite{Ko.LH.09,Fi.Go.Ba.Bi.Le.Bl.Wa.08} In contrast to cold atomic
gases, where the interaction and/or hopping strength can be varied, the
phase transition may be observed by changing the detuning between the
two-level system and the resonator. Analysis of coupled cavity
models is fruitful in its own right, as a detailed understanding of the
corresponding models offers insight into strongly correlated polariton
systems. An important aspect of such studies is the extent to which such
systems resemble the familiar Bose-Hubbard physics.

From the above examples and many more in the literature, it is
apparent that interacting boson systems on a lattice are of great interest
for the progress of both theory and experiment.
Compared to Bose fluids, the lattice changes the physics in several aspects.
Although long-range phase coherence still gives rise to phonon
excitations---despite the breaking of translational symmetry---the quenching
of the kinetic energy makes the system much more strongly
correlated.\cite{Zw03} Besides, the lattice allows the formation of
incompressible MI states with the same integer particle number at each site.

A large amount of work has been devoted to detailed studies of the
Bose-Hubbard model, leading to a wealth of knowledge with and without additional
complications such as trapping potentials or disorder. However, the dynamical
properties and excitations
in particular of the SF phase in the vicinity of the quantum phase transition, are still not completely understood. 
A number of authors have addressed the dynamics of the Bose-Hubbard model in different dimensions,
\cite{
0295-5075-22-4-004, %49 RPA '93
PhysRevB.59.12184, %34 SC Mott '99 
Ku.Wh.Mo.00,    %33 DMRG  1d, sigma
Ro.Bu.04,       %53 ED 1d, S
Ba.As.Sc.De.05, %54 qmc 1d, S
Se.Du.05,       % 47 SC (1d)
KoDu06,         %51 CPT   1d Mott, A
Hu.Al.Bu.Bl.07, % 6  Schwinger
CaSa.GuSo.Pr.Sv.07,        %40 qmc 2d
capogrosso-sansone:134302, %41 qmc 3d '08
Oh.Pe.08,       % 39 SC (3d)
Me.Tr.08} %        % 7  RPA 1d
with results providing
valuable information about the underlying physics,
while corresponding work on coupled cavity models has just
begun.\cite{Ai.Ho.Ta.Li.08,Sc.Bl.09}
The two most important dynamic observables are the dynamic structure factor
and the single-particle spectral function, which are also at the heart of
theoretical and experimental works on Bose fluids.\cite{Griffin93}
Experimentally, the dynamic structure factor may be measured by Bragg
spectroscopy or lattice modulation (in cold atomic gases) as well as by
neutron scattering (in liquid helium), and single-particle excitations of
optical solid-state systems are accessible by means of photoluminescence
measurements. 

Whereas the standard Bose-Hubbard model only supports MI and SF phases, the
physics of the polariton models is slightly richer. Owing to the composite
nature of the conserved particles (polaritons), these phases can either be of
polaritonic, excitonic or photonic
character\cite{Ai.Ho.Ta.Li.08,irish_polaritonic_2008,Le.Li.08,Ir.09} with
distinct dynamic properties. Which of the cases is realized depends on the
value of the detuning between the cavity mode and the transition frequency of
the atoms that mediate polariton repulsion. Very recently it has been
proposed that the fractional quantum Hall effect may also be realized in
coupled cavity arrays.\cite{Ch.An.Bo.08}

In general, accurate and unbiased results are very hard to obtain. Most
existing work on spectral properties in the Mott phase is based on mean-field
and/or strong-coupling approximations, in which fluctuations of the particle
numbers are more or less restricted.  Results of extensive strong coupling
expansions for the phase diagram \cite{Fr.Mo.96,PhysRevB.59.12184} do,
however, agree very well with precise density-matrix renormalization group
(DMRG)\cite{Ku.Wh.Mo.00} and quantum Monte Carlo (QMC)
results.\cite{capogrosso-sansone:134302,CaSa.GuSo.Pr.Sv.07} Bogoliubov type
descriptions have been found to accurately describe the SF phase only in the
limit of weak interaction, and fail to account for the transition to a MI and
correlation features in the SF close to the transition. Hence the most interesting
(and most difficult) regime is that near the quantum phase transition, where quantum fluctuations
and correlation effects cannot be neglected.

In one dimension (1D), quantum fluctuation effects are particularly
pronounced and mean-field methods are in general insufficient.  Notable
exceptions include situations where coupling to additional degrees of freedom
provides an effective long-range interaction.\cite{BhHoSiSi08} An interesting
aspect of 1D is that for strong (repulsive) interaction, fermions and bosons
behave in a very similar way, and that the low-energy, long-wavelength
physics is described by the Luttinger liquid
model.\cite{PhysRevLett.93.210401}

In the present paper we employ the directed loop quantum Monte Carlo
method,\cite{SySa02} which is exact and therefore yields unbiased results
also in difficult parameter regimes.  Importantly, our simulations preserve
the full quantum dynamics.

Few nonperturbative results are available for the spectra in the Bose-Hubbard model, namely
for the dynamical conductivity,\cite{Ku.Wh.Mo.00} for the dynamic structure
factor $S(k,\omega)$\cite{Ro.Bu.04,Ba.As.Sc.De.05} on small systems, and for
the single-boson spectral function $A(k,\om)$ in the Mott phase deduced from
small systems,\cite{KoDu06} all in 1D.  For the polariton model considered
here, only $A(k,\om)$ in the Mott phase has been calculated.\cite{Ai.Ho.Ta.Li.08}

The focus of our work is therefore on the calculation of excitation spectra
for both the Bose-Hubbard model and the polariton model within and around the first Mott lobe (\ie, the
lobe with density one), for which comparison to recent analytical and
numerical results is made. Other issues addressed include the sound velocity
in the SF phase, particle and hole masses, as well as temperature and
detuning effects for the case of the polariton model.

Our simulations are performed at low but finite temperatures. On one
hand, this complicates the analysis of the results, but on the other hand it
matches the experimental situation.\cite{Griffin93,Griffin98}

The paper is organized as follows. In Sec.~\ref{sec:model} we introduce the
two models considered. Section~\ref{sec:method} contains some details about
the method. Results are discussed in Sec.~\ref{sec:results}, and in
Sec.~\ref{sec:conclusions} we present our conclusions.

\section{Models}\label{sec:model}

The polariton model we consider is the simplest among several recent
proposals.\cite{GrTaCoHo06,HaBrPl06,AnSaBo07,hartmann:070602,Ha.Br.Pl.08}
It describes an array of $L$ optical microcavities, each of which contains
a single two-level atom with states $\ket{\DO}$, $\ket{\UP}$ separated by
energy $\epsilon$. Within the rotating wave approximation one such cavity
is represented by the Jaynes-Cummings Hamiltonian\cite{Ja.Cu.63} ($\hbar=1$)
\begin{eqnarray}\label{eq:JC}\nonumber
  \hat{H}^{\text{JC}}_i
  &=&
  \epsilon \ket{\UP_i}\bra{\UP_i}
  +
  \omega_0 a^\dag_i a^\nag_i
  \\
  &&+
  g (\ket{\UP_i}\bra{\DO_i} a^\nag_i + \ket{\DO_i}\bra{\UP_i} a^\dag_i)
  \,.
\end{eqnarray}
Here $\omega_0$ is the cavity photon energy, and $\Delta=\epsilon-\omega_0$
defines the detuning.  The atom-photon coupling $g$ ($a^\dag_i$, $a^\nag_i$
are photon creation and annihilation operators) gives rise to formation of
polaritons (combined atom-photon or exciton-photon excitations). Allowing for
nearest-neighbor photon hopping between cavities with amplitude $t$ leads to
the lattice Hamiltonian
\begin{eqnarray}\label{eq:ham_PM}
  \hat{H}^{\text{PM}}
  &=&
  -t\sum_{\las i,j\ras} a^\dag_i a^\nag_j + \sum_i \hat{H}^{\text{JC}}_i
  - \mu \hat{N}_\text{p}
  \,.
\end{eqnarray}
The conserved polariton number $\hat{N}_\text{p}=\sum_i
\hat{n}_{\text{p},i}$, with $\hat{n}_{\text{p},i}= a^\dag_i a^\nag_i +
\ket{\UP_i}\bra{\UP_i}$, is determined by the chemical potential
$\mu$.\cite{Ma.Co.Ta.Ho.Gr.07} Polaritons experience an effective repulsion
$U_\text{eff}(n_\mathrm{p})$ [see Eq.~(\ref{eq:Ueff})] due to the nonlinear
dependence of the single-site energy on the local occupation number
$n_\mathrm{p}$. We use $g$ as the unit of energy and set $\omega_0/g$,
$k_\text{B}$ and the lattice constant equal to unity. The rotating wave
approximation becomes unjustified for $g$ comparable to $\en$. The motivation
for setting $g=\en$ is direct comparison to previous work. The
Hamiltonian~(\ref{eq:ham_PM}) has been
studied in [\onlinecite{GrTaCoHo06,Ma.Co.Ta.Ho.Gr.07,AnSaBo07,Ro.Fa.07,Ai.Ho.Ta.Li.08,rossini_photon_2008,Ha.Br.Pl.08,Le.Li.08,irish_polaritonic_2008,Ir.09,Sc.Bl.09,Ko.LH.09}].

We also consider the Bose-Hubbard Hamiltonian
\begin{equation}\label{eq:ham_BHM}
  \hat{H}^{\text{BHM}}
  =
  -t\sum_{\las i,j\ras} b^\dag_i b^\nag_j  
  +
  \frac{U}{2}\sum_i n_i (n_i-1)
  - \mu \hat{N}
  \,,
\end{equation}
describing soft-core bosons with repulsion $U$ and hopping $t$.
Here $\hat{N}=\sum_i \hat{n}_i=\sum_i b^\dag_i b^\nag_i$, is the
total number of bosons, and we use $U$ as the unit of energy.

As an alternative to the spin language used here, the polariton
model~(\ref{eq:ham_PM}) can be written as a two-band Bose-Hubbard
Hamiltonian;\cite{koch:042319} one boson species is itinerant, whereas the other is
immobile (corresponding to localized excitons) with a hard-core
constraint. This correspondence provides a direct
connection to recent work on cold atomic gases in optical lattices, with the
natural extension to the case where the excitons are mobile as
well.\cite{BhHoSiSi08}

We shall see below that owing to the composite nature of the bosonic
particles in the polariton model, it is generally easier to understand the features
of the Bose-Hubbard model first, and then explore similarities to the
polariton model. Moreover,
analytical approximations are more readily available for the Bose-Hubbard model and provide
insight into the numerical data. Periodic boundary conditions in
real space are applied in all simulations, and the system size is denoted as $L$.

\section{Method}\label{sec:method}

We use the directed loop method,\cite{SySa02} a generalization of the loop
algorithm,\cite{loop_evertz_93,HGE03} which has no systematic errors and is
efficient (low autocorrelations), facilitating the simulation of large
systems at low temperatures.  We make use of the ALPS
library~\cite{ALPS_I,ALPS_II} and of the ALPS
applications,\cite{ALPS_DIRLOOP} which use the stochastic series
expansion (SSE) representation\cite{SandvikSSE} of worldline path integrals.  We have
verified that we obtain the correct phase boundary in 1D for selected points
in parameter space.

In contrast to most previous QMC calculations of the Bose-Hubbard model, the focus of the
current paper is on dynamical properties.  The SSE representation has the
drawback that dynamical correlation functions in imaginary time, which we
need to obtain spectra, are very inefficient to calculate, since they involve
a convolution of Green functions at different SSE
distances.\cite{DorneichT01} On the other hand, Green functions can be
measured easily in an imaginary time representation.
For this reason we revert a mapping from continuous time to
SSE\cite{interaction_representation__sandvik__PRB97} when measuring Green
functions.  To each operator in a given SSE operator string we associate a
time $\tau \in [0, \beta]$ which is stochastically sampled out of a uniform
distribution.  This maps the SSE configuration into a worldline configuration
in continuous imaginary time.\cite{Spin_peierls_franzi_07}

Correlation functions of diagonal operators can then be measured directly.
For example, in the case of $\langle\hat{\rho}_i(\tau)\hat{\rho}_j(0)\rangle$ 
we evaluate the density $\rho_i(\tau)$ on a fine time grid. This
time discretization limits the high energy range of the Green function, but
does not introduce any discretization error to the QMC algorithm itself.
With the Fourier transformation of the density $\mathcal{F}(\hat{\rho}_i(\tau)) =
\hat{\rho}_{k,\omega}$, we measure the correlation function
$\mathcal{F}(\braket{\hat{\rho}_i(\tau)\hat{\rho}_j(0)}) =
\braket{\hat{\rho}_{k,\om}\hat{\rho}_{-k,\om}}$ using
fast Fourier transforms.

The evaluation of off-diagonal single-particle correlation functions 
of the form $\langle\psi_i^\nag(\tau) \psi_0^\dag(0)\rangle$ 
requires some care. We again make use of the worldline picture, 
in which two operators $\psi^\dag$ and $\psi$ are inserted whenever a new loop update
starts. Let us assume that $\psi$ moves around (loop head) while
$\psi^\dagger$ is pinned (loop tail). The time and position of the loop tail
are set as the new origin of our coordinate system, and we store the values
$\langle\alpha | \psi_i^\nag(\tau) \psi_0^\dag(0) |\beta\rangle$ whenever the loop head
$\psi_i(\tau)$ crosses a point on the time grid with distance $(i,\tau)$ from the new
origin. Here $\ket{\alpha}$, $\ket{\beta}$ are the states in the world line
configuration prior to the arrival of the loop head.
We then again use fast Fourier transformation to evaluate the correlation functions in Fourier space.

Let us now define the observables of interest. The quantum phase transition can be detected
by calculating the superfluid density $\rho_\text{s}$, measured in the
simulations in terms of the spatial winding number $w$ as $\rho_\mathrm{s} =
L\las w^2\ras/\beta$,\cite{PhysRevB.36.8343,prokofev_two_2000} $\beta=1/kT$ being the inverse
temperature.
Another important observable in the context of the MI-SF transition is the
total density, $n=\las\hat{N}\ras/L$ in the Bose-Hubbard model, and
$n_\mathrm{p}=\las\hat{N}_\mathrm{p}\ras/L$ in the polariton model. 

Concerning dynamical properties, we compute the dynamic structure factor
$S(k,\om)$ and the single-particle spectral function $A(k,\om)$.
The dynamic structure factor at momentum $k$ and energy $\om$ is given by
\begin{eqnarray}
  S(k,\om)
  &=& 
  \frac{1}{2\pi L}\int_{-\infty}^\infty d \tau e^{\rmi \om \tau}
  \braket{\hat{\rho}_k(\tau){\hat{\rho}^{\dag}_k}(0)}
  \\\nonumber
  &=&
  \frac{1}{ L}
  \sum_{n,m}
  \frac{e^{-\beta E_n}}{Z} \left|\bra{m}{\hat{\rho}^{\dag}}_k\ket{n}\right|^2
  \delta[\om-(E_m-E_n)]
  \,,
\end{eqnarray}
with the grand-canonical partition function $Z$ and the energy of the $n$th
eigenstate $E_n$. In our simulations, $S(k,\om)$ is obtained from
\begin{equation}\label{eq:skw}
  \braket{\hat{\rho}_k(\tau) \hat{\rho}_{-k}(0)} 
  = 
  \int  d\omega S(k,\omega) 
  \frac{e^{-\tau \omega}}{1 + e^{-\omega \beta}}
\end{equation}
by means of the maximum entropy method.%\cite{MaxEnt}

For the Bose-Hubbard model, the density operator $\hat{\rho}_i=\hat{n}_i$, and
$\rho^\dag_k=\sum_q b^\dag_{q+k} b_q$. For the
polariton model, we can calculate the dynamic structure factor for photons
[$S^\text{ph}(k,\om)$], atoms [$S^\text{at}(k,\om)$] or polaritons
[$S(k,\om)$] by using 
\begin{equation}
\hat{\rho}_i =
\begin{cases}
 a^\dag_i a^\nag_i & \text{for\;photons}\,,\\
\ket{\UP_i}\bra{\UP_i} & \text{for\;atoms}\,,\\
 a^\dag_i a^\nag_i + \ket{\UP_i}\bra{\UP_i} & \text{for\;polaritons}\,,
\end{cases}
\end{equation}
respectively.

The single-particle spectral function is defined as
\begin{eqnarray}
  A(k,\om) 
  &=&
  -\frac{1}{\pi} \Im\, \las\las \hat{\psi}^\nag_k;\hat{\psi}_k^\dag\ras\ras_\om
  \\\nonumber
  &=&
  \sum_{n,m}
  \frac{e^{-\beta E_n}}{Z} \left|\bra{m}{\hat{\psi}^{\dag}}_k\ket{n}\right|^2
  \delta[\om-(E_m-E_n)]
  \,,
\end{eqnarray}
where the real-space operator $\hat{\psi}_i$ entering the Green function is given
by $\hat{\psi}_i=b_i$ for the Bose-Hubbard model, and by $\hat{\psi}_i=a_i$
for the polariton model.
maximum entropy is again used to map to real frequencies.

The QMC algorithm samples the partition function in the grand canonical
ensemble. However, using only those configurations which have a given number
of polaritons enables us to measure observables in the canonical ensemble as
well. Here this simple but powerful trick permits us to study the
fixed-density phase transition which occurs in the polariton model as a function of $t/g$.

The SSE representation requires to set a maximum boson number per site.
In the Bose-Hubbard model, we allow a maximum of six bosons per site.
In the polariton model we allow from six (Mott insulator, fixed-density transition) up to
16 (SF phase) photons per site. Convergence has been monitored by plotting
histograms of the photon number distribution, and the cut-offs have been
chosen generously such that there was no truncation error.

\section{Results}\label{sec:results}

We begin with a review of the properties of the Bose-Hubbard model and the
polariton model as they emerge
from previous work. Whereas a substantial literature exists for the Bose-Hubbard model, work
on the polariton model began only recently, based on mean-field theory,\cite{GrTaCoHo06,Ko.LH.09}
exact diagonalization,\cite{Ma.Co.Ta.Ho.Gr.07} the DMRG,\cite{Ro.Fa.07} the
variational cluster approach,\cite{Ai.Ho.Ta.Li.08}, QMC\cite{Zh.Sa.Ue.08} and
strong coupling theory.\cite{Sc.Bl.09}
Our discussion focuses on 1D, and follows Fisher \etal\cite{PhysRevB.40.546}
and K\"uhner \etal\cite{Ku.Wh.Mo.00}

The {\em Bose-Hubbard model} describes the competition of kinetic energy and
local, repulsive interaction. Depending on the ratio $t/U$ and the density of
bosons $n$ (the system is superfluid for any $t>0$ if $n$ is not integer),
the Bose-Hubbard model at temperature $T=0$ is either in a MI state or in a SF state. The MI
is characterized by an integer particle density,
phase fluctuations and a gap in the single-particle excitation spectrum. In
the SF phase, we have significant density fluctuations, phase coherence, and
nonzero superfluid density $\rho_\text{s}$, as well as gapless (phonon)
excitations with linear dispersion at small $k$.

\begin{figure} % 1
  \includegraphics[width=0.4\textwidth]{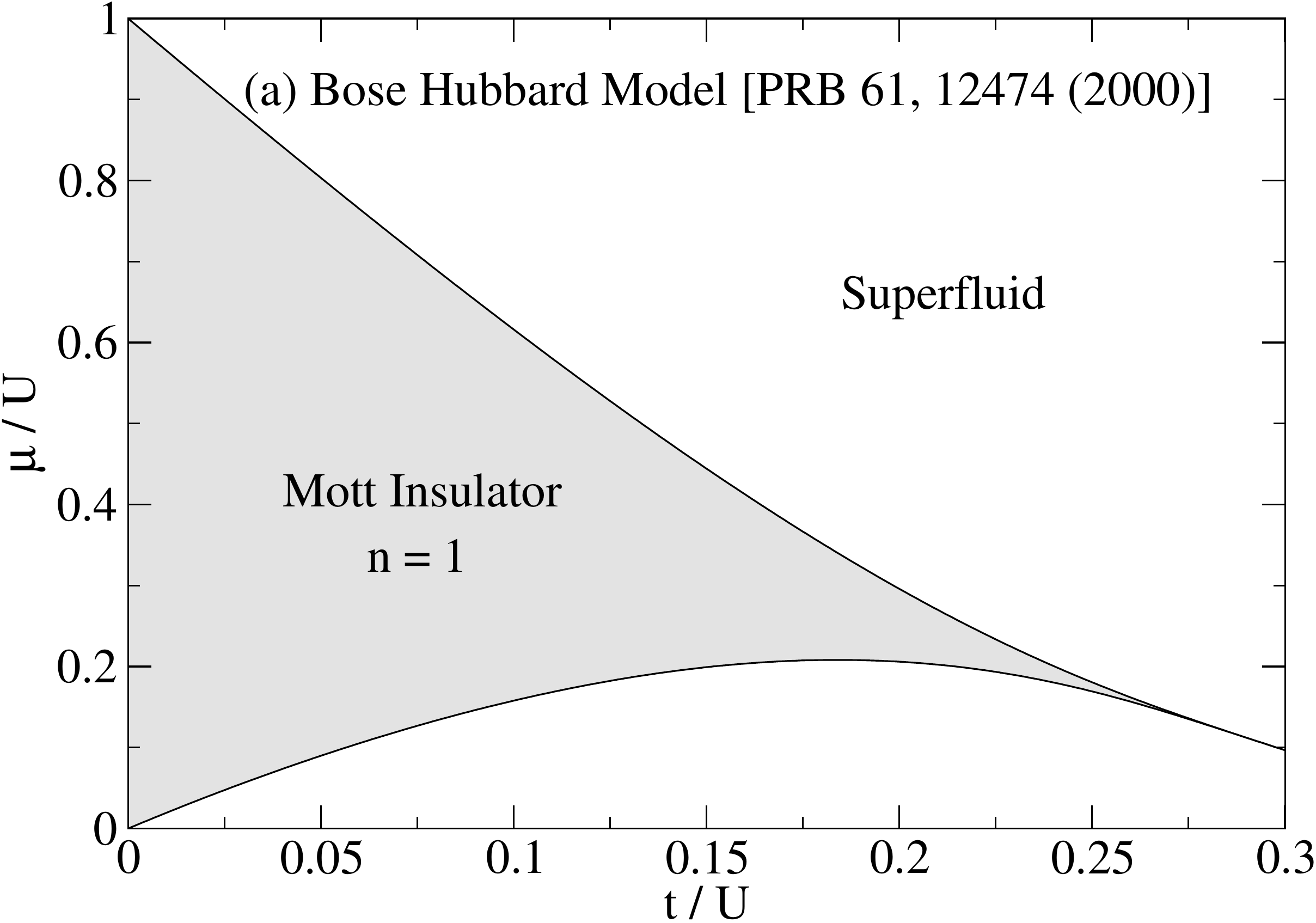}\\
  \includegraphics[width=0.4\textwidth]{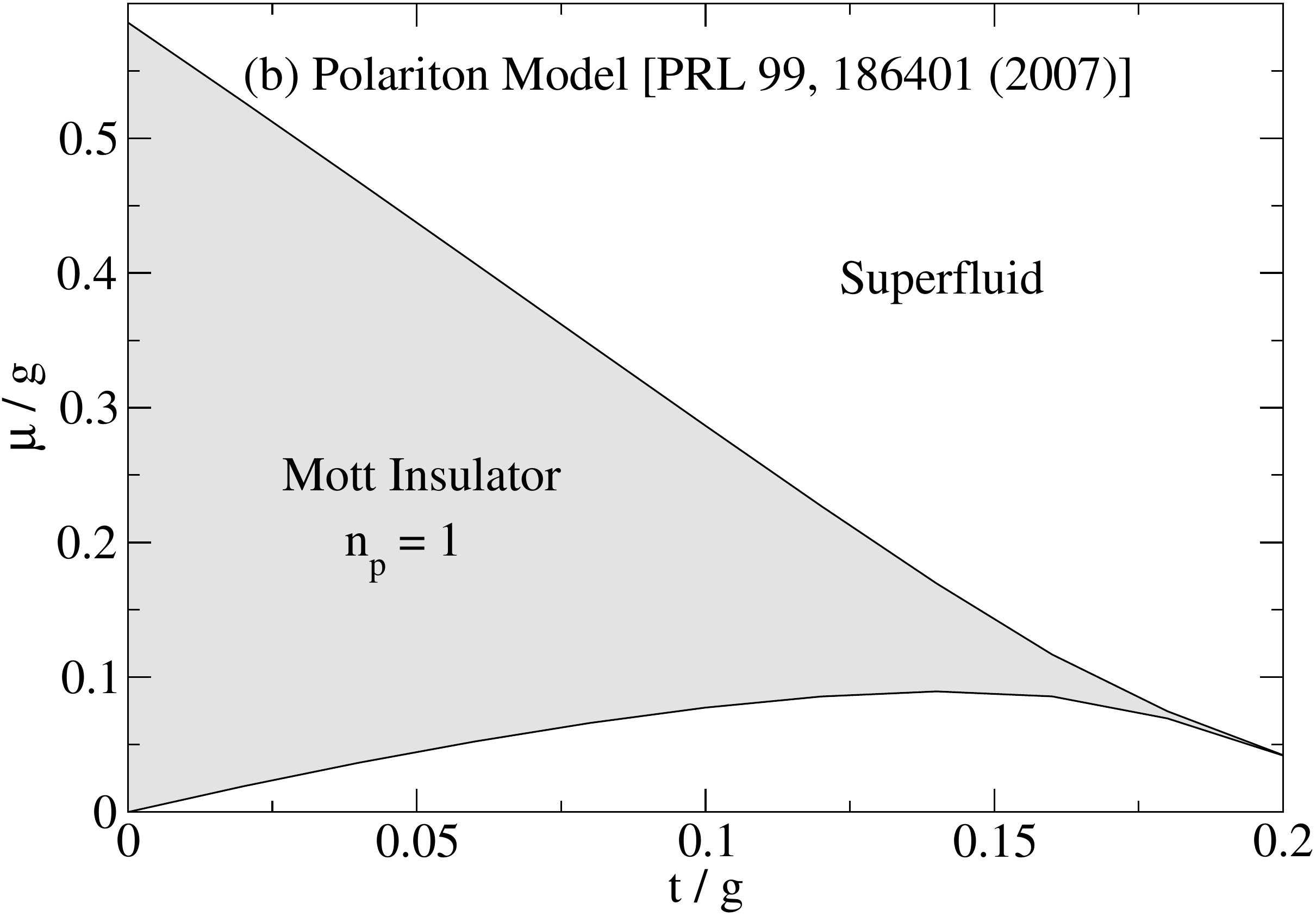} 
\caption{\label{fig:phasediagrams}
  Zero-temperature phase diagram for (a) the Bose-Hubbard model and (b) the polariton
  model in 1D. We only show the Mott lobes with density one. These DMRG results
  were obtained by (a) K\"uhner \etal\cite{Ku.Wh.Mo.00} and (b) Rossini
  \etal\cite{Ro.Fa.07}}
\end{figure}

For the case of one dimension considered here, a precise zero-temperature
phase diagram in the $\mu/U,t/U$ plane has been determined by K\"uhner
\etal,\cite{Ku.Wh.Mo.00} and these data are shown in
Fig.~\ref{fig:phasediagrams}(a). There exists a Mott lobe inside which the
density $n=1$ (higher lobes with integer $n>1$ are not shown), and which is
surrounded by the SF phase.

There are two qualitatively different ways to make a transition from the MI
to the SF.\cite{PhysRevB.40.546} The generic MI-SF transition is driven by
addition or subtraction of small numbers of particles to the incompressible
MI phase, the total energy cost for which is given by the distance in
$\mu$-direction from the nearest phase boundary. Since additional particles
or holes (which Bose condense at $T=0$) can move freely, the gain in kinetic
energy can outweigh the interaction energy, leading to the MI-SF
transition. Across the generic transition, which is mean-field like in
character, the density varies continuously and the single-particle gap closes
linearly as a function of the distance from the phase boundary,
$E_\text{g}\propto\delta$, where $\delta=t-t_\text{c}$ or $\mu-\mu_\text{c}$
is the distance from the phase boundary.\cite{PhysRevB.40.546}

There also exists a MI-SF transition at fixed density, driven by the onset
of boson hopping due to the increase of the ratio $t/U$, \ie by quantum fluctuations. It has been shown
that this transition occurs at the tip of the Mott lobe, and that it has a
different universality class than the generic transition.\cite{PhysRevB.40.546}
In $d$ dimensions, the universality class is that of the $(d+1)$ dimensional
$XY$ model, so that in 1D there is a Kosterlitz-Thouless phase transition
at the multicritical point. For this case, the Mott gap
$E_\text{g}\propto\exp(\text{-const}/\sqrt{t_\mathrm{c}-t})$ closes
exponentially (\ie, very slowly) as a function of the distance from the lobe
tip,\cite{PhysRevB.59.12184} and strong
deviations from the parabolic lobes predicted by mean-field
theory\cite{PhysRevB.40.546} are observed in both
strong-coupling\cite{0295-5075-26-7-012,Fr.Mo.96,PhysRevB.59.12184} and DMRG
results.\cite{PhysRevB.58.R14741} Another remarkable aspect of the 1D case is
the occurrence of multiple MI-SF transitions along lines of constant chemical
potential over an extended range of $\mu\lesssim0.2$ [see
Fig.~\ref{fig:phasediagrams}(a)].\cite{PhysRevB.58.R14741,Ku.Wh.Mo.00}

The {\em polariton model} also shows a series of Mott lobes, in
which the polariton density $n_\text{p}$ is pinned to an integer 
(see Fig.~\ref{fig:phasediagrams}(b) for the phase boundaries of the
$n_\text{p}=1$ lobe obtained by DMRG\cite{Ro.Fa.07}). Even for pinned $n_\text{p}$ the
photon and exciton densities can fluctuate. Deep in the Mott phase and for $n_\text{p}\geq1$, we can
approximate the ground state by a product over single sites, each of which
is described by the Jaynes-Cummings eigenstates (see, \eg, [\onlinecite{Sc.Bl.09}])  
\begin{eqnarray}\nonumber\label{eq:eigenstates}
\ket{n_\text{p},-}
&=&
\cos\theta(n_\text{p})\ket{n_\text{p},\DO}-\sin\theta(n_\text{p})\ket{n_\text{p}-1,\UP}\,,\\
\ket{n_\text{p},+}
&=&
\sin\theta(n_\text{p})\ket{n_\text{p},\DO}+\cos\theta(n_\text{p})\ket{n_\text{p}-1,\UP}\,,
\end{eqnarray}
where $\tan\theta(n_\text{p})=2g\sqrt{n_\text{p}}/[2\chi(n_\text{p})-\Delta]$,
$\chi(n_\text{p})=\sqrt{g^2n_\text{p}+\Delta^2/4}$, and with eigenvalues
$E^\pm(n_\text{p})=-(\mu-\om_0)n_\text{p}+\Delta/2\pm\chi(n_\text{p})$. Hence for
fixed polariton number $n_\text{p}$, the ground state $\ket{n_\text{p},-}$ is
a coherent superposition of two states which differ by the state of the atom
(or spin) as well as the number of photons; this hybridization provides the
connection to exciton polaritons.

The extent of the lobes in both the
$\mu$ and $t$ directions diminishes quickly with increasing $n_\text{p}$ due to
the reduced polariton-polariton repulsion $U_\mathrm{eff}(n_\mathrm{p})$;
the $t=0$ vertical width of the lobes in the Bose-Hubbard model is always
$U$. At large values $\zeta t>\om-\mu$ ($\zeta$ being the coordination number), beyond
those considered in the present work, the polariton model shows an instability.\cite{Ko.LH.09}

In this work we restrict our discussion to the region in the phase diagram in
or close to the Mott lobes with density $n_\mathrm{p}=1$ or $n=1$. This lobe
is the largest in the polariton model with zero detuning, and quantum effects are most
pronounced. A density of one is also the most interesting case for
experimental realizations.\cite{GrTaCoHo06,Ai.Ho.Ta.Li.08}

All the discussion so far has been for $T=0$. Both experiments and our
simulations are carried out at low but finite temperatures, with several
important consequences. Strictly speaking, there is no true MI at $T>0$ due
to thermal excitations. However, there exist quasi-MI regions which have
finite but very small compressibility (see also the discussion of temperature
effects later). As long as the density remains close to an integer, these
regions may be regarded as Mott insulating. Corresponding ``phase diagrams''
at finite $T$ have been obtained for both the polariton and the Bose-Hubbard
model.\cite{gerbier:120405,Ai.Ho.Ta.Li.08,Oh.Pe.08} Except for our analysis
of temperature effects in Sec.~\ref{sec:results}, the simulations have been
carried out at values of $\beta=3L$, large enough to ensure that we have an
(almost) integer density in the Mott phase.

The Bose-Hubbard model in more than one dimension (and most likely the
polariton model as well) exhibits
a phase transition from a SF to a normal state (gapless with no phase
coherence), related to the well-known $\lambda$ transition in liquid helium,
at a temperature
$T_\lambda$.\cite{CaSa.GuSo.Pr.Sv.07,capogrosso-sansone:134302} This gives
rise to an intervening normal region in the phase diagram, between the MI (at
small $t/U$) and the SF (at large $t/U$).\cite{gerbier:120405} In the 1D case
considered here we have $T_\lambda=0$, so that for any $T\neq0$ only quasi-MI
and normal states exist in the thermodynamic limit. However, when the temperature
is so low that the SF correlation length in the thermodynamic limit far
exceeds the system size $L$, results will be representative of the SF state.
Making use of finite size and finite temperature effects, a scaling analysis
in fact yields accurate results for the $T=0$ phase
boundaries.\cite{alet:024513,Zh.Sa.Ue.08} Remarkably, interacting 1D bosons
can be realized using cold atomic gases (the Tonks-Girardeau
gas)\cite{Pa.Wi.Mu.Va.Ma.Fo.Ci.Sh.Ha.Bl.04,Fa.Cl.Fa.Fo.Mo.vdS.In.09} and are
described by the Bose-Hubbard model at low but finite
temperatures.\cite{PhysRevLett.93.210401}

Similar to Bose fluids, the low-energy excitations in the SF phase are
phonons. Within Bogoliubov theory,\cite{Bogolyubov_superfluidity_1947} these
quasiparticles are described by a creation operator $\psi^\dag_k =
\mathsf{u}_k b^\dag_k + \mathsf{v}_k b^\nag_{-k}$, and they have been
observed experimentally in ultracold atom
systems.\cite{PhysRevLett.88.060402} As some of our results can be understood
in terms of Bogoliubov theory, let us state some key results for the Bose-Hubbard model.
The coefficients of the coherent superpositions of particle and hole
excitations are given by\cite{rey_bogoliubov_2003}
\begin{equation}
  \label{eq:boson_weights}
\begin{split}
  |\mathsf{u}_k|^2 = & \frac{K(k)+n_0 U + \omega_k}{2 \omega_k}\,, \\
  |\mathsf{v}_k|^2 = & \frac{K(k)+n_0 U - \omega_k}{2 \omega_k} = |\mathsf{u}_k|^2-1\,,
\end{split}
\end{equation}
with excitation energy
\begin{align}
  \label{eq:boson_energies}
  \omega_k & = \sqrt{K(k)(2 n_0 U + K(k))}\,, \\
  K(k) & = 4 t \sin^2(k/2)\,. \nonumber
\end{align}
Here $n_0$ is the condensate fraction, equal to $n_0 = (\mu+t)/U$ in the
simple Bogoliubov approach at $T=0$.\cite{Me.Tr.08} For small $k\approx0$, we
have a linear dispersion $\omega_k \approx \pm \sqrt{2 n_0 t U} k$, and both
$|\mathsf{u}_k|$ and $|\mathsf{v}_k|$ are nonzero. For large $k\approx\pi$,
the energy dispersion is $\pm(-c k^2 + 2 \sqrt{4 t^2+2 n_0 U t})$ and thus free
particle like. If we assume $t \gg U$, which is the parameter region where
Bogoliubov theory is valid, then $|\mathsf{u}_k|^2 \approx 1$ and
$|\mathsf{v}_k|^2 \approx 0$ for $k\gg0$, \ie only one excitation branch is
populated at large momenta. This also holds true for the parameters studied
numerically in this work. We further compare to the higher order approximation proposed in
Ref.~\onlinecite{rey_bogoliubov_2003}. For the Bose-Hubbard model, the latter
yields the same equations for $|\mathsf{u}_k|^2$, $|\mathsf{v}_k|^2$ and
$\om_k$, but $n_0$ is determined self-consistently, allowing for depletion
effects.

In the case of free bosons at $T=0$, all particles condense in the same $k=0$
state. However, finite temperature and/or interactions cause a certain
fraction of these particles to occupy states of higher energy. Indeed, both
for $U\rightarrow0$ (noninteracting bosons) and $n_0\rightarrow0$ (high
temperature limit) we have $|\mathsf{u}_k|^2=1$, $|\mathsf{v}_k|^2=0$.
Moreover, with decreasing $U$ or $n_0$, $|\mathsf{v}_k|^2$ approaches zero
most quickly at large $k$ since in this case $K(k)\gg n_0U$ so that
$\omega_k\approx K(k)$, canceling the term $-\om_k$ in the expression for
$|\mathsf{v}_k|^2$. This will explain the temperature evolution of the
single-particle spectrum shown in Sec.~\ref{sec:results}.

\subsection{Bose-Hubbard model}

\begin{figure} % 2
\subfigure{
  \includegraphics[width=0.46\linewidth]{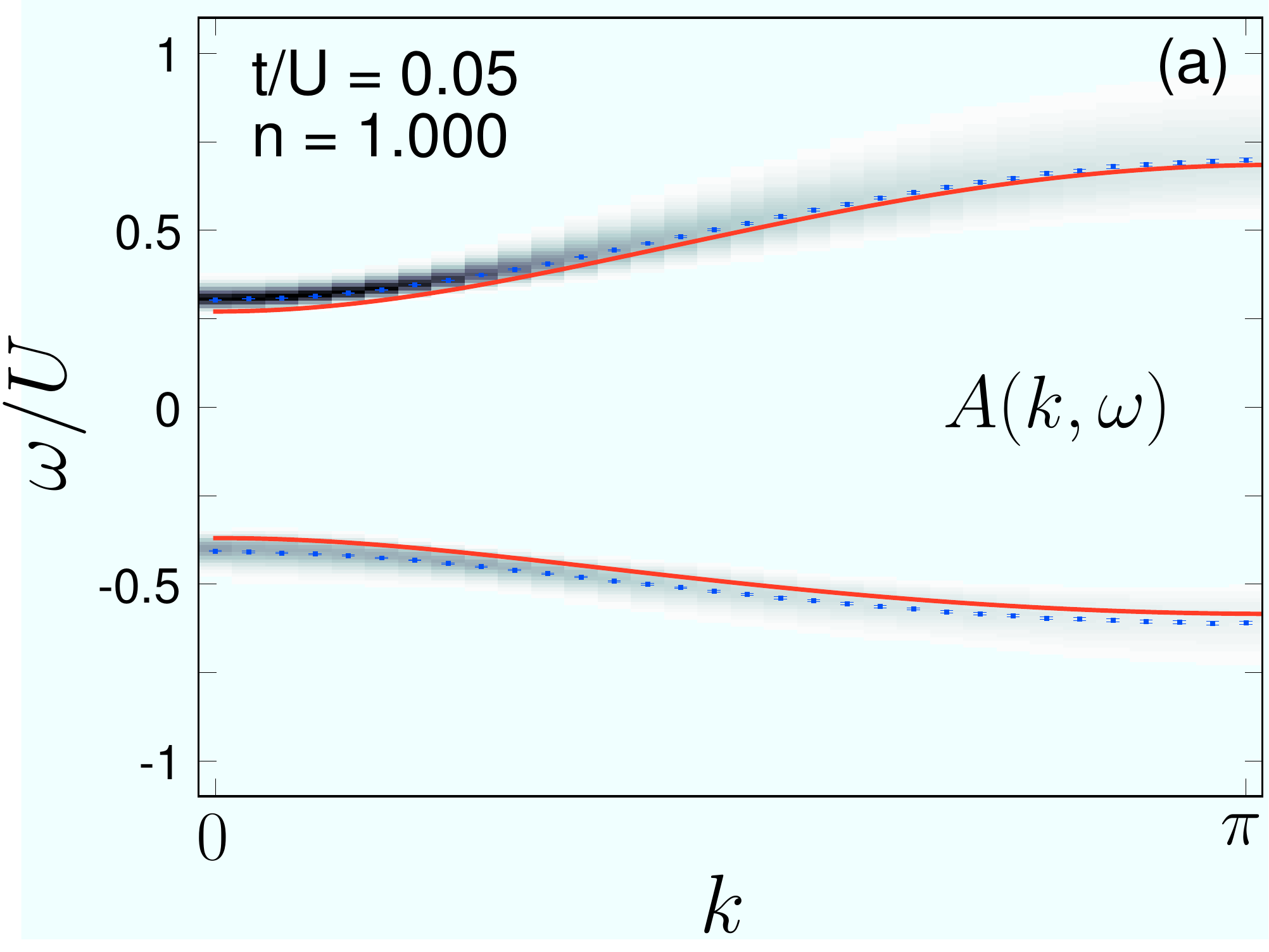}
  }
\subfigure{
  \includegraphics[width=0.46\linewidth]{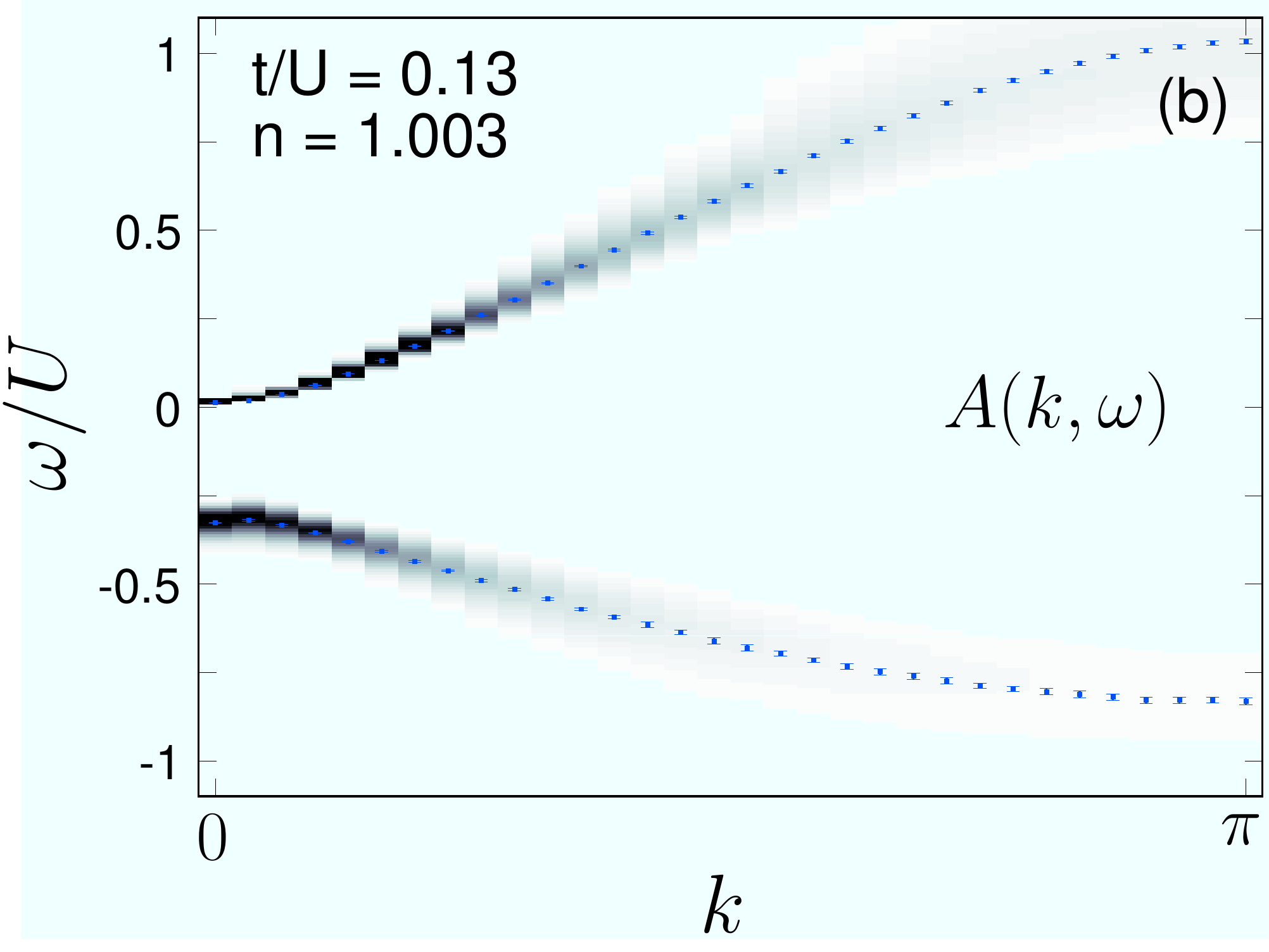}
  }
\subfigure{
  \includegraphics[width=0.46\linewidth]{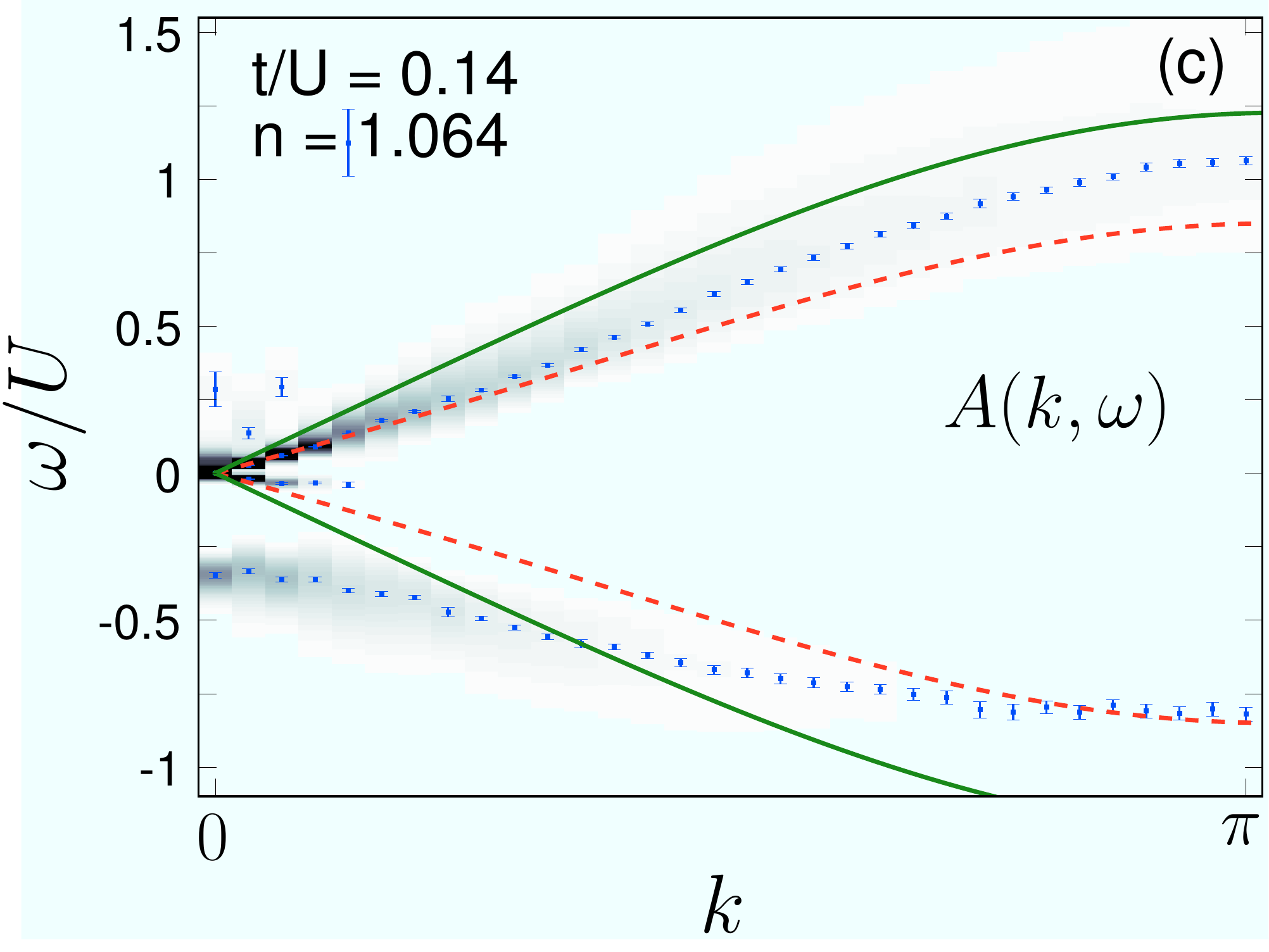}
}
\subfigure{
  \includegraphics[width=0.46\linewidth]{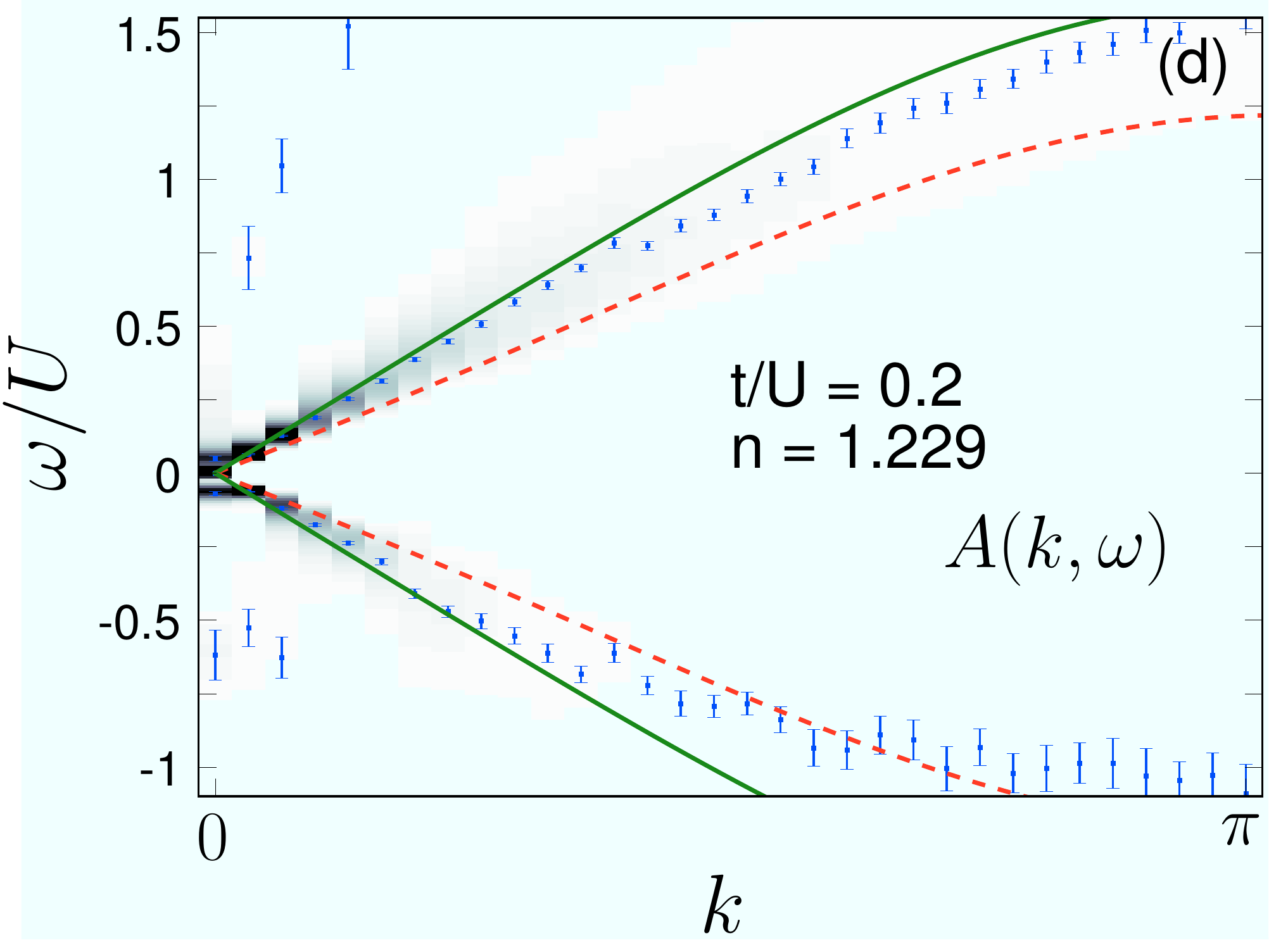}
}
\caption{\label{fig:BH_green} (color online)
  Single-boson spectral function $A(k,\om)$ of the 1D Bose-Hubbard model, for different
  hoppings $t$ (and total density $n$), corresponding to (a) the MI phase, (b) just below the
  MI-SF transition, (c) just above the transition, and (d) the SF
  phase. Here $\mu/U = 0.5$, $L = 64$ and $\beta U = 3L$.
  Here and in
  subsequent spectra, the symbols and errorbars indicate the maxima of the
  peaks and the associated errors obtained by the maximum entropy method.
  As discussed in Sec.~\ref{sec:bhm_skw}, features with very small spectral
  weight are difficult to determine accurately.
  The solid red lines in (a) are  mean field results.\cite{PhysRevA.63.053601}
  The solid  lines in (c) and (d) are the Bogoliubov results, while the dashed
  lines are a fourth order approximation (see
  text).\cite{rey_bogoliubov_2003} 
}
\end{figure}

Despite the extensive literature on this model, there are few nonperturbative
results available for the spectra, as mentioned in Sec.~\ref{sec:introduction}.
Therefore, we investigate the
single-boson spectral function $A(k,\om)$ and the dynamic structure factor
$S(k,\om)$, with results shown in Figs.~\ref{fig:BH_green} and \ref{fig:BH_szsz}.

%\begin{figure}[Htb]
%  \centering
%  \includegraphics[width=0.95\linewidth]{bh_sound_velocity_mu0_5}
%  \caption{Sound velocity $v_\text{s}$ of the phonon mode in the
%    1D Bose Hubbard model along the line of constant $\mu/U=0.5$. We evaluated $v_\text{s}$
%    for different system sizes $L$ and time steps $\Delta \tau$. The (generic) phase
%    transition occurs at $t_\text{c}/U\approx0.14$. The solid line is the prediction
%    from Bogoliubov theory ($n_0=(\mu+t)/U$) which agrees with RPA,\cite{Me.Tr.08} while the dashed and
%    dot-dashed lines are obtained by using the particle density $n$
%    respectively the superfluid density $\rho_\mathrm{s}$ from QMC.}
%  \label{fig:sound_velocity_BH}
%\end{figure}

\begin{figure}[Htb] % 3
  \centering
  \includegraphics[width=0.95\linewidth]{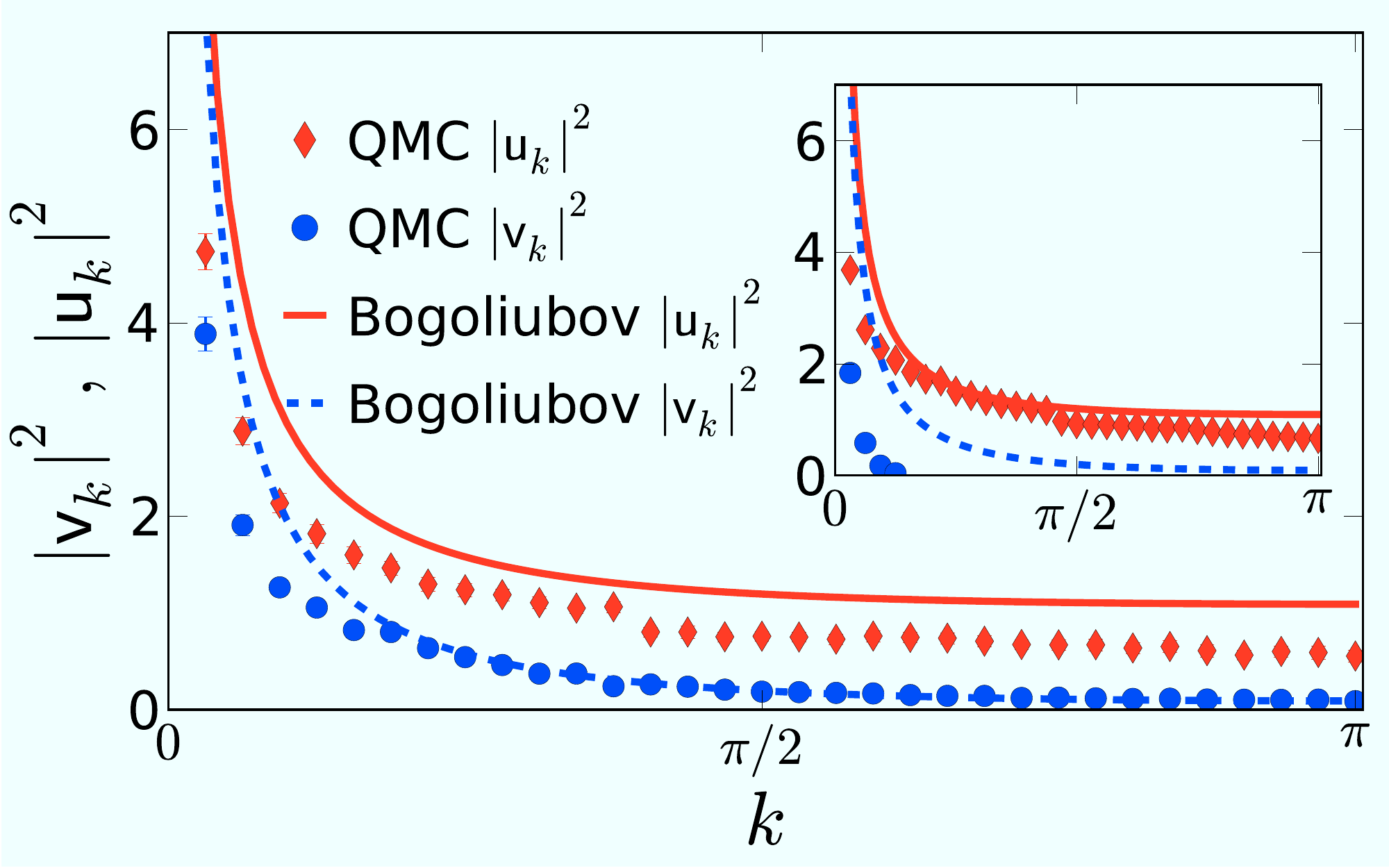}
  \caption{(color online) Quasiparticle weights $\mathsf{u}_k$ and $\mathsf{v}_k$ of the
    gapless modes at $t/U=0.2$.  The symbols are integrated intensities from
    QMC and maximum entropy, the lines are the predictions from
    Bogoliubov-theory.  The inset shows data at $t/U=0.14$.  Again,
    $\mu=0.5$, $L=64$ and $\beta = 3 L$.  }
  \label{fig:quasiparticle_weight_BH}
\end{figure}

\subsubsection{Single-particle spectrum}

Menotti and Trivedi reviewed previous work on the single-particle spectrum, and
presented results from a random phase approximation.\cite{Me.Tr.08} Their main
findings are as follows. For large $t/U$, a weakly interacting SF exists, and
the spectrum consists of the usual two gapless phonon modes which exhaust the
sum rule for $A(k,\om)$. Reducing $t/U$, two additional gapped modes appear
at small $k$ whose spectral weight increases upon approaching the quantum
phase transition. At the transition, one of the phonon modes evolves into the particle or hole mode
(depending on which of the gaps $E_\text{g,p}$, $E_\text{g,h}$ is smaller),
whereas one of the gapped modes in the SF becomes a gapped mode in the MI.
Menotti and Trivedi\cite{Me.Tr.08} argued that the appearance of gapped modes
and the redistribution of spectral weight from coherent phonon modes to
incoherent gapped modes indicate the strongly correlated nature of the SF
state near the transition. Let us point out that particle and hole dispersions in
the MI have been calculated by several authors before,
\cite{Me.Tr.08,PhysRevB.59.12184,Oh.Pe.08,PhysRevA.63.053601,lu:043607,Hu.Al.Bu.Bl.07,0295-5075-22-4-004,Se.Du.05,0953-4075-40-1-013,KoDu06}
whereas the full spectral function of the MI (which also reveals the spectral
weight and the width of the excitations) was only shown in [\onlinecite{KoDu06,Me.Tr.08}].

Our numerical results for the single-particle spectral function $A(k,\om)$
are shown in Fig.~\ref{fig:BH_green}.  The four different values of the ratio
$t/U$ cover the range in which the generic MI-SF transition takes place.
According to Fig.~\ref{fig:phasediagrams}(a), for the chosen value of
$\mu/U=0.5$ the transition occurs at $t/U\approx0.14$. In each panel we also
report the total density $n$ to three decimal places, although our
simulations provide much higher accuracy. The MI [(a) and (b)] exhibits the familiar gapped
particle and hole bands.\cite{PhysRevB.59.12184}  The additional particles exhibit a free-particle
dispersion since the energy penalty for double occupation is the same at
every site. In particular, we see in Fig.~\ref{fig:BH_green}(a), (b) that the
particle band width is $8t$ (the factor of two arising from the fact that
particle hopping involves a doubly occupied site), whereas the hole bandwidth
is $4t$. The Mott gap decreases with increasing $t$ and a symmetry of
particle and hole bands
emerges.\cite{PhysRevB.40.546,Ai.Ho.Ta.Li.08,CaSa.GuSo.Pr.Sv.07} In addition
to our QMC results we plot the mean-field
dispersion~\cite{PhysRevA.63.053601} in Fig.~\ref{fig:BH_green}(a). For
larger $t/U=0.13$, mean-field theory already predicts a superfluid, although
the critical hopping in 1D is $t_\text{c}/U\approx 0.14$.

In the SF phase [Fig.~\ref{fig:BH_green}(c),(d)], we obtain the expected
Goldstone modes with linear dispersion at small $k$. Additionally, we see two
gapped signals which we relate to the gapped modes discussed by other
authors.\cite{Se.Du.05,Hu.Al.Bu.Bl.07,Me.Tr.08} Whereas the negative-energy
gapped mode is clearly visible in Fig.~\ref{fig:BH_green}(c) just above
$t_\mathrm{c}$, the gapped modes have almost disappeared in
Fig.~\ref{fig:BH_green}(d). Since we approach the phase
transition above the lobe tip ($\mu/U = 0.5$) the particle band becomes
the gapless mode and carries more spectral weight, while the gapped hole band
evolves into a gapped mode in the SF. This agrees
well with the findings of Menotti and Trivedi.\cite{Me.Tr.08}  In
accordance with Bogoliubov theory, the excitations in the SF phase are
free-particle like for large $k$.
The bandwidths of the excitations both in the MI and the SF
phase scale roughly linearly with $t$. 

In Figs.~\ref{fig:BH_green}(c) and (d) we also show results for the phonon
dispersion $\pm\om_k$ (without taking into account the weights $|\mathsf{u}_k|$, $|\mathsf{v}_k|$)
from Bogoliubov theory as well as the higher-order approximation of
Ref.~\onlinecite{rey_bogoliubov_2003}. Whereas the simple Bogoliubov approach
(neglecting depletion of the condensate) agrees quite well with our data
despite the rather small value of $t/U$, we do not find the higher order approach to
be systematically better. In particular, at large $k$, the phonon bandwidth
is noticeably underestimated, which may be a result of an overestimate of
depletion effects (these are most visible at large $k$). The agreement with
Bogoliubov theory at small $k$ coincides with the findings of Menotti and
Trivedi.\cite{Me.Tr.08} Rey \etal\cite{rey_bogoliubov_2003} found the higher
order approximation to be consistent with numerical results for other
observables but do not show the spectra seen in
Fig.~\ref{fig:BH_green}. Note that these authors consider larger
particle densities $n\geq5$ where the Bogoliubov-type approximations are more reliable.
Finally, we tried to use our QMC results for the superfluid fraction for
$n_0$ in the expressions obtained from Bogoliubov theory, but the results are
worse than for $n_0=n$. 

The spectral weight of the excitations decreases with increasing $k$ in all
spectra of Fig.~\ref{fig:BH_green}, although this is more pronounced in the
SF phase than in the MI.
In Fig.~\ref{fig:quasiparticle_weight_BH} we show the quasiparticle weights
of the massless modes in the SF phase, obtained by integrating over the
quasiparticles peaks in the spectra, and compare them to Bogoliubov theory
(Eqs.~\ref{eq:boson_weights} and~\ref{eq:boson_energies}).  We verified that
the QMC spectra satisfy the sum rule.  The spectral weight of the
lower branch decreases more quickly, consistent with the Bogoliubov picture.
However, Bogoliubov theory overestimates the quasiparticle weights,
especially at small $k$.  Besides, there is a significant broadening of the
peaks on approaching the zone boundary.  At strong coupling close to the
phase transition (inset of Fig.~\ref{fig:quasiparticle_weight_BH}), the
quasiparticle weight of the lower branch decays much more quickly than
Bogoliubov theory would predict.

\begin{figure} % 4
\subfigure{
  \includegraphics[width=0.46\linewidth]{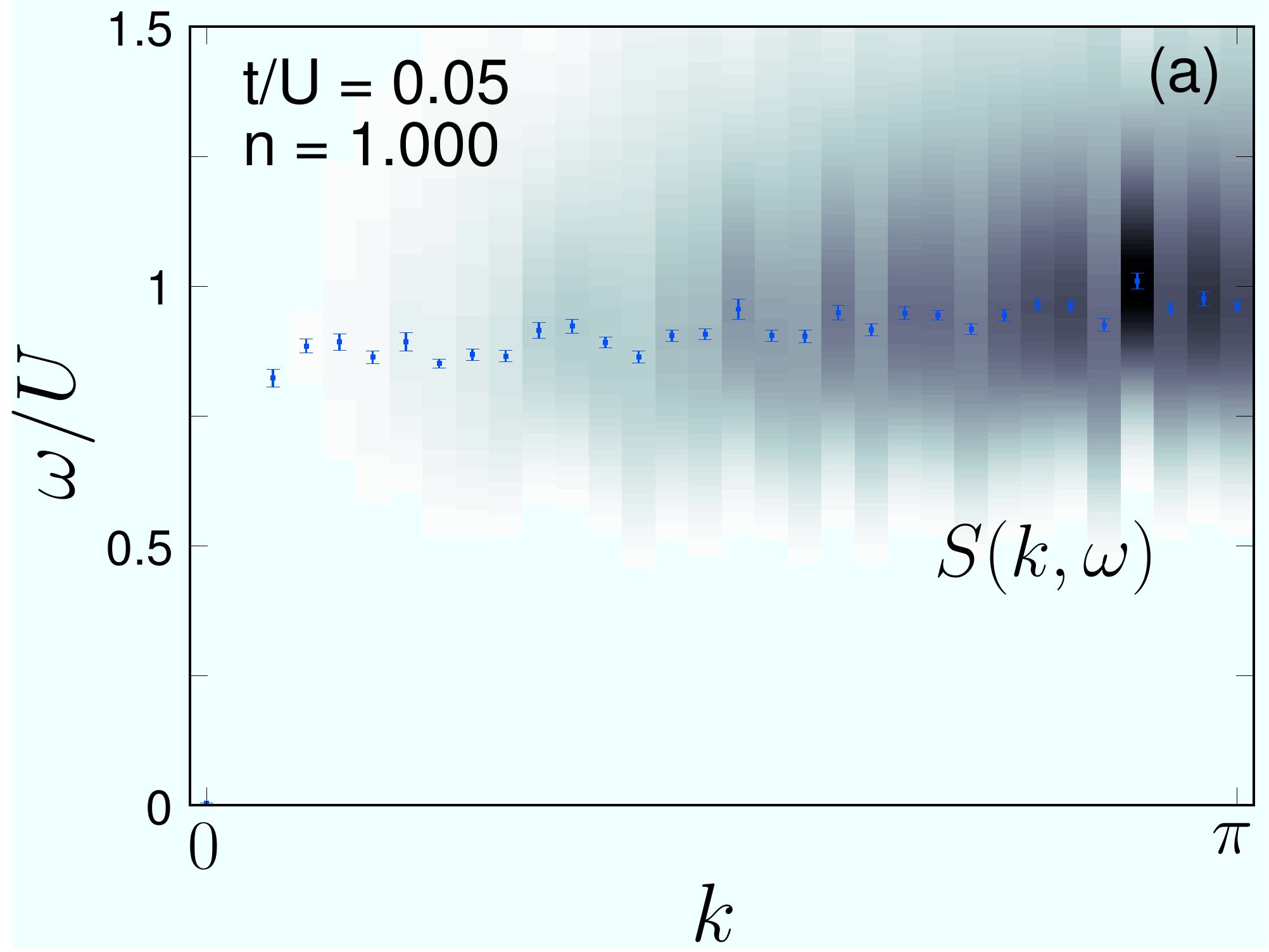}
  }
\subfigure{
  \includegraphics[width=0.46\linewidth]{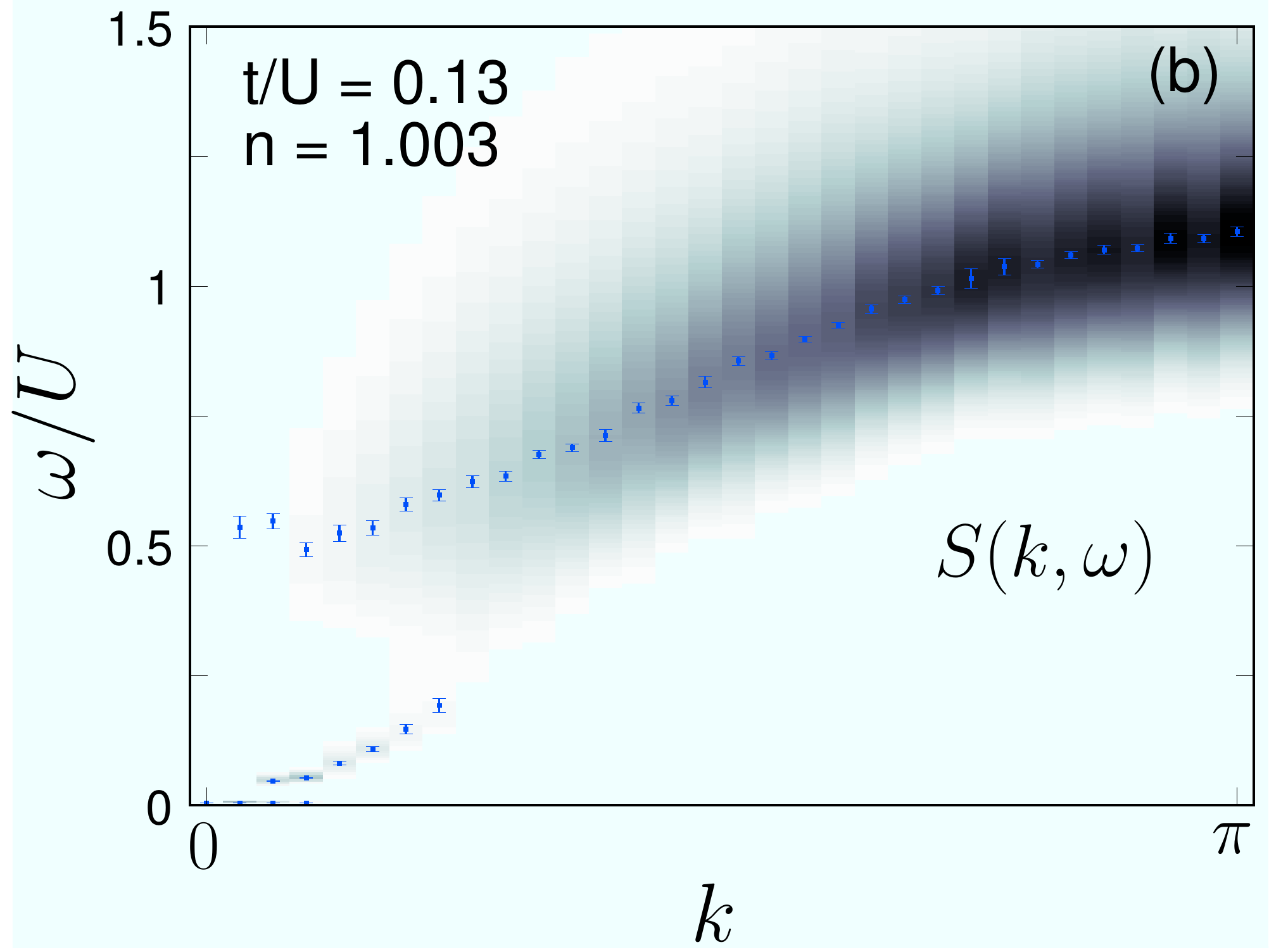}
  }
\subfigure{
  \includegraphics[width=0.46\linewidth]{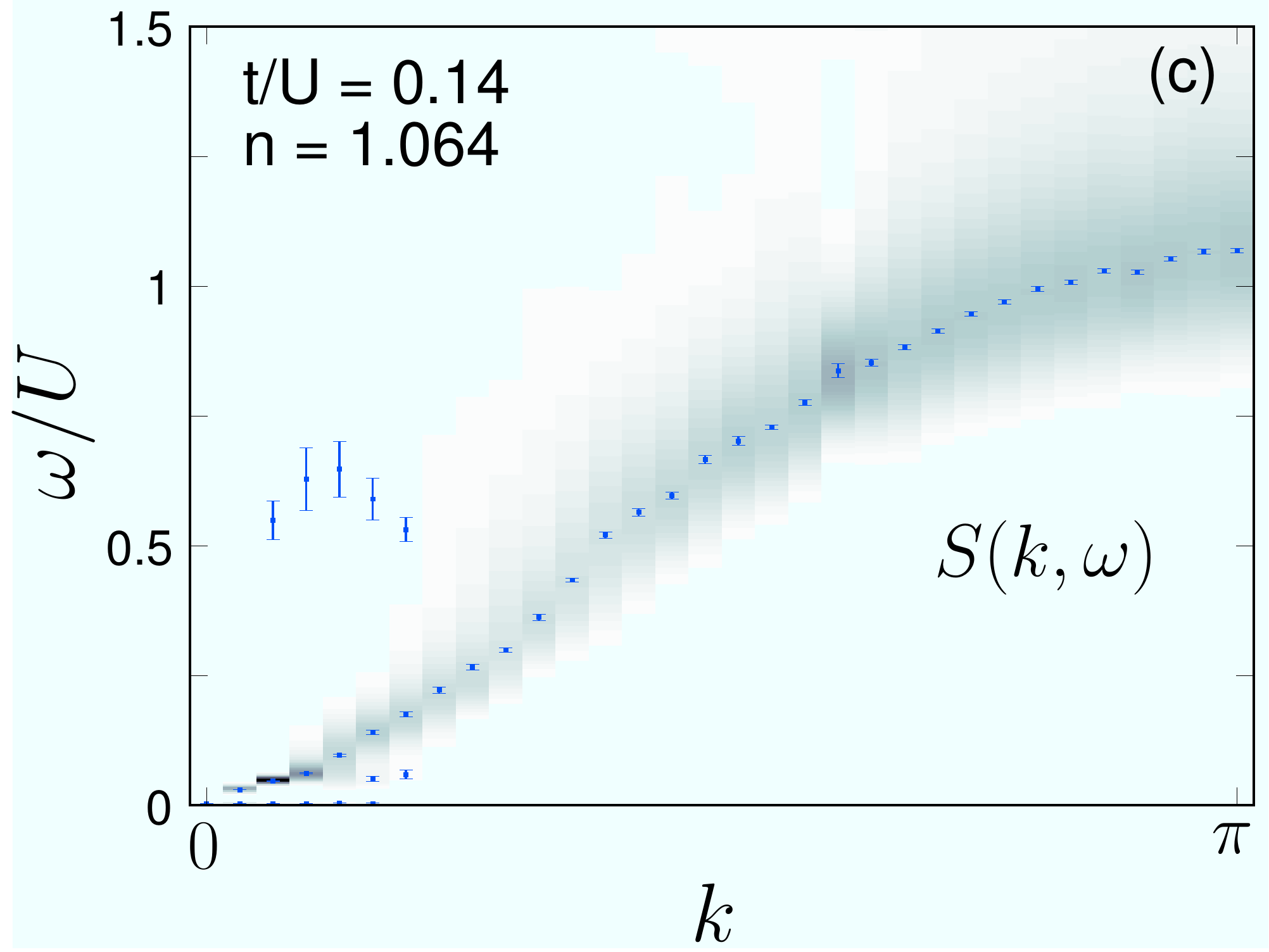}
}
\subfigure{
  \includegraphics[width=0.46\linewidth]{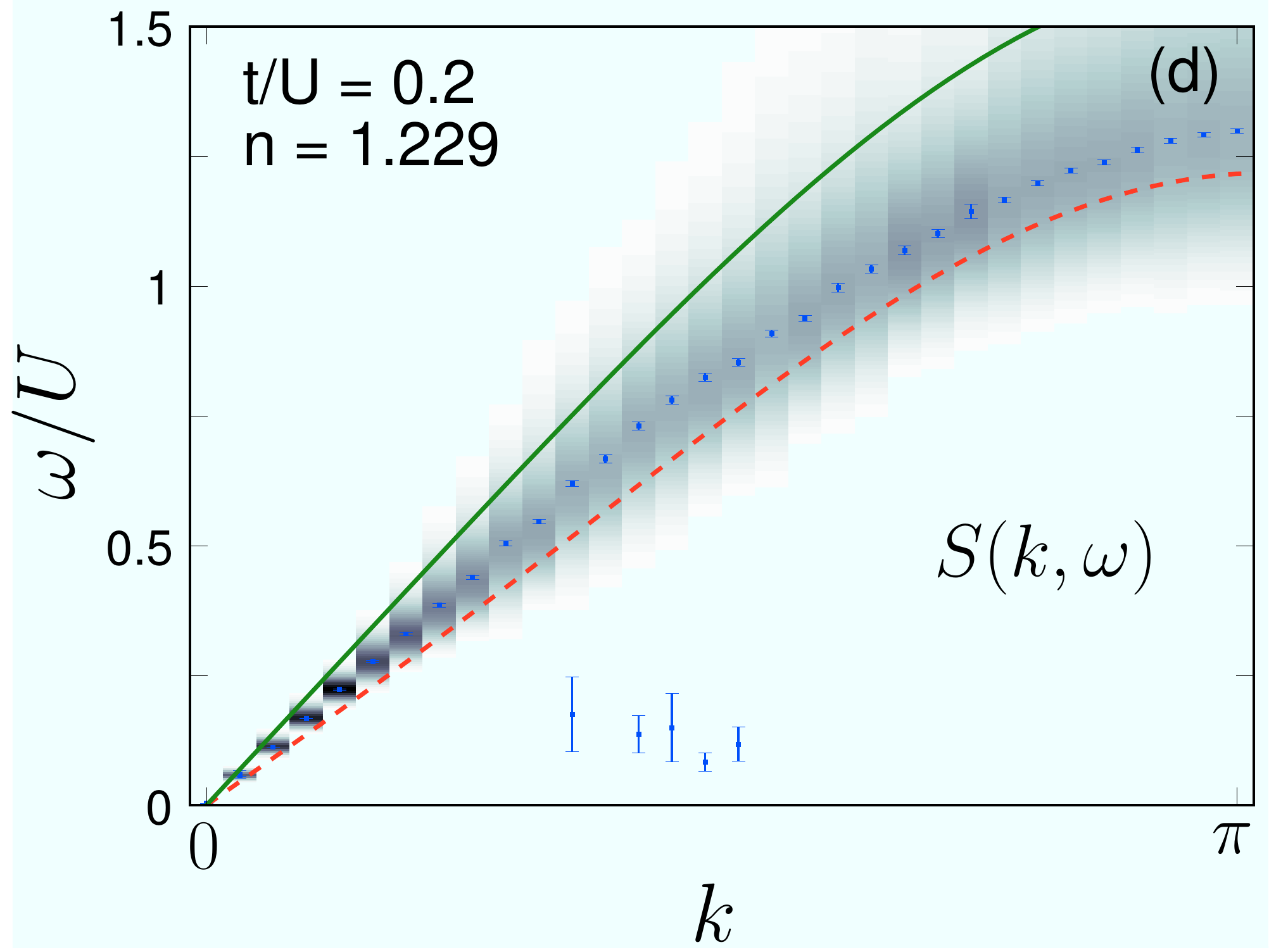}
}
\caption{\label{fig:BH_szsz} (color online)
  Dynamic structure factor  $S(k,\om)$ of the Bose-Hubbard model for
  the same parameters as in Fig.~\ref{fig:BH_green}.  Panel (d) includes
  the same analytical approximations as Fig.~\ref{fig:BH_green}(d).
  }
\end{figure}

{\em Sound velocity.}
The sound velocity $v_\text{s} = \frac{\partial \om_k}{\partial k}|_{k \to
0}$ of the phonon excitations in the SF phase was calculated for the
Bose-Hubbard model by Menotti and Trivedi using a random phase
approximation.\cite{Me.Tr.08} They concluded that $v_\text{s}$ vanishes at the
generic transition, but remains nonzero when crossing the multicritical
point.\cite{Me.Tr.08,Hu.Al.Bu.Bl.07} In their results, there is a very sharp
downturn of $v_\text{s}$ toward zero close to $t_\text{c}$. We are not aware
of any calculations of $v_\text{s}$ for the polariton model.

From our QMC simulations, we can determine $v_\text{s}$ from linear fits to
the spectrum. Apart from the limited accuracy of the maximum entropy
inversion, this works quite well away from $t_\text{c}$. In agreement with
Bogoliubov theory, we find for the Bose-Hubbard model a linear dependence
$v_\text{s}\propto |t-t_\text{c}|$ and good agreement of results for $L=32$ and 64. 

Determining the behavior of $v_\text{s}$ as $t\rightarrow t_\text{c}$ is
more difficult for two reasons. First, the phonon spectrum becomes nonlinear
due to finite-temperature effects (see discussion below), rendering linear fits
ill-defined. Second, the position of the phase transition changes with system
size, so that no reliable finite-size scaling of $v_\text{s}$ can be carried
out. The situation is similar for the polariton model, and we therefore do
not show results for $v_\text{s}$ here, leaving this as an interesting
issue for future work.

\subsubsection{Dynamic structure factor}\label{sec:bhm_skw}

The single-particle spectral function provides information about
the energy and lifetime of particles or holes added to the interacting
ground state. In contrast, the dynamic structure factor---corresponding
to the imaginary part of the dressed particle-hole propagator---yields
insight into the density fluctuations in the ground state. In general the two
quantities do not exhibit the same features. However, for broken U(1) gauge
symmetry in the SF phase, they are both dominated by the same
single-particle excitations (phonons).\cite{Griffin93} We find this
statement to hold in 1D even though no symmetry breaking occurs.

The density operator in the Bose-Hubbard model is $\rho^\dag_k=\sum_l e^{-i k l} \on_l$.
For $k=0$, we have $\rho^\dag_k=\sum_l \on_l$, and $S(k,\om)$ has a trivial
contribution at $\om=0$ which we dismiss by considering
$\tilde{\rho}^\dag_k=\sum_l e^{-i k l} (\on_l-\las \on_l\ras)$. The
above-mentioned relation to particle-hole excitations becomes evident by
rewriting the density operator as $\rho^\dag_k=\sum_q b^\dag_{q+k} b_q$.

We show results for $S(k,\om)$ in Fig.~\ref{fig:BH_szsz}.  According to
Huber \etal,\cite{Hu.Al.Bu.Bl.07} $S(k,\om)$ in the MI phase should exhibit a
continuum of particle-hole excitations, starting at $\omega=E_\text{g}$ due
to the Mott gap in the single-particle spectrum (see
Fig.~\ref{fig:BH_green}). For the parameters in Fig.~\ref{fig:BH_szsz}(a),
$E_\text{g}/U\approx0.7$. The dispersion of the particle and hole bands is
very weak, Note that we find no agreement with the two single-particle
excitations $E^\text{p}_\text{g}+\epsilon_\text{h}(k)$,
$E^\text{h}_\text{g}+\epsilon_\text{p}(k)$ discussed by Huber \etal This may
be a result of their mean-field treatment of the two-dimensional case.  Our
results do agree qualitatively with exact numerical results on small
clusters.\cite{Ro.Bu.04}

For larger $t/U$, the Mott state contains nontrivial density fluctuations, and
the upper band in $S(k,\om)$ acquires some $k$ dependence. The energy of the
excitations in $S(k,\om)$ [following $\sum_q
\{\epsilon_\text{h}(q)+\epsilon_\text{p}(k-q)\}$] \cite{Hu.Al.Bu.Bl.07}
generally increases with increasing $k$. This is obvious from the momentum
dependence of the particle and hole bands in $A(k,\om)$, and also agrees with
the expectation that long-wavelength density fluctuations in a Mott state
require less energy than fluctuations with short periods in real space.

For $t\lesssim t_\mathrm{c}$ in Fig.~\ref{fig:BH_szsz}(b), we find a
low-energy mode with nonlinear dispersion, which we interpret as a precursor of
the linear excitations of the SF phase [see panel (d)]. Even for $t\gtrsim
t_\mathrm{c}$ [Fig.~\ref{fig:BH_szsz}(c)], the gapless low-energy mode in our
numerical results is not linear. A linear spectrum is a result of the
condensation of bosons in the SF phase, but is not expected in the normal phase. Since our
simulations are done at finite temperature, and because the phase coherence
length is small close to $t_\mathrm{c}$, we can understand the absence of a
clear, linear signature in Fig.~\ref{fig:BH_szsz}(c). Going to larger
$t_\mathrm{c}$, we indeed see linear excitations near $k=0$
[Fig.~\ref{fig:BH_szsz}(d)]. Similar effects are expected for the
single-particle excitations, but are difficult to see on the scale of
Fig.~\ref{fig:BH_green}. Coming back to Fig.~\ref{fig:BH_szsz}(c), away
from $k=0$, we find a free-particle like contribution, similar to the case of
the MI. This excitation carries negligible spectral weight near $k=0$. 

Apart from finite-temperature effects, these features are qualitatively
similar to the excitations discussed by Huber \etal,\cite{Hu.Al.Bu.Bl.07}
namely a gapless sound mode (related to phase and density modulations)
dominant at small $k$, and a massive mode (corresponding to exchange between
condensate and noncondensate at fixed density) acquiring spectral weight at
$k>0$. Additionally, we see in Fig.~\ref{fig:BH_szsz}(c) the (weak) signature
of a gapped mode at small $k$, the nature of which we cannot determine from
our present simulations. For $t/U=0.2$ [Fig.~\ref{fig:BH_szsz}(d)] the
excitation ``band'' in $S(k,\om)$ follows closely the Bogoliubov mode, in
accordance with the discussion at the beginning of this section.

At this point, a comment concerning the accuracy of the spectra
obtained from the maximum entropy inversion is in order. The spectral weight
of the features visible in density plots such as Fig.~\ref{fig:BH_szsz}(d)
varies over orders of magnitude. Some very weak signals, such as the group of
points located at around $k=\pi/2$ below the main excitation band (with a
weight that is a factor 10000 smaller than that of the dominant features), 
are expected to be artifacts. We shall see below that in the polariton model,
there actually exist real excitations with very small spectral weight which
are easy to miss in the maximum entropy inversion. To reliably study such
excitations, analytical approaches (if available) are clearly
superior.\cite{Sc.Bl.09}

Our findings for the dynamic structure factor are consistent with previous numerical
results on small systems $(L=10,20)$.\cite{Ba.As.Sc.De.05,Ro.Bu.04}
We can confirm the broadening of the excitations with increasing $k$ in the
SF phase,\cite{Ba.As.Sc.De.05} related to two-particle continua.\cite{Hu.Al.Bu.Bl.07}
However, the maximum entropy method is not capable of resolving fine structures as (generically) seen in exact
diagonalization results for small clusters.\cite{Ro.Bu.04}

\subsection{Polariton model}

For the polariton model, the only published results on dynamic properties are for the
single-particle spectrum of the MI phase at zero
temperature.\cite{Ai.Ho.Ta.Li.08,Sc.Bl.09} As pointed out before, the nature
of the conserved particles in the polariton model is determined by the
detuning. We start by discussing the case $\Delta=0$ for which the
polaritonic character of the excitations is most pronounced. This can readily
be seen from Eq.~(\ref{eq:eigenstates}), where $\ket{n_\text{p},\DO}$ and
$\ket{n_\text{p}-1,\UP}$ contribute with equal weight.

\begin{figure} % 5
\subfigure{
  \includegraphics[width=0.46\linewidth]{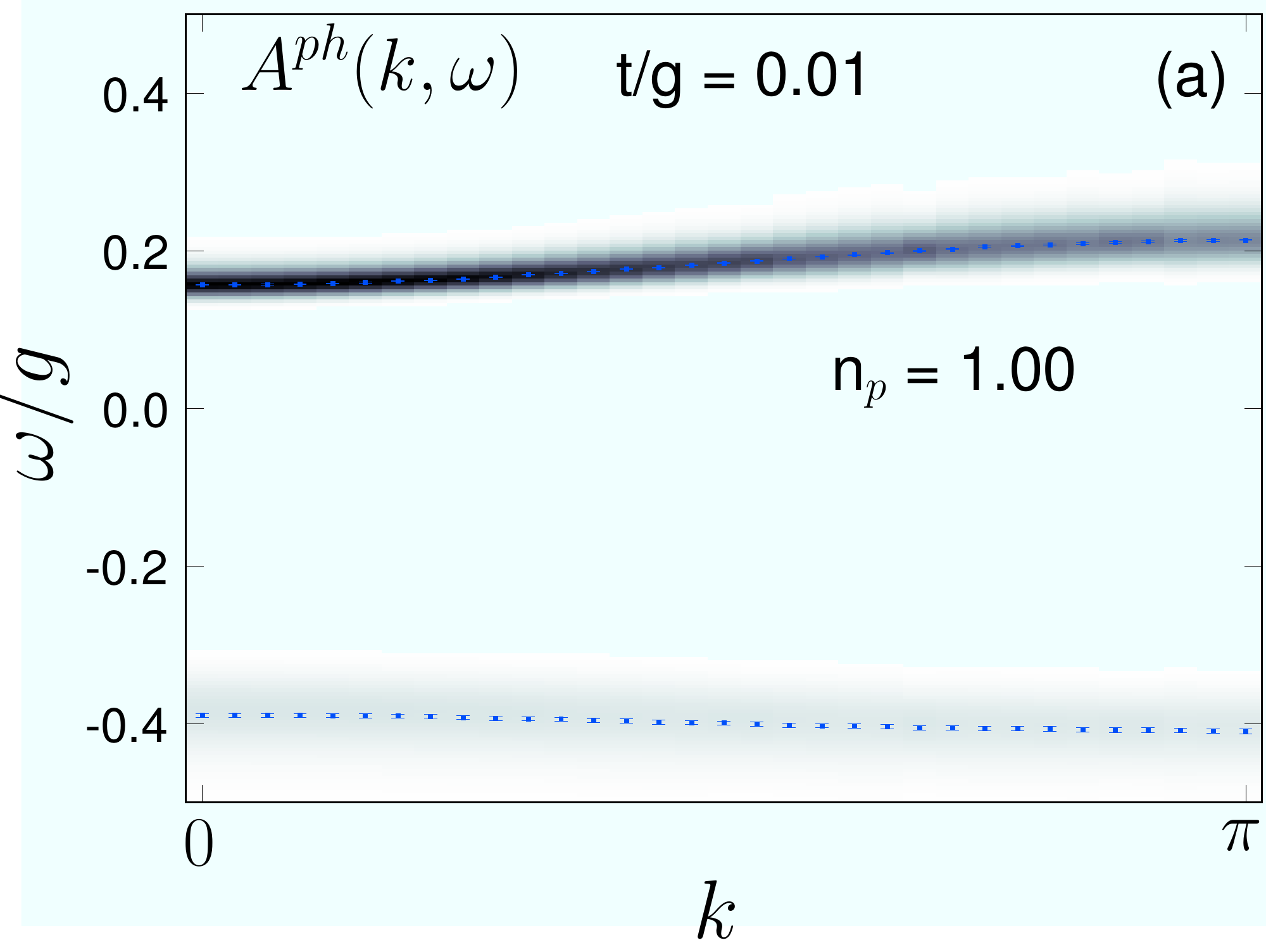}
  }
\subfigure{
  \includegraphics[width=0.46\linewidth]{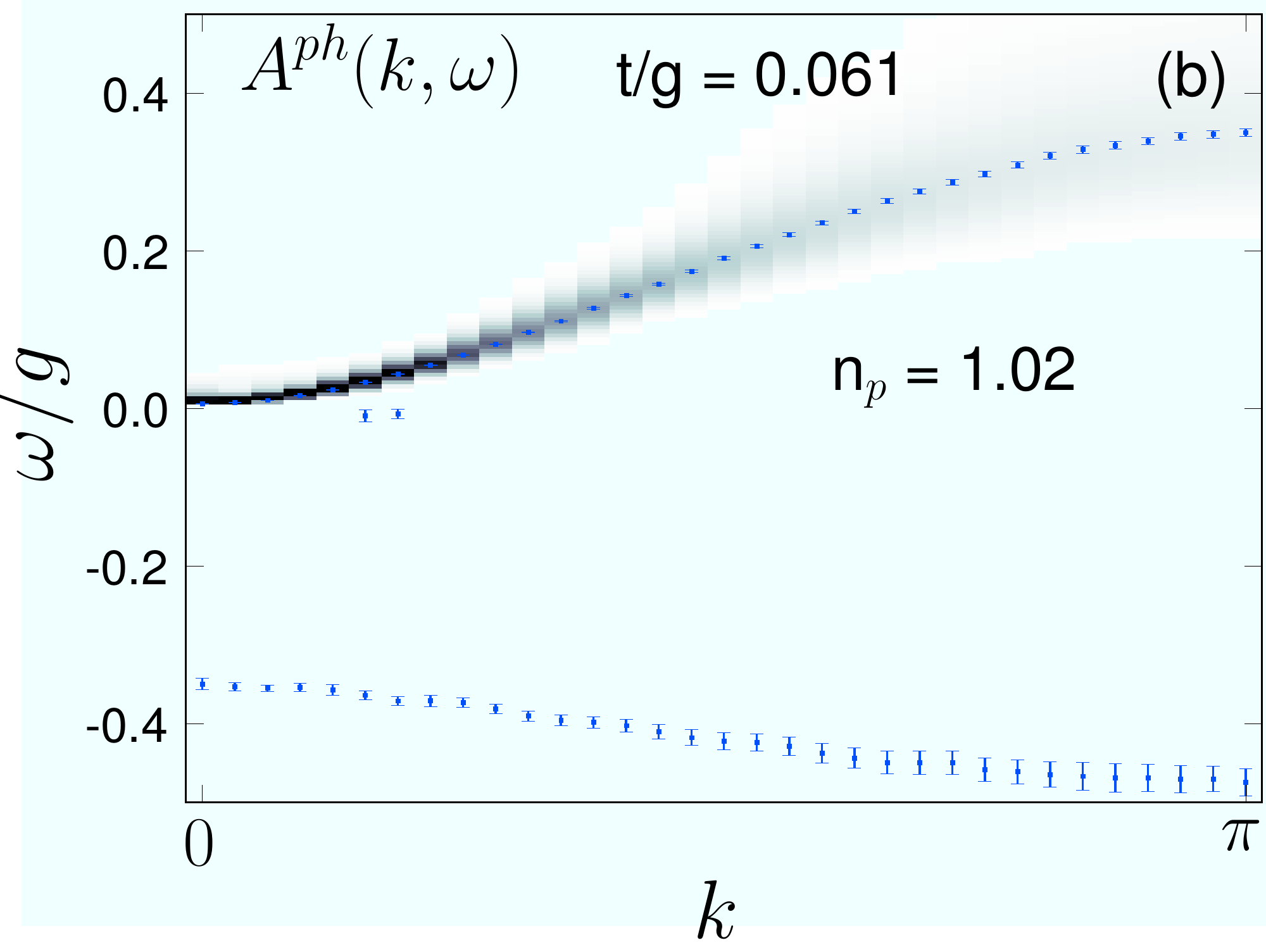}
  }
\subfigure{
  \includegraphics[width=0.46\linewidth]{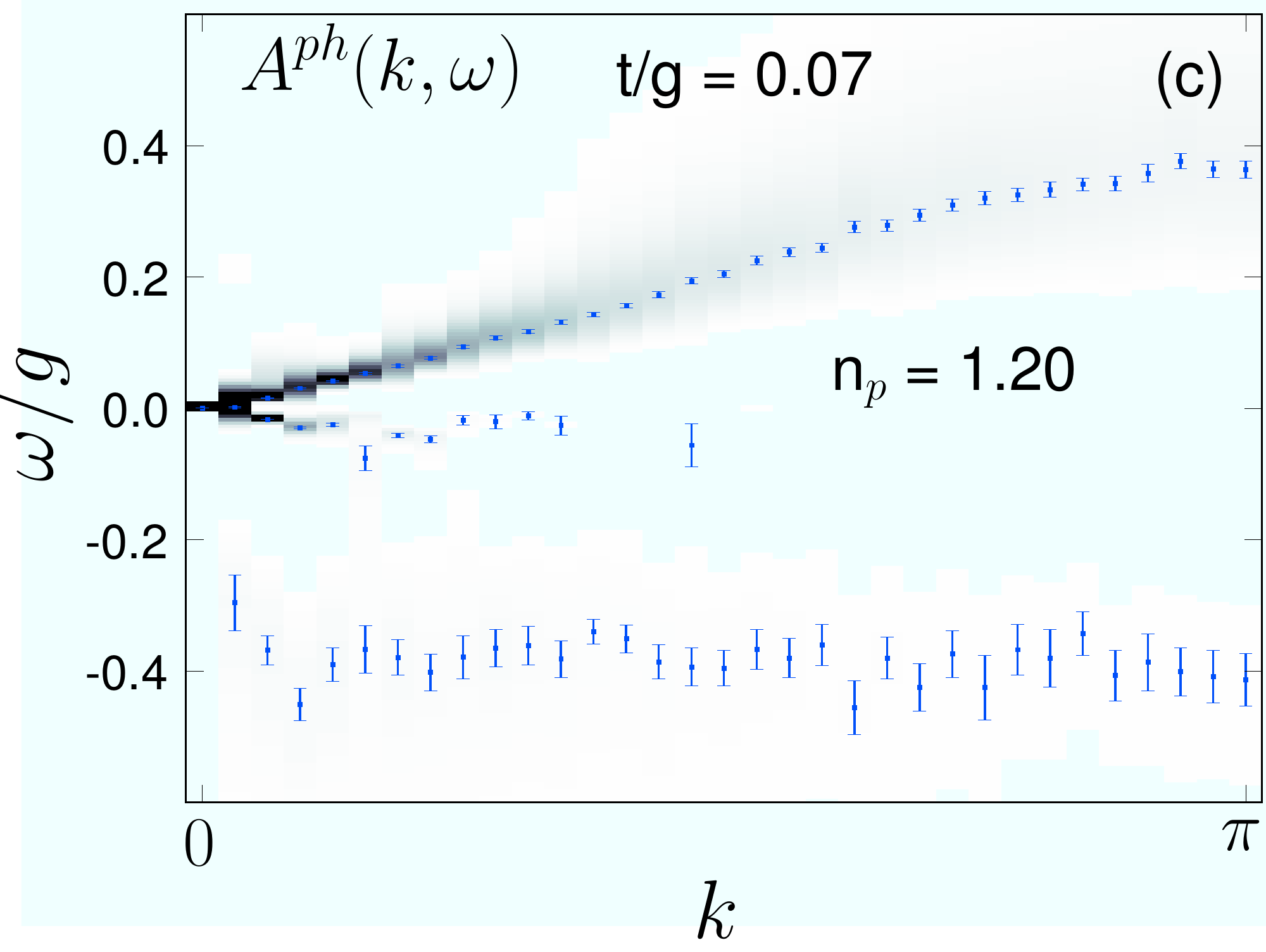}
  }
\subfigure{
  \includegraphics[width=0.46\linewidth]{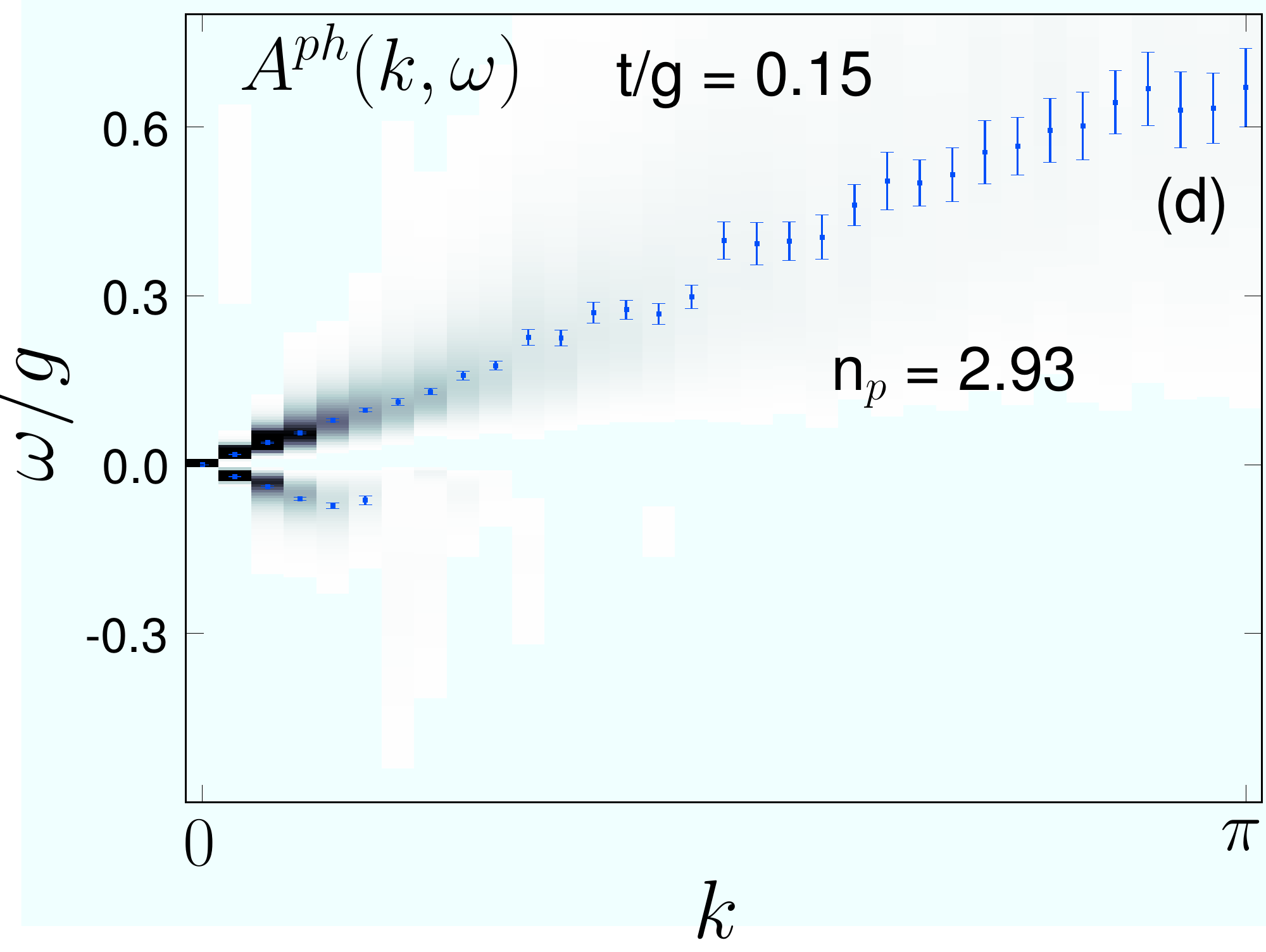}
  }
\caption{\label{fig:PT_greens} 
  Single-photon spectral function $A^\text{ph}(k,\om)$ of the 1D polariton model
  at $\mu/g = 0.4$ for different hoppings $t$, corresponding to (a) deep in the
  MI, (b) just below the MI-SF transition, (c) just above the
  transition, and (d) in the SF phase. Here $\beta g = 3L$ and $L = 64$.
  With increasing $t$, the density plots are more and more
  ``overexposed'' to see less dominant features.}
\end{figure}

\begin{figure} % 6
\subfigure{
  \includegraphics[width=0.46\linewidth]{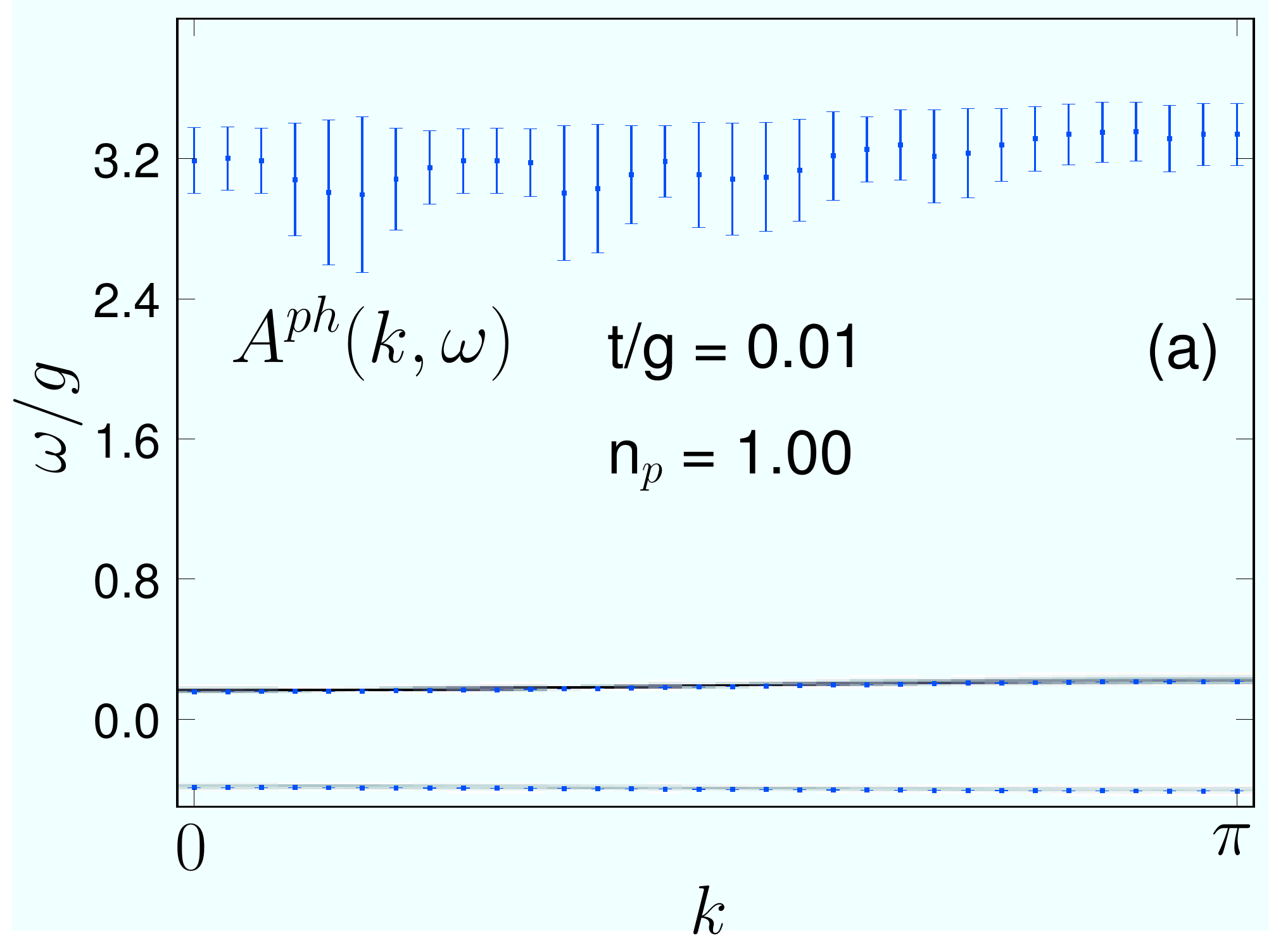}
  }
\subfigure{
  \includegraphics[width=0.46\linewidth]{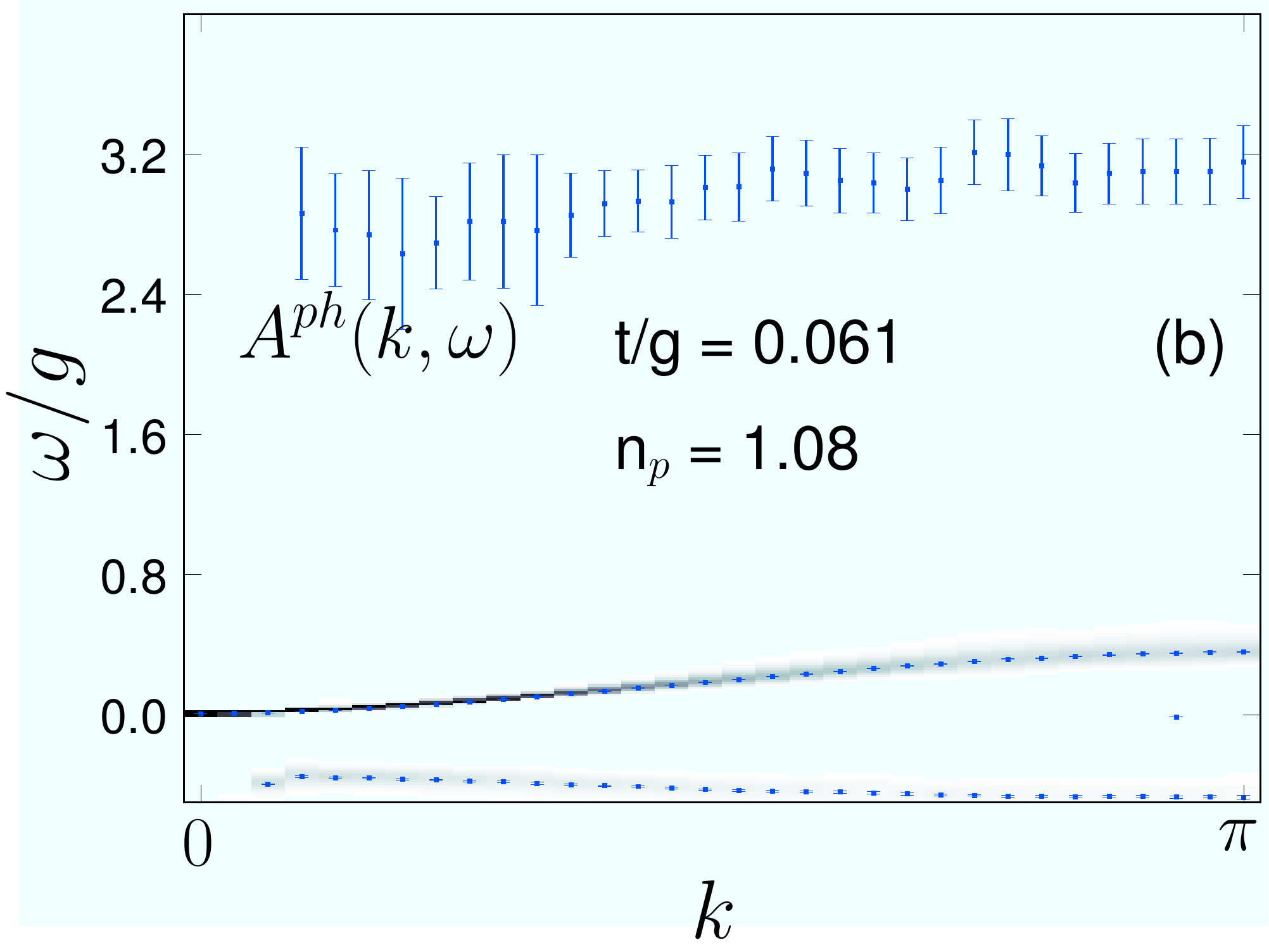}
  }
\caption{\label{fig:PT_greens2} 
  Single-photon spectral function in the Mott phase for the same parameters as in
  Fig.~\ref{fig:PT_greens}, showing additional excitations at higher energies.}
\end{figure}

\subsubsection{Single-particle spectrum}

\begin{figure}[Htb] % 7
  \centering
  \includegraphics[width=0.95\linewidth]{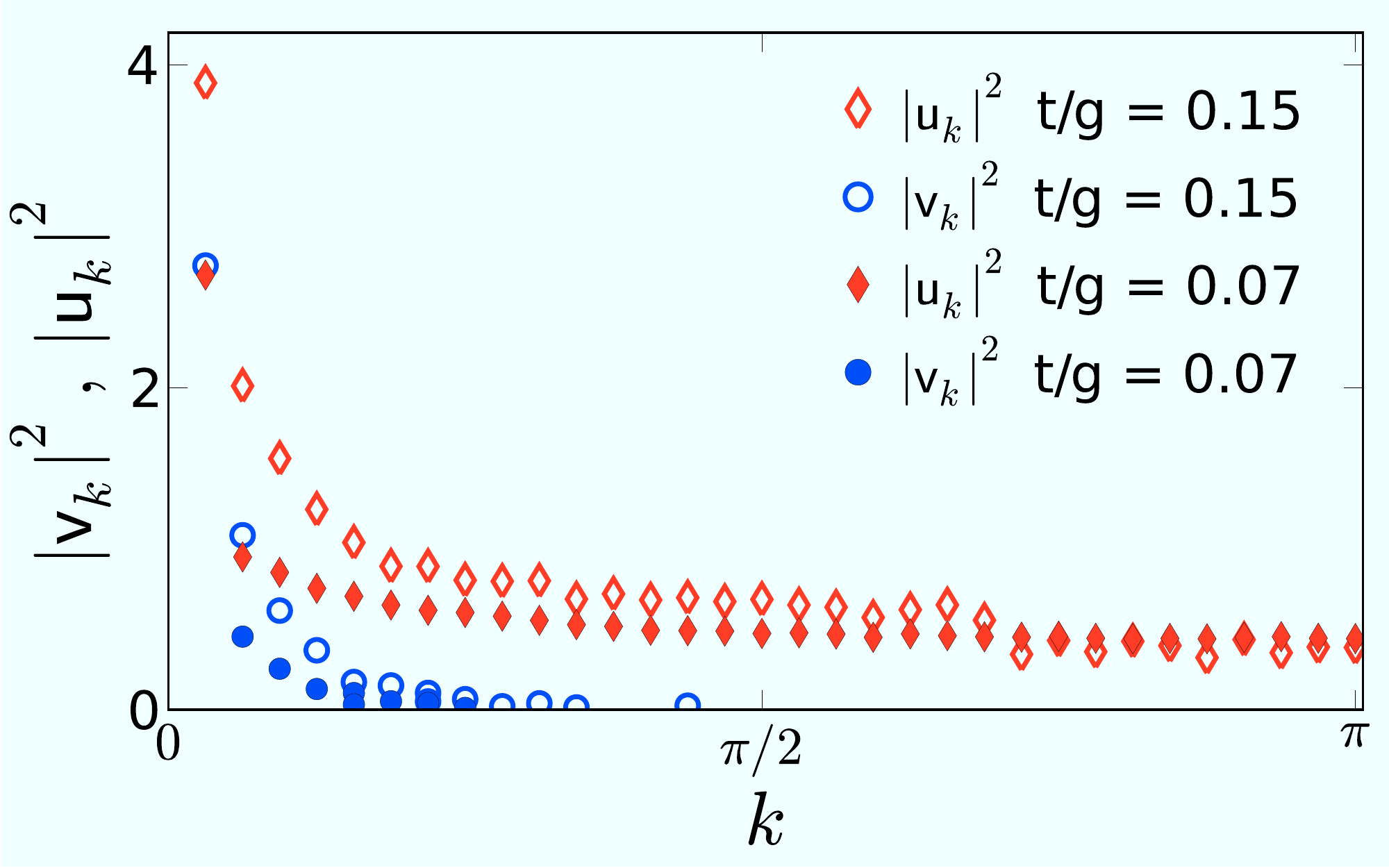}
  \caption{(color online) Quasiparticle weights $\mathsf{u}_k$ and
    $\mathsf{v}_k$ of the gapless modes of the polariton model, similar to
    Fig.~\ref{fig:quasiparticle_weight_BH}.  }
  \label{fig:quasiparticle_weight_pm}
\end{figure}

In Fig.~\ref{fig:PT_greens} we show our QMC results for the single-photon
spectral function. As for the Bose-Hubbard model, the values of the ratio $t/g$ range from
deep in the Mott phase across the generic transition well into the SF phase.
According to a finite size scaling analysis for $\mu/g=0.4$, the phase
transition occurs at $t_\mathrm{c}/g =0.0626(1)$ (see
Fig.~\ref{fig:fss_mu0_4}), in agreement with Fig.~\ref{fig:phasediagrams}(b).
Hence panels (a) and (b) are for the MI regime, whereas (c) and (d) are for
the SF phase.

The results in the MI shown in Fig.~\ref{fig:PT_greens}(a), (b) agree well
with previous numerical work.\cite{Ai.Ho.Ta.Li.08} Similar to the Bose-Hubbard model, there exist
particle and hole bands, separated by the Mott gap. It is important to stress
that although we add bare photons to the system, the particle and hole
excitations reflect the properties of the polaritons in the system.  Whereas
the ratio of particle and hole bandwidths is two to one in the Bose-Hubbard model, it
depends on the character of the quasiparticles (polaritons) in the polariton model and
varies with detuning.\cite{Ai.Ho.Ta.Li.08} With increasing $t/g$, the gap
closes and the bandwidths of excitations increase (effective masses decrease). 

Recent analytical work revealed the existence of so-called upper
polariton modes at higher energies, which represent an important difference between
the Bose-Hubbard model and the polariton model.\cite{Sc.Bl.09} For the Mott
lobe with $n_\text{p}=1$, only one such (particle) band exists, corresponding (for
small enough $t/g$) to a transition between the ground state
$\ket{n_\text{p}=1,-}$ and the state $\ket{n_\text{p}=2,+}$ (see
Eq.~(\ref{eq:eigenstates})). The weight of this high-energy excitation is
very small compared to the dominant particle and hole modes discussed above
(0.04 as compared to 1.46 for the $k=0$ atomic-limit results in
[\onlinecite{Sc.Bl.09}]); with increasing $t/g$ the weight difference becomes
even larger.\cite{Sc.Bl.09} The energy splitting between the $-$ and $+$ branches
of eigenstates increases further for detuning $\Delta\neq0$ (Fig.~2 in
[\onlinecite{GrTaCoHo06}]).

\begin{figure} % 8
  \centering \subfigure{
    \includegraphics[width=0.46\linewidth]{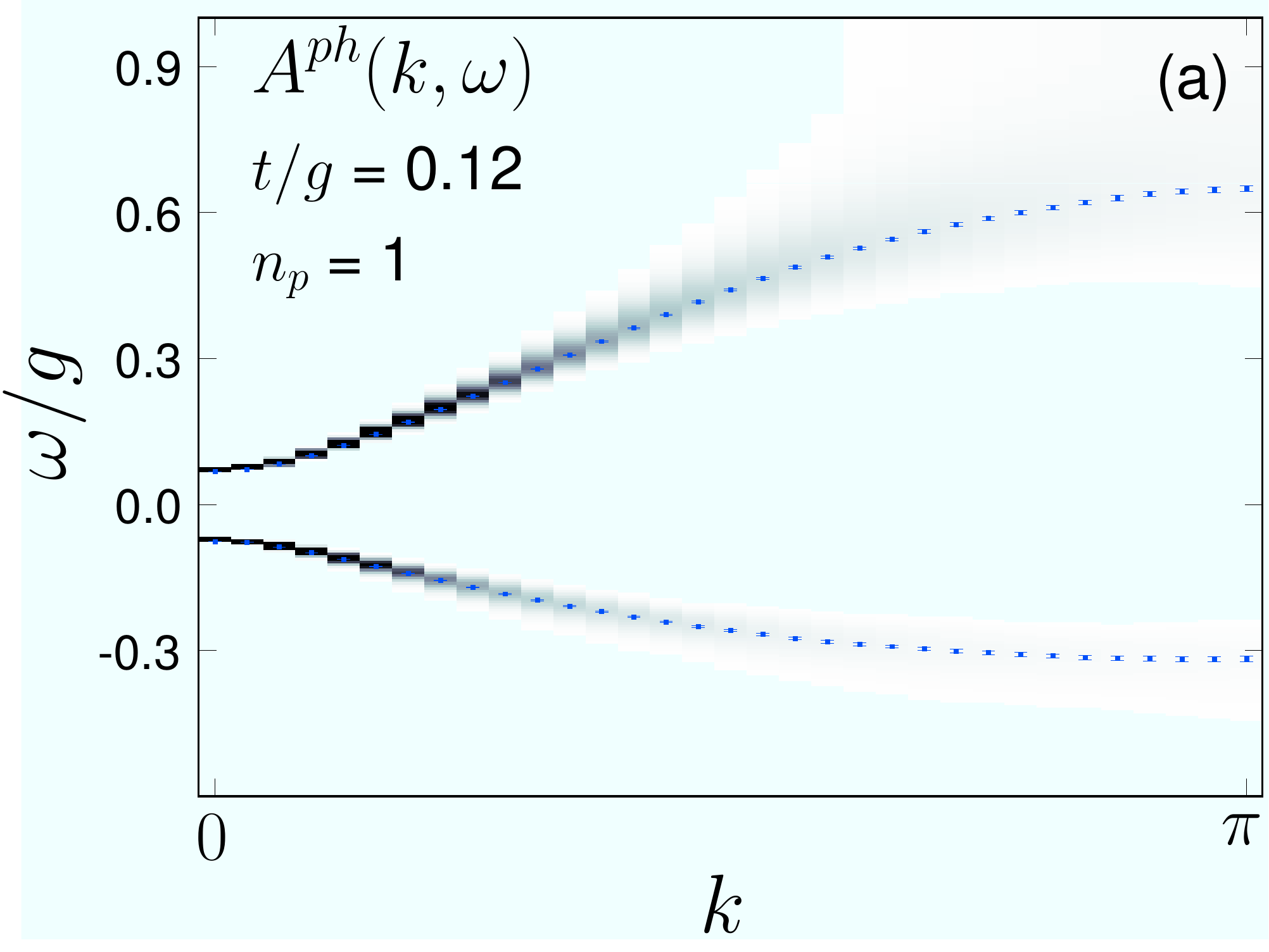}
  } \subfigure{
    \includegraphics[width=0.46\linewidth]{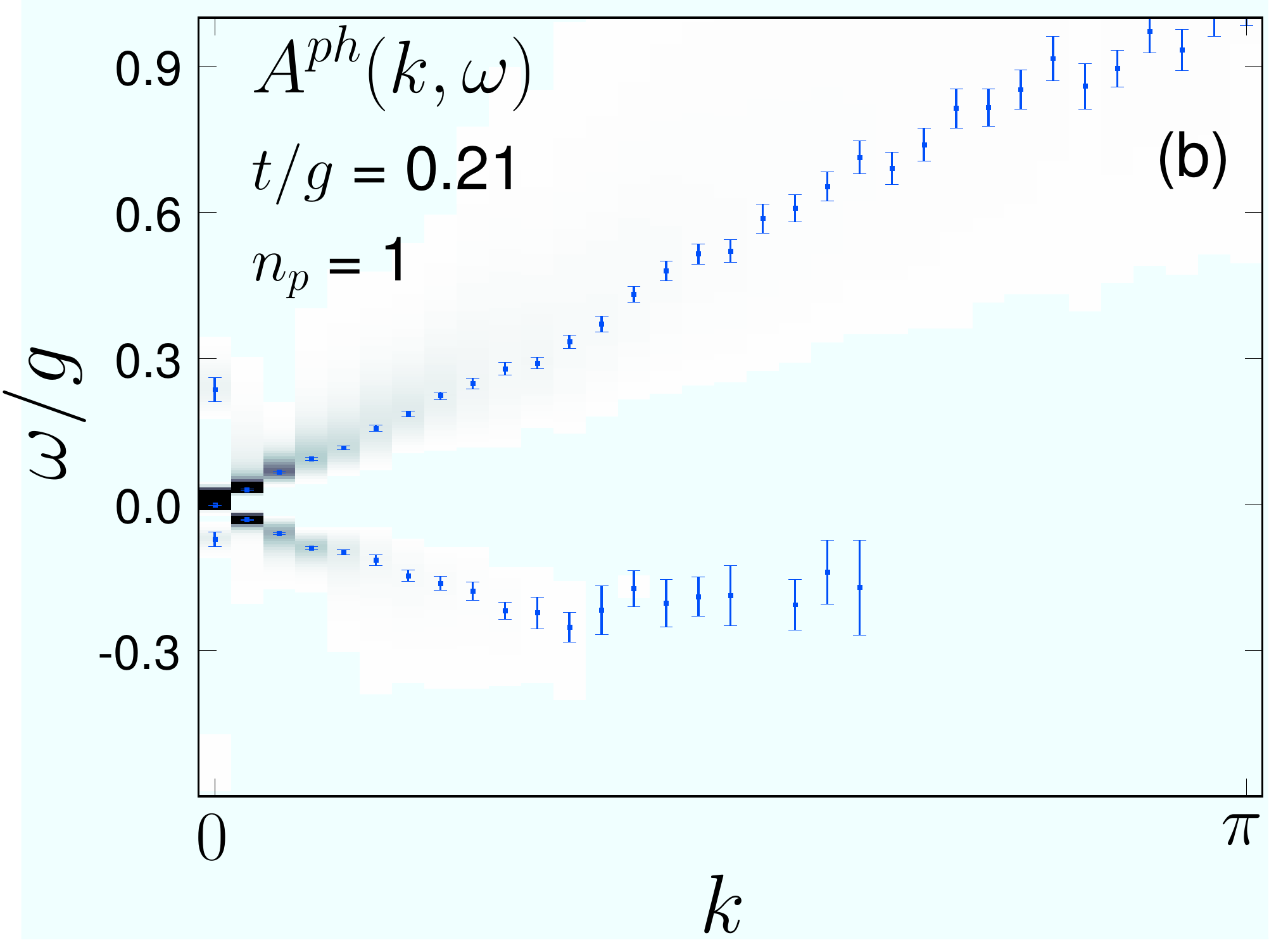}
  }
  \caption{Single-photon spectrum $A^\text{ph}(k,\om)$ of the polariton model along the
    line $n_\mathrm{p}=1$ crossing the Kosterlitz-Thouless transition. Here $L=64$ and $\beta g =
    3 L$.}
  \label{KT_A_ph}
\end{figure}

The upper polariton mode is not visible in Figs.~\ref{fig:PT_greens}(a) or
(b). Excitations with small spectral weight are notoriously difficult to see using
QMC in combination with maximum entropy. In the present case, this is
aggravated by the fact that the resolution of maximum entropy decreases at
high energy. Nevertheless, we see a signature of the upper polariton band in
Fig.~\ref{fig:PT_greens2}, and the latter is also present (but not shown) in
the high-temperature data of Fig.~\ref{fig:finite_T_greens}(a); high-energy
features are easier to resolve in QMC/maximum entropy at higher temperatures.
From the eigenvalues of the states~(\ref{eq:eigenstates})
we can determine the excitation energy of the upper mode in the
atomic limit as $w^+_p/g=-(\mu-\om_0)/g+(\sqrt{2}+1)\approx3$ for the parameters
of Fig.~\ref{fig:PT_greens2}, in reasonable agreement with our results in
Fig.~\ref{fig:PT_greens2}(a) given the ill-conditioned nature of the problem under
consideration. Note that the upper polariton mode can be seen even close to
the phase transition in Fig.~\ref{fig:PT_greens2}(b). The weight of the upper
mode in Fig.~\ref{fig:PT_greens2} is about a factor of 100 smaller than that of
the conventional particle and hole excitations. Although the upper polariton
mode exists also in other results for the single-particle spectrum in the
Mott phase (Figs.~\ref{KT_A_ph},~\ref{fig:finite_T_greens} and~\ref{fig:detuning_A_ph}),
we focus on the low-energy conventional modes with large spectral weight.
The latter can be determined accurately from our simulations, and will be
the dominant feature in experiments.

Figures~\ref{fig:PT_greens}(c) and (d) contain the first spectra of the
polariton model in
the SF phase. There is a clear signature of the gapless phonon modes starting
at $k=\om=0$, with linear dispersion at small $k$. In the SF phase but close
to the transition, we see an additional gapped mode at $\om<0$
[Fig.~\ref{fig:PT_greens}(c)]. Our results at these and at further couplings
$t/g$ (and $t/U$) suggest that these gapped modes disappear more quickly with
increasing $t/g$ than for the Bose-Hubbard model, which can be explained in terms of the
photonic SF expected for the present parameters (see below).\cite{Ir.09} Note
that a simple Bogoliubov type theory for the polariton model does not exist, due to the
composite nature of polariton excitations. 

Figure~\ref{fig:quasiparticle_weight_pm} shows the quasiparticle weights.
The general shapes resemble the Bose-Hubbard model case (Fig.~\ref{fig:quasiparticle_weight_BH}),
but the lower branch decays very quickly in the polariton model even at $t/g=0.15$ quite
far from the phase transition. This may be attributed to the
fact that the energy cost for particle and hole excitations is different due
to the dependence of $U_\text{eff}$ on $n_\text{p}$. Again, most of the
spectral weight in the SF phase is found at small $k$.

\begin{figure}%[H] % 9
  \centering
  \includegraphics[width=0.8\linewidth]{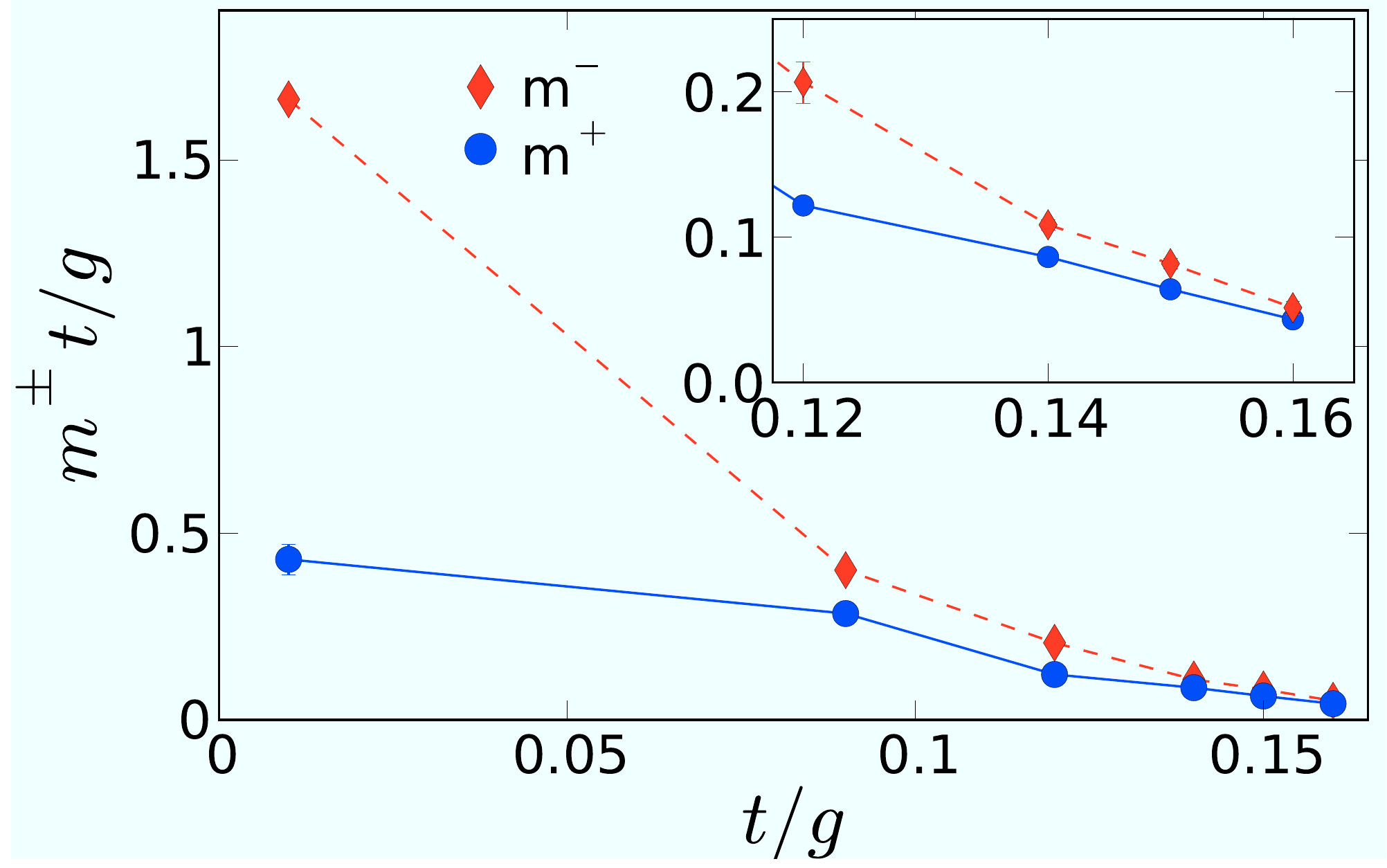}
  \caption{(color online) Effective particle ($m^+$) and hole masses ($m^-$)
    along the line $n_\textrm{p}=1$, as obtained from fits to the bands in
    $A^\text{ph}(k,\om)$ near $k=0$.}
  \label{PM_meff}
\end{figure}

{\em Fixed density.}
In Fig.~\ref{KT_A_ph} we show the single-photon spectrum across
the fixed density transition ($n_\mathrm{p} = 1$),
obtained by selecting configurations only at that density. In
Ref.~\onlinecite{rossini_photon_2008}, the critical hopping was determined as
$t_\mathrm{c}/g = 0.198$ [cf Fig.~\ref{fig:phasediagrams}(b)]. The spectra in
both the MI and the SF look very similar to those across the generic
transition shown above. This may be different very close to the multicritical
point, but this regime is most demanding numerically if the results are to be
used in a maximum entropy inversion.

From the spectra obtained at constant density in the MI, we can estimate the
effective particle and hole masses by fitting a quadratic dispersion to the
bands in the vicinity of $k=0$. In the Bose-Hubbard model, there is an emergent
particle-hole symmetry on approaching the lobe
tip,\cite{PhysRevB.40.546,CaSa.GuSo.Pr.Sv.07} and similar behavior is
suggested by the evolution of the particle and hole bands with increasing
$t/g$ also in the polariton model. For fixed polariton density, the two
masses approach each other and vanish at the phase transition. This has been
demonstrated in 2D based on a strong-coupling approach.\cite{Sc.Bl.09}
In the region not too close to the phase transition, where stable fits
can be obtained, Fig.~\ref{PM_meff} confirms this observation also in 1D.

\begin{figure} % 10
\subfigure{
  \includegraphics[width=0.46\linewidth]{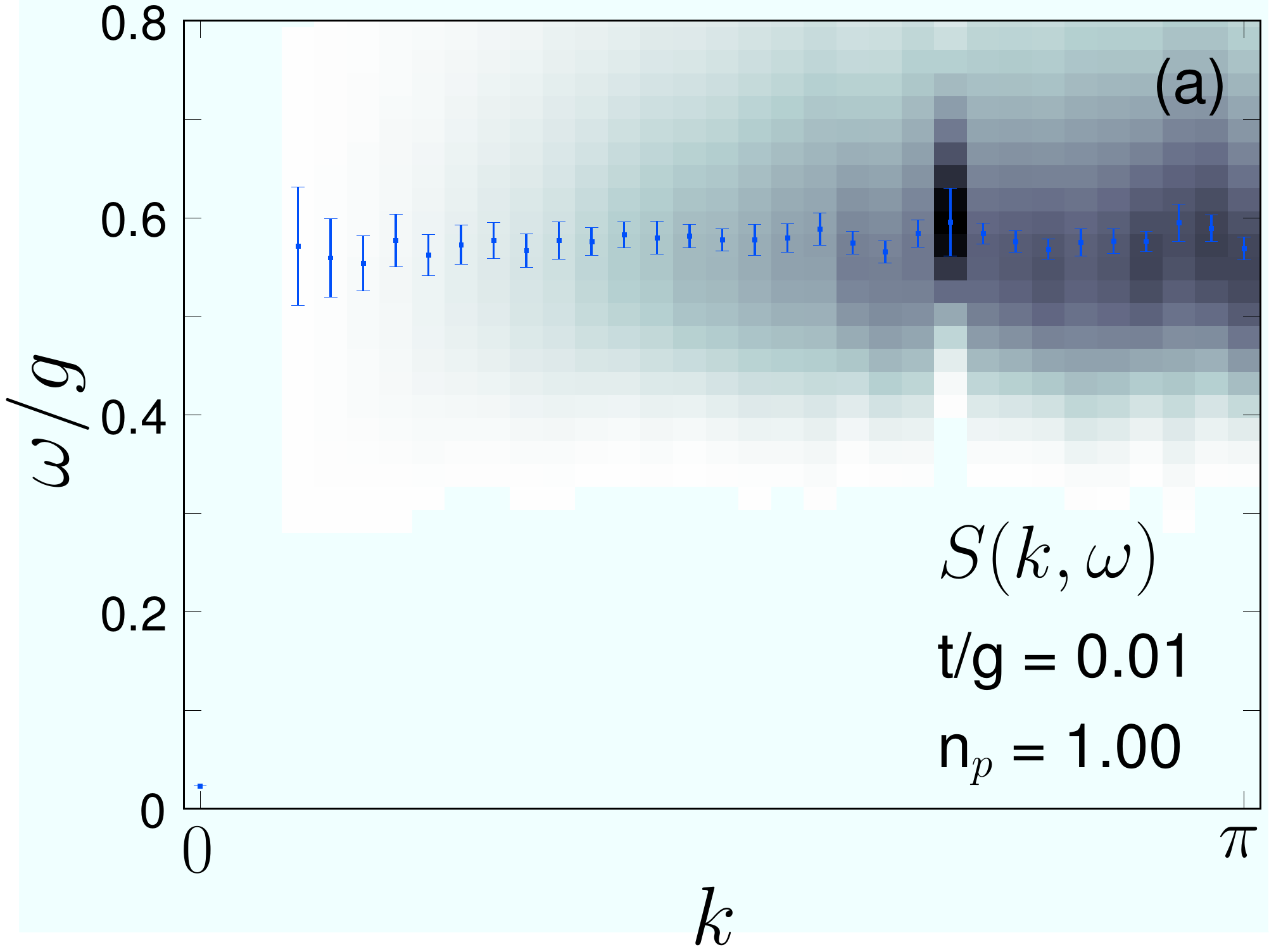}
 }
\subfigure{
  \includegraphics[width=0.46\linewidth]{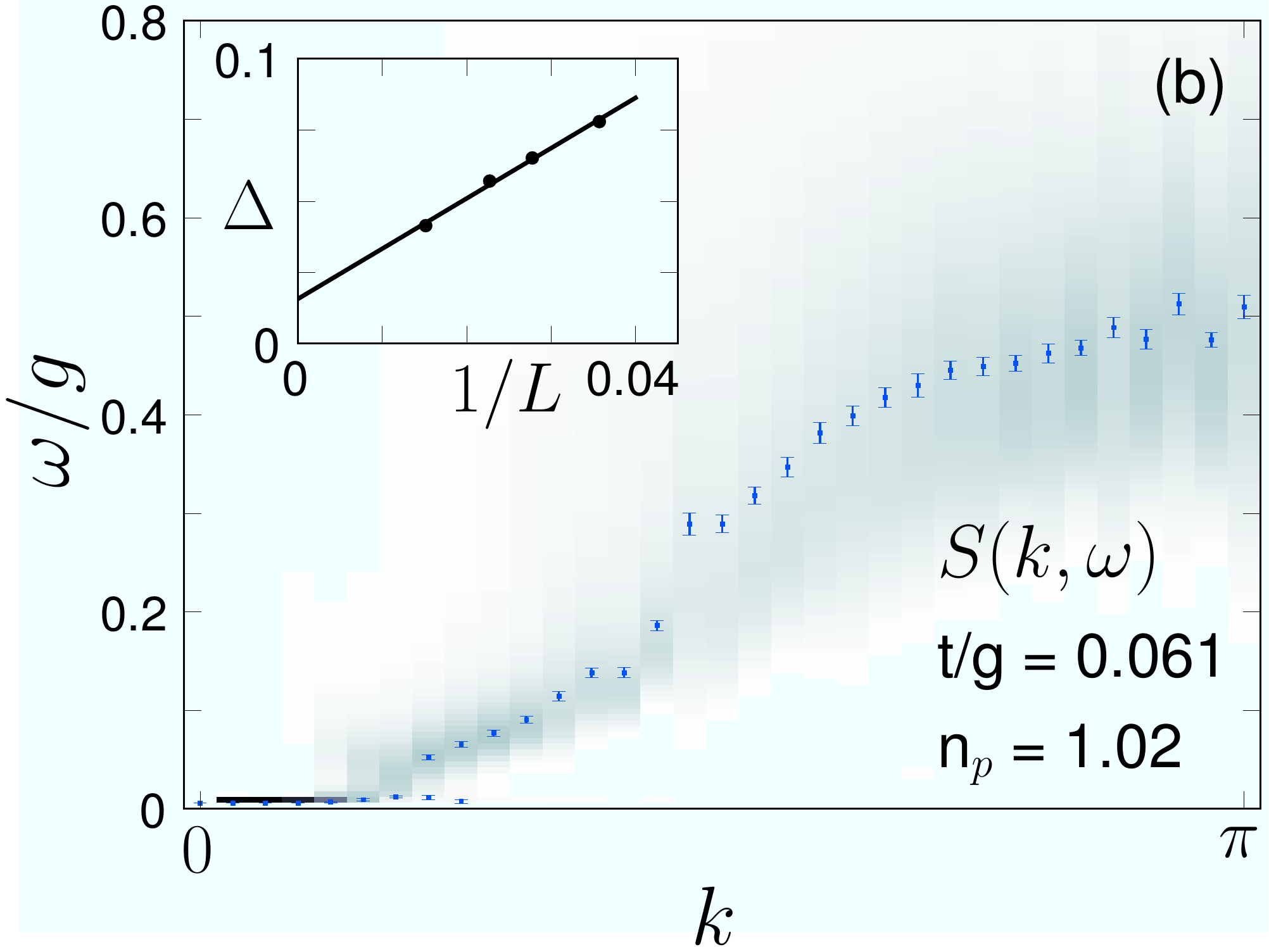}
  }
\subfigure{
  \includegraphics[width=0.46\linewidth]{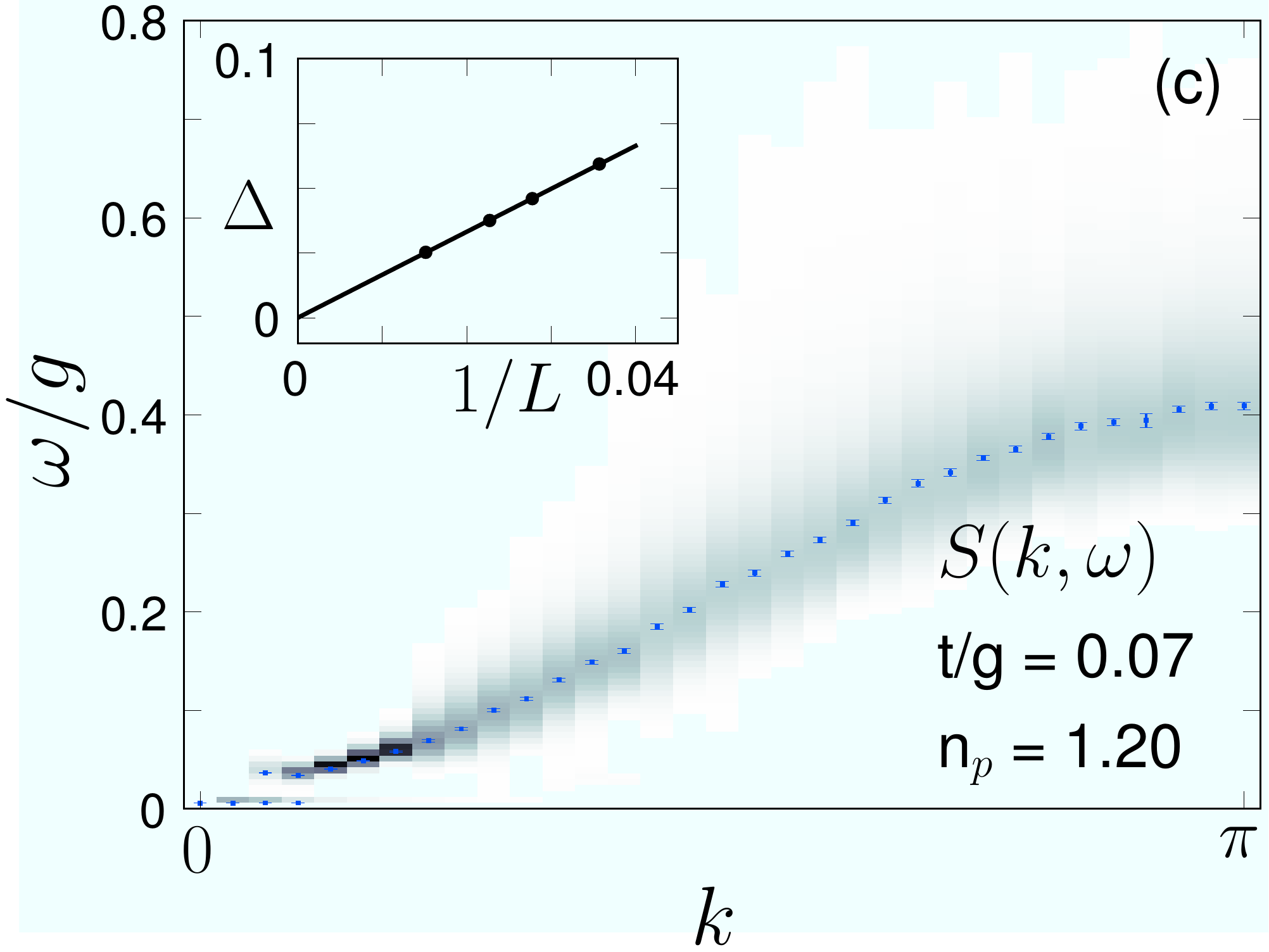}
}
\subfigure{
  \includegraphics[width=0.46\linewidth]{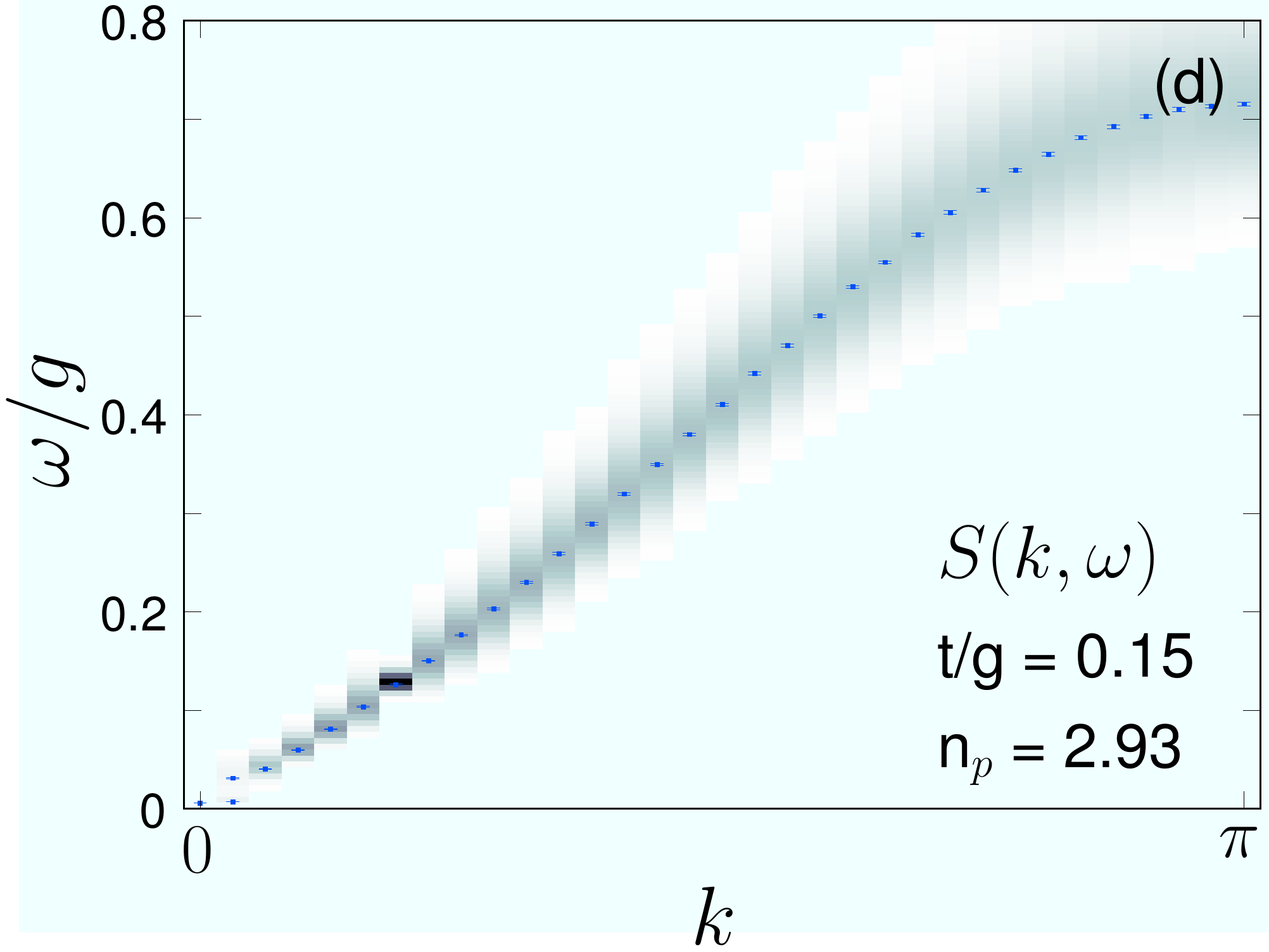}
}
\caption{\label{fig:PT_szsz}
  Polariton dynamic structure factor $S(k,\om)$
  for the same parameters as in Fig.~\ref{fig:PT_greens}.  The insets show an
  extrapolation of the Mott gap at small $k$ to $L \to \infty$.}
\end{figure}

\subsubsection{Dynamic structure factor}

The evolution of the polariton dynamic structure factor $S(k,\om)$ across the
MI-SF transition is shown in Fig.~\ref{fig:PT_szsz}. Remarkably, the results
look very similar to those for the Bose-Hubbard model. Close to the atomic limit [$t/g=0.01$
in Fig.~\ref{fig:PT_szsz}(a)] we see a gapped, almost flat feature with
energy $\om\approx 0.6g$. A look at the corresponding single-particle
spectrum in Fig.~\ref{KT_A_ph}(a) reveals that this value is identical to
the Mott gap. The almost flat particle and hole bands cause a very weak
dispersion also for the particle-hole excitations visible in $S(k,\om)$.
It is useful to remember
that it is the effective polariton-polariton repulsion mediated by the atom-photon
coupling that determines the Mott gap. For a single site and $n_\mathrm{p}=1$
(\ie, for the case of adding a second polariton),
\begin{equation}\label{eq:Ueff}
  U_\mathrm{eff}(1) = 2 \sqrt{g^2 + (\Delta/2)^2} - \sqrt{2g^2 +
      (\Delta/2)^2} - \Delta/2\,.
\end{equation}
For zero detuning ($\Delta=0$), $U_\mathrm{eff}(1)/g=2-\sqrt{2}\approx0.59$.

As for the Bose-Hubbard model, the excitations in $S(k,\om)$ acquire a noticeable dispersion
with increasing $t/g$, and the $k=0$ gap closes.
Figures~\ref{fig:PT_szsz}(b) and (c) are both close to the phase transition.
An inspection of the $k=0$ region shows a weak linear mode with very small
slope $O(0.01)$, corresponding to the small superfluid density existing in
both $L=64$ systems.  The massive mode extends to $k=0$ with a tiny intensity
(thus not visible in the figure).  An extrapolation of the gap to $L=\infty$
(insets) shows that it scales to zero in Fig.~\ref{fig:PT_szsz}(c), but stays
finite at the smaller hopping in Fig.~\ref{fig:PT_szsz}(b).  Indeed, a finite
size scaling of the superfluid density (discussed later) implies that
Fig.~\ref{fig:PT_szsz}(b) is just below the phase transition. In addition to finite size effects, we
again observe finite temperature effects in the form of deviations from the
expected linear spectrum close to $t_\text{c}$ (see discussion for the Bose-Hubbard model). For even
larger $t/g$, the spectrum exhibits a single linear mode at small $k$.
Similar to the spectral function, gapped modes seem to be suppressed quickly
in the SF phase.

Clearly, the polariton dynamic structure factor $S(k,\om)$ represents a
useful probe to distinguish between the MI and the SF phases. We have argued
before that the polariton MI has fluctuations in the photon and exciton
density, whereas the polariton density is pinned.

\begin{figure}%[H] % 11
  \centering
  \subfigure{
    \includegraphics[width=0.46\linewidth]{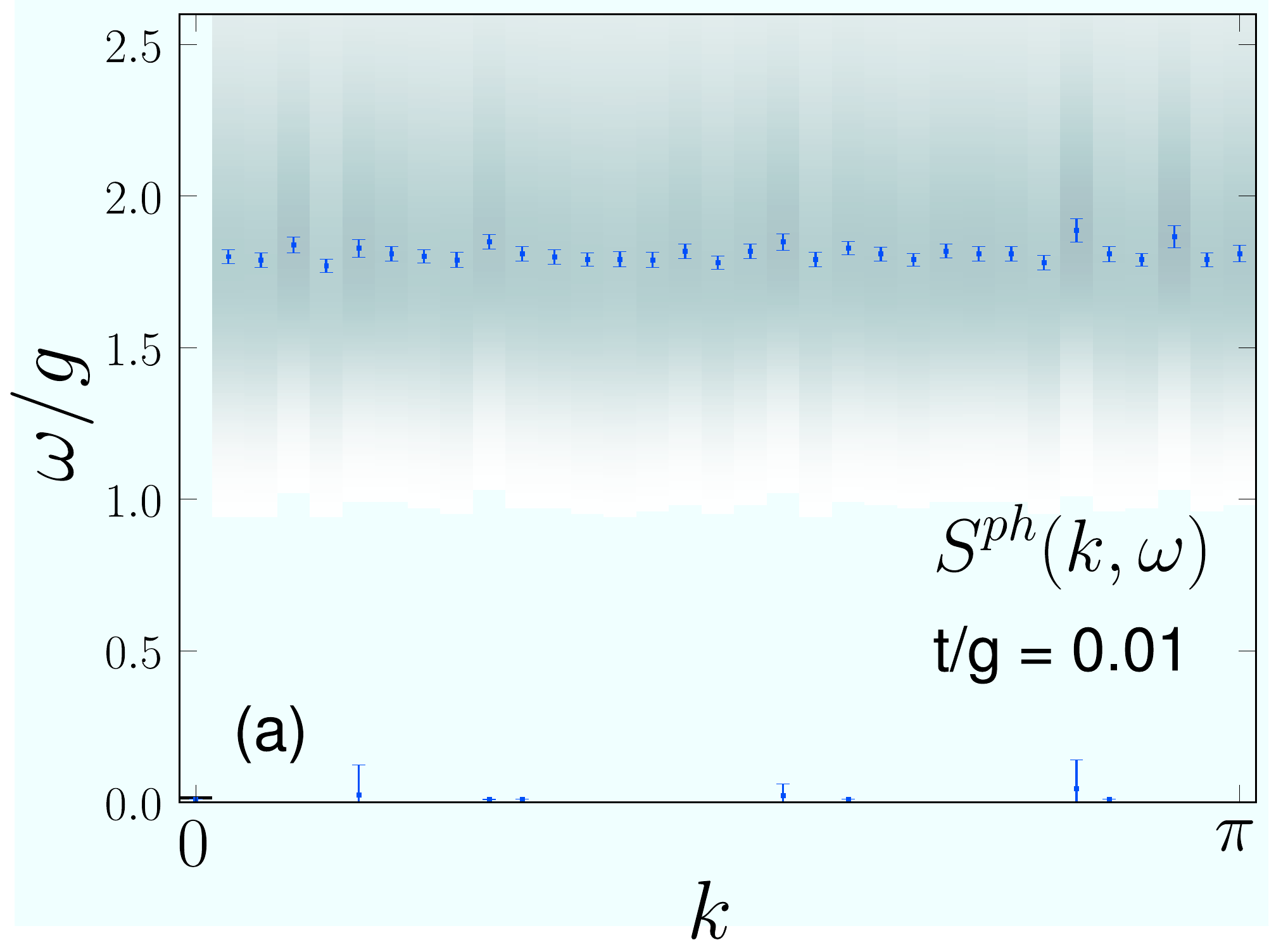}
  }
  \subfigure{
    \includegraphics[width=0.46\linewidth]{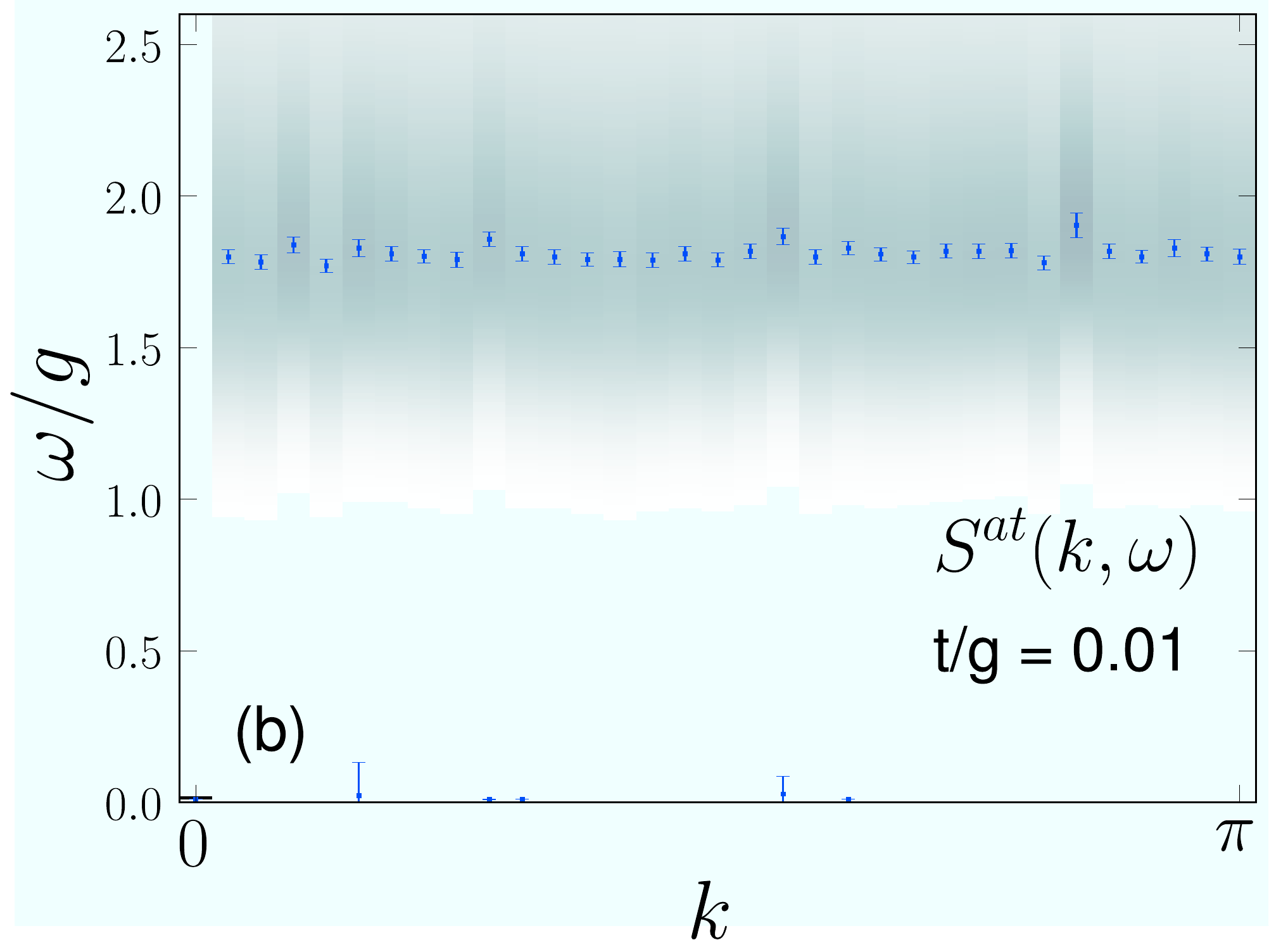}
  }
  \caption{Dynamic structure factor for excitons ($S^\mathrm{at}$) and
    photons ($S^\mathrm{ph}$) for the same
    parameters as Fig.~\ref{fig:PT_szsz}(a).} 
  \label{mott_S_ph_S_at}
\end{figure}

\begin{figure} % 12
  \subfigure{
    \includegraphics[width=0.46\linewidth]{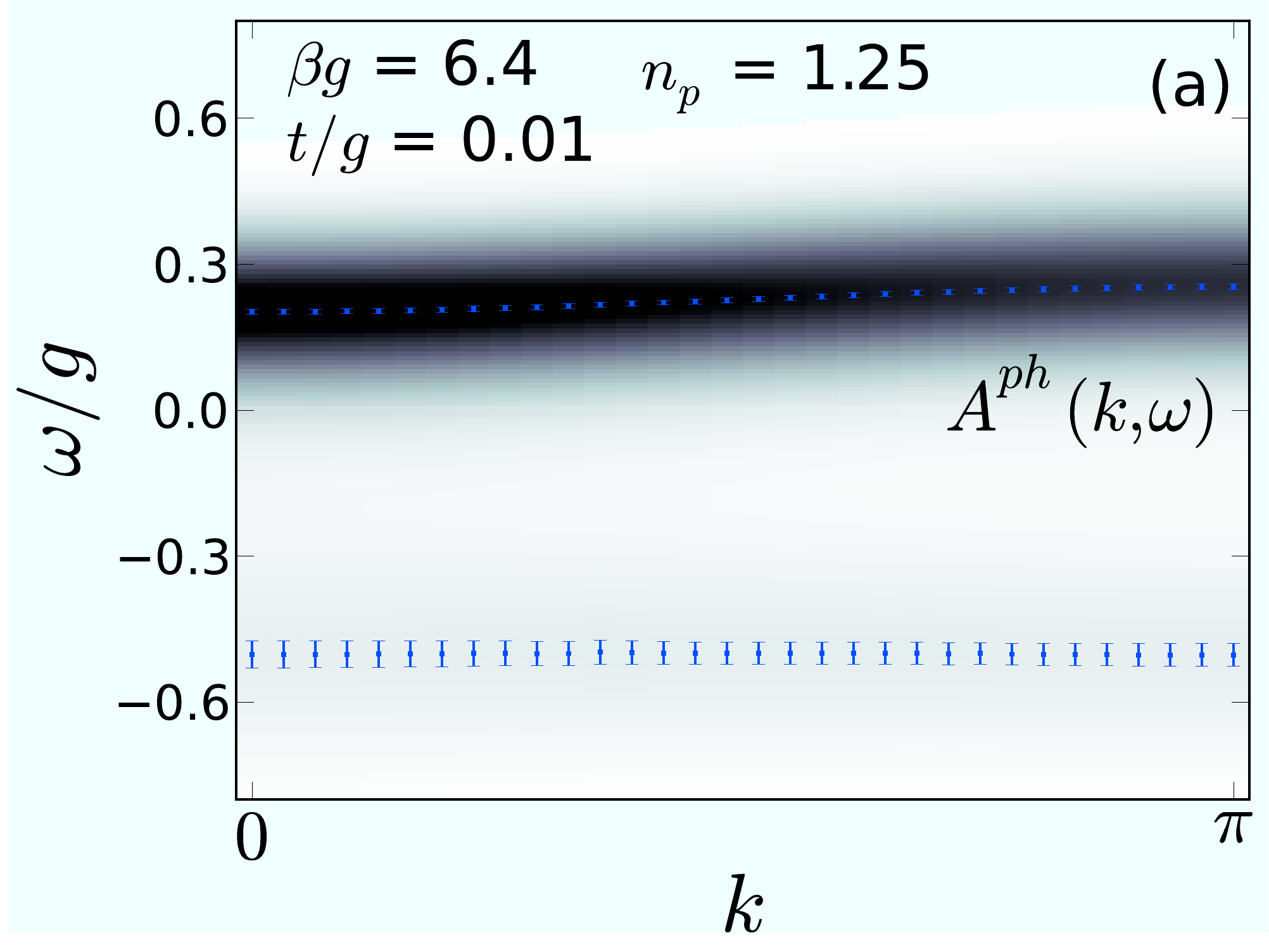}
  }
  \subfigure{
    \includegraphics[width=0.46\linewidth]{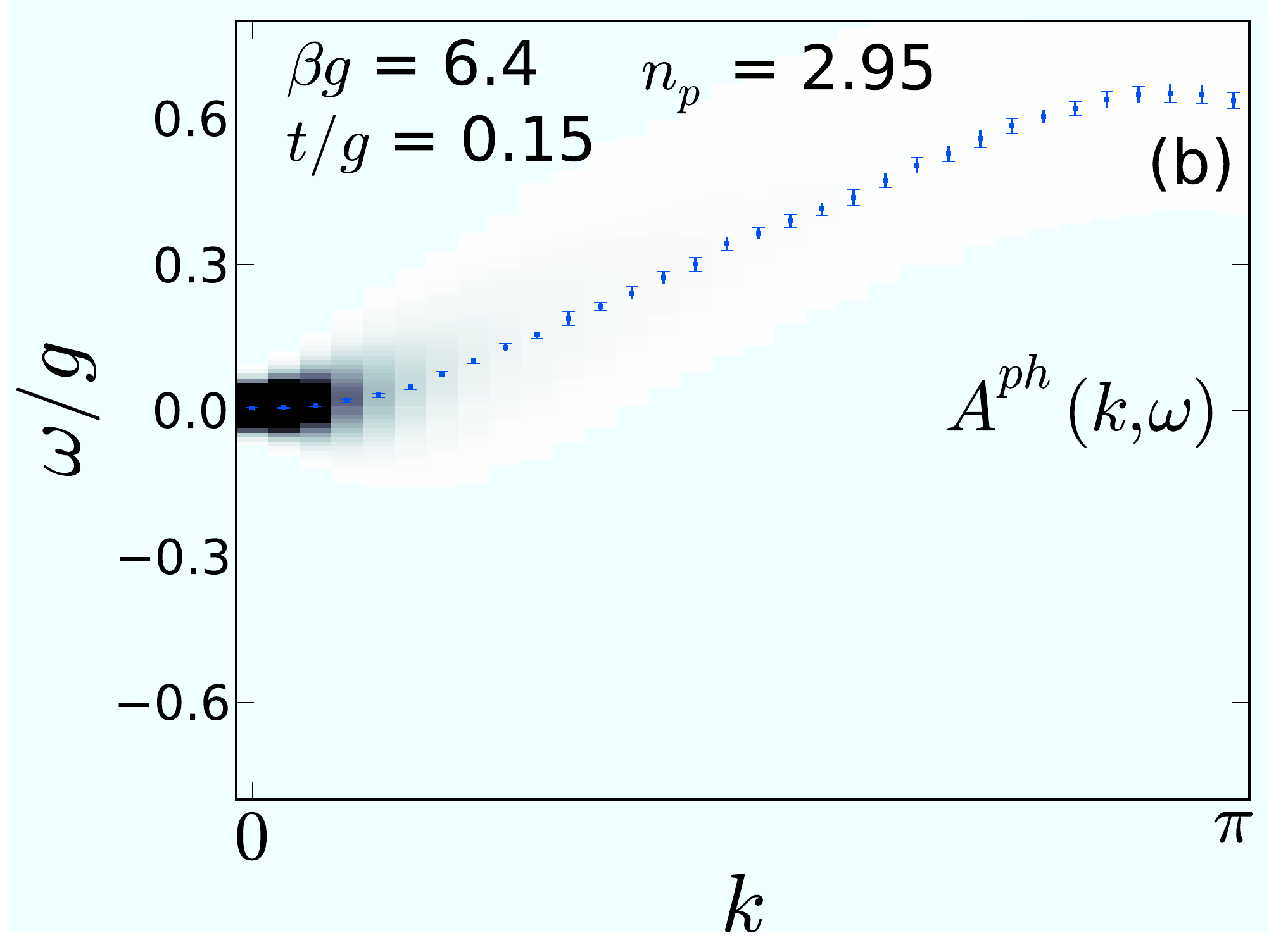}
  }
  \subfigure{
    \includegraphics[width=0.46\linewidth]{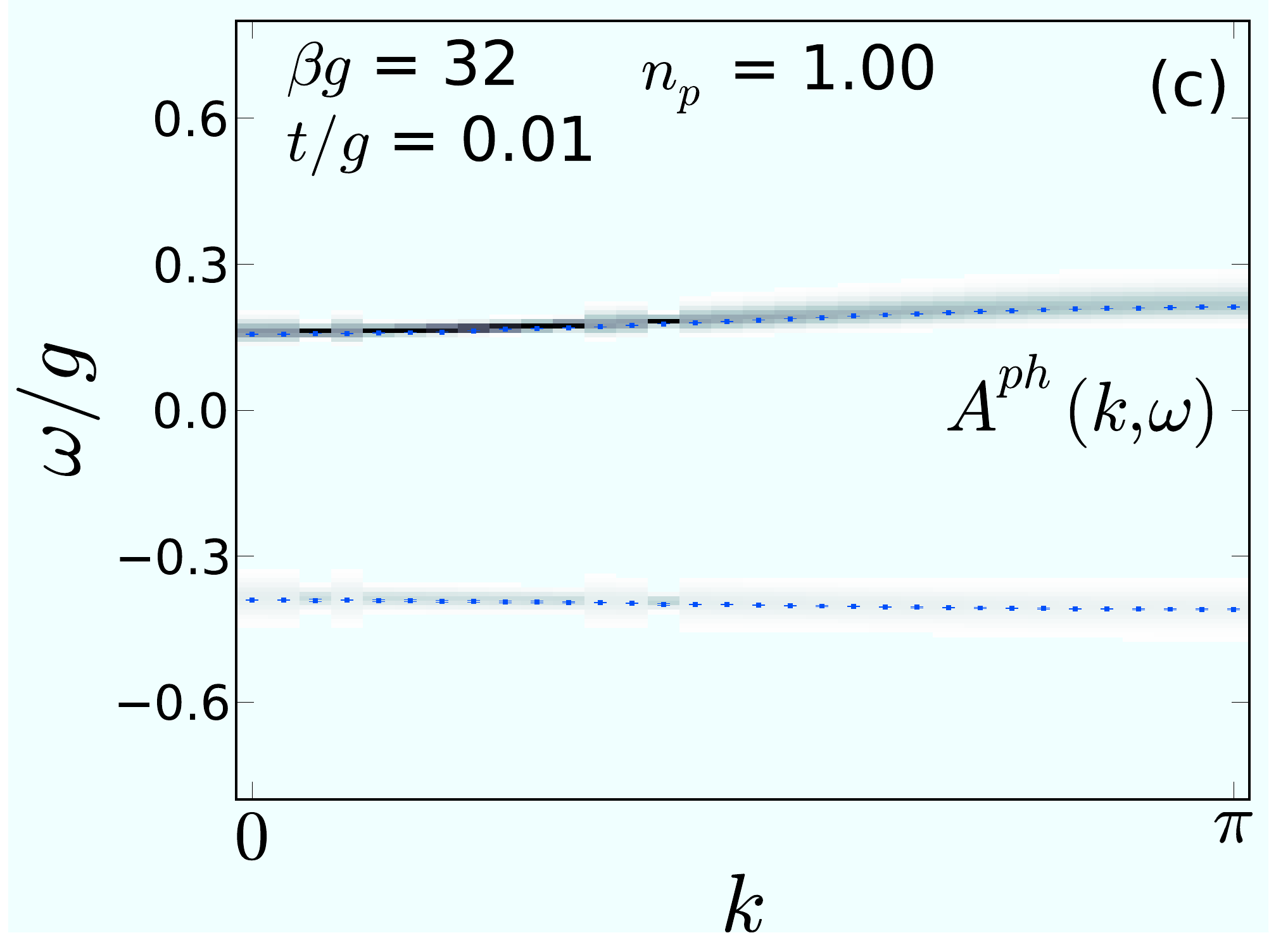}
  }
  \subfigure{
    \includegraphics[width=0.46\linewidth]{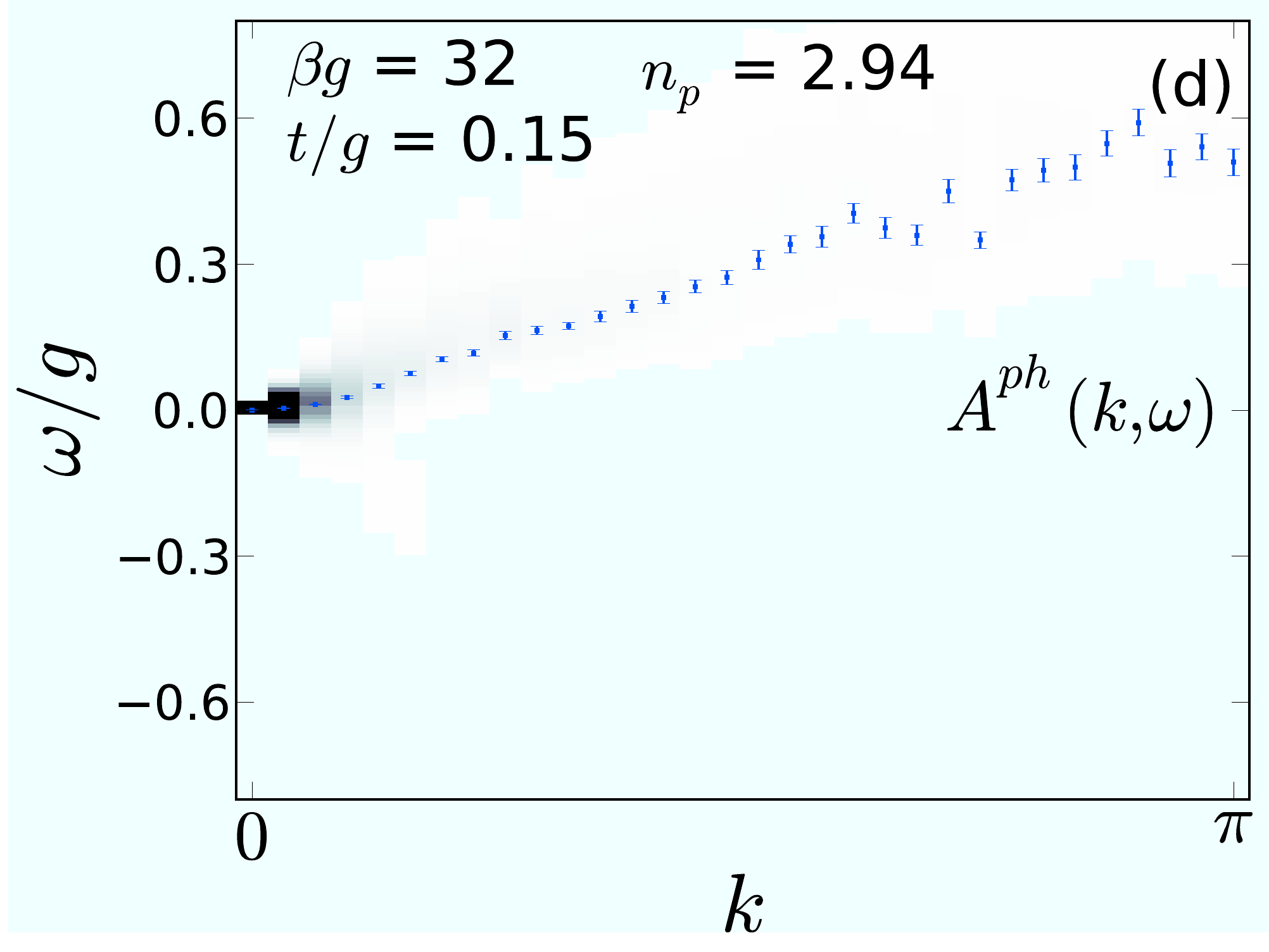}
  }
\caption{\label{fig:finite_T_greens}
  Temperature dependence of $A^\text{ph}(k,\om)$ in the polariton model at $\mu/g =
  0.4$, in the MI (left) and in the SF phase (right). Here again $L =
  64$. Corresponding results at $\beta g =192$ can be found in
  Figs.~\ref{fig:PT_greens}(a),(d).}
\end{figure}

We now demonstrate that the exciton (atom) and photon structure factors,
$S^\mathrm{at}(k,\om)$ and $S^\mathrm{ph}(k,\om)$, shown in
Fig.~\ref{mott_S_ph_S_at}, do not reflect this fact, and therefore cannot be
used to characterize the nature of the Mott state.  To this end, it is
important to notice that the Jaynes-Cummings Hamiltonian has two branches of eigenstates
$\ket{n_\text{p},+}$ and $\ket{n_\text{p},-}$ (the latter containing the
ground state, see also Eq.~(\ref{eq:eigenstates})) with the same polariton number but different
energy.\cite{Ja.Cu.63} In the atomic limit, the dynamic structure factor for
photons and excitons [Eq.~(\ref{eq:skw})] is dominated by the contributions
$\bra{n_\text{p},-} \hat{\rho}^\dag_k\ket{n_\text{p},-}$ (with $\om=0$) and
$\bra{n_\text{p},+} \hat{\rho}^\dag_k \ket{n_\text{p},-}$ (with energy $\om=2
\sqrt{n g^2 + \Delta^2/4}$, equal to 2 in Fig.~\ref{mott_S_ph_S_at}). Any
additional peaks at finite $t/g$ have much smaller spectral weight and cannot
be accurately resolved by our method. However, since the matrix elements of
exciton and photon density operators for the combination $\bra{n_\text{p},+}
\hat{\rho}^\dag_k \ket{n_\text{p},-}$  have the same modulus but opposite
sign, the dominant contributions to $S^\mathrm{at}(k,\om)$ and
$S^\mathrm{ph}(k,\om)$ cancel in the case of the polariton structure factor
$S(k,\om)$. The dispersionless excitations near $\om/g=2$ seen in
Fig.~\ref{mott_S_ph_S_at} are therefore absent in $S(k,\om)$, as confirmed by our data.
Hence, while the upper polariton modes in the single-particle
spectrum have small but finite weight, here their contribution is
zero. Consequently, the polariton dynamic structure factor closely resembles
$S(k,\om)$ of the Bose-Hubbard model.

\subsubsection{Temperature effects}

Experimental realizations of Bose-Hubbard models using cold atomic gases
are usually prepared very close to zero temperature (nK range). In contrast, due
to the strong matter-light coupling achievable in cavities, realizations of
polariton models offer a chance of operation at significantly higher
temperatures. The critical temperature at which the MI state starts to lose
its characteristic integer density has been estimated for the polariton model
as $T^*/g\approx0.03$.\cite{Ai.Ho.Ta.Li.08,Ma.Co.Ta.Ho.Gr.07} For feasible
values of the coupling $g$, $T^*$ falls into the mK range. Generally,
Mott-like physics is expected as long as the Mott gap is significantly larger
than the thermal energy (and the number of particle-hole excitations is
small). The finite-temperature physics of the Bose-Hubbard model has been
analyzed by several
groups.\cite{gerbier:120405,PhysRevA.68.043623,PhysRevA.70.013611,lu:043607}
Here we consider the effect of low but finite temperatures on the excitation
spectra of the polariton model. This also provides information about the
sensitivity of the results to the (necessarily finite) value of $\beta$ used in our
simulations. 

The present method permits calculation of spectra also outside the MI, \ie,
in the SF and the normal phase.  We have pointed out above that,
strictly speaking, there is no SF phase at $T\neq0$ in 1D. Nevertheless, SF
like properties can be seen for $T$ sufficiently small. The results in
Fig.~\ref{fig:finite_T_greens} underline the discussion of
finite-temperature effects on the dispersion in the SF near $k=0$. The
particle excitation is obviously not linear in panels (b) and (d), for which
the temperature is higher than in Fig.~\ref{fig:PT_greens}.

Results for $A^\text{ph}(k,\om)$ are shown in Fig.~\ref{fig:finite_T_greens}.
At finite but low temperature [Fig.~\ref{fig:finite_T_greens}(c,d)] they
still closely resemble the results at $T\approx 0$ in
Figs.~\ref{fig:PT_greens}(a) and (d).
At high temperature [Fig.~\ref{fig:finite_T_greens}(a,b)], we observe strong
broadening of the particle band at all $k$, and strongly suppressed spectral
weight for hole excitations.  Existing work for the Bose-Hubbard model finds that at finite
temperature additional multi-particle and hole bands
arise.\cite{PhysRevA.68.043623} We see an additional excitation for
$\beta g = 4.4$ and $t/g = 0.01$ at an energy of $\om/g \sim 3.1$.
The weight of that excitation is about $50$ times smaller than the weight of
the main peak with energy $\om/g \sim 0.2$ and thus not shown in
Fig.~\ref{fig:finite_T_greens}~(a). This excitation is consistent
with the upper polariton mode discussed above.

We note that a broadened ``gapped'' spectrum is compatible with a density
that deviates from the integer value characteristic of the MI, and this has
to be kept in mind for potential applications relying on integer density.
The numerical results for the total density are shown in each of the panels,
demonstrating that despite the large particle-hole gap the polariton density
deviates significantly from the low-temperature value $n_\text{p}=1$ for the
parameters of Fig.~\ref{fig:finite_T_greens}(a).

Some of the features observed in the SF phase can be explained by means of
Bogoliubov theory for the Bose-Hubbard model. In particular, we have discussed above that
with increasing temperature (where the condensate fraction $n_0\to0$) the
spectral weight of the negative energy branch vanishes first at large $k$, in
agreement with our numerical results.  In addition the broadened positive
energy branch no longer has a clear linear behavior at small
$k$. 

We would like to point out that not only finite temperature but also disorder is an
inevitable feature of experimental realizations of coupled cavity systems. Although not
studied here directly, it has been stated\cite{Ma.Co.Ta.Ho.Gr.07} that the effect of
disorder (in the form of local variations of the parameters $\om_0$, $g$ and
$t$) has similar consequences as finite temperature.

\begin{figure}\hspace*{0.25em} % 13
  %\subfigure{
    \includegraphics[width=0.46\linewidth]{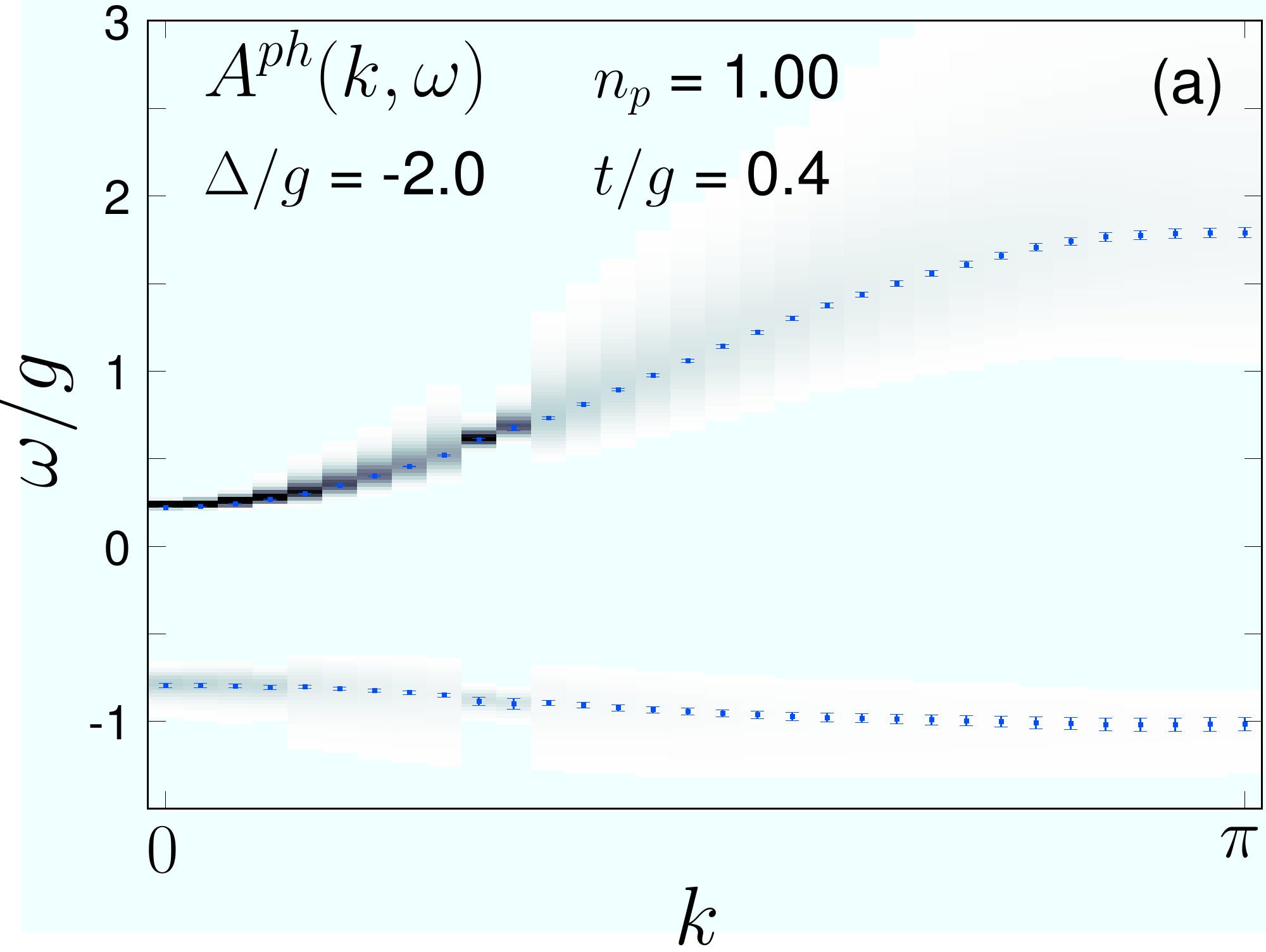}
  %}  
  %\subfigure{
    \hspace*{0.3em}
    \includegraphics[width=0.46\linewidth]{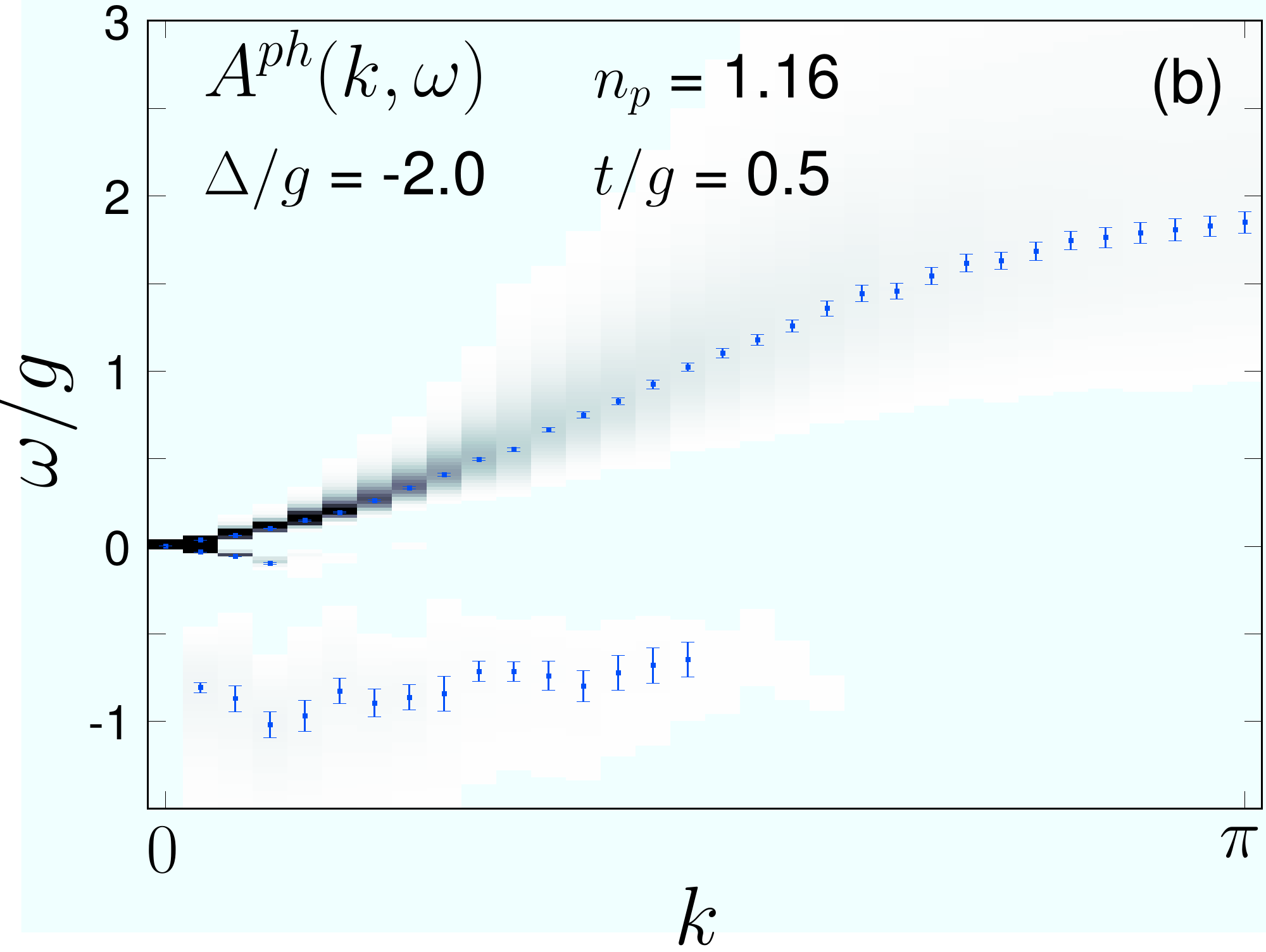}\\\vspace*{1em}
  %}
  %\subfigure{
    \includegraphics[width=0.49\linewidth]{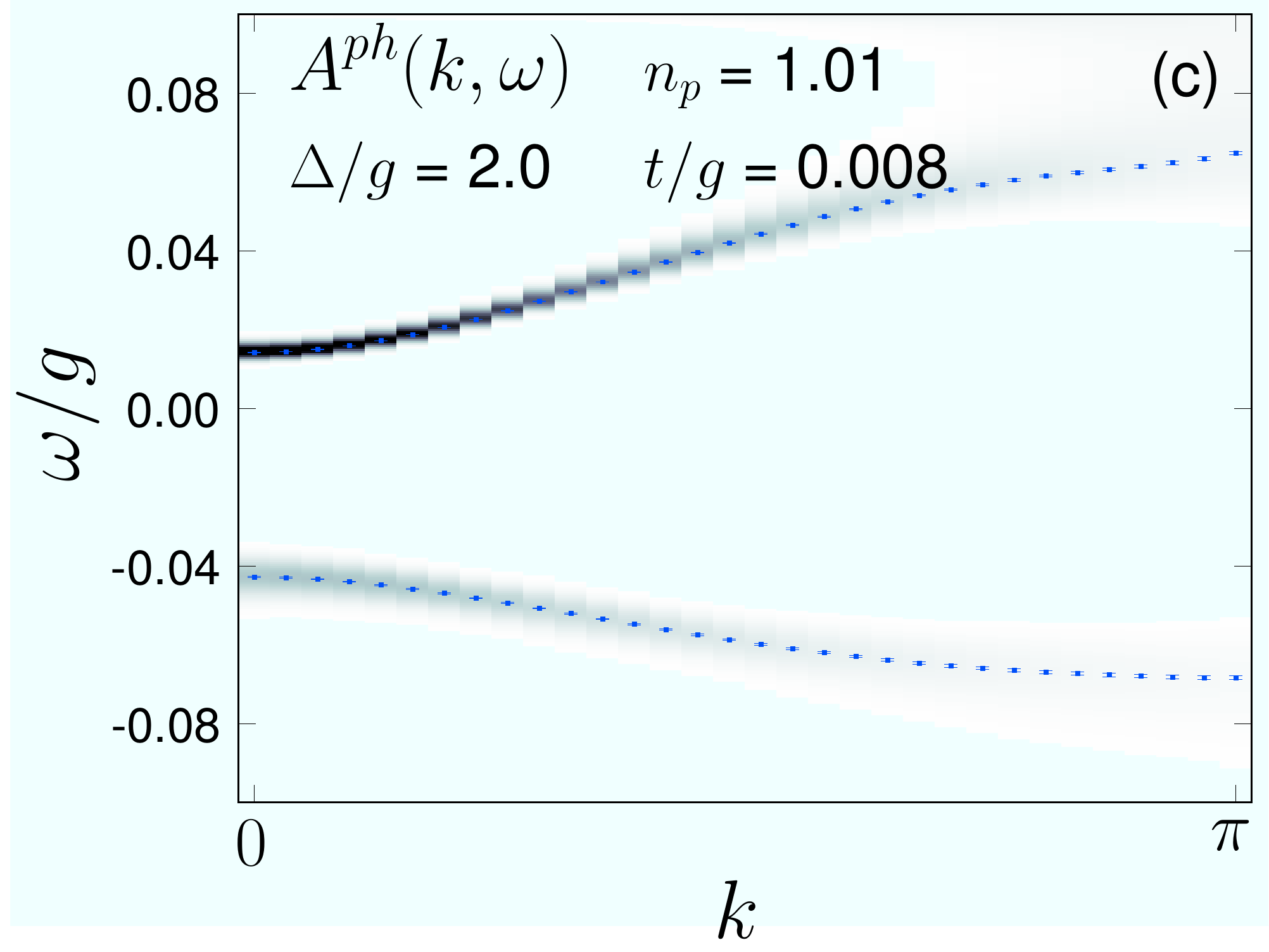}
  %}
  %\subfigure{
    \includegraphics[width=0.49\linewidth]{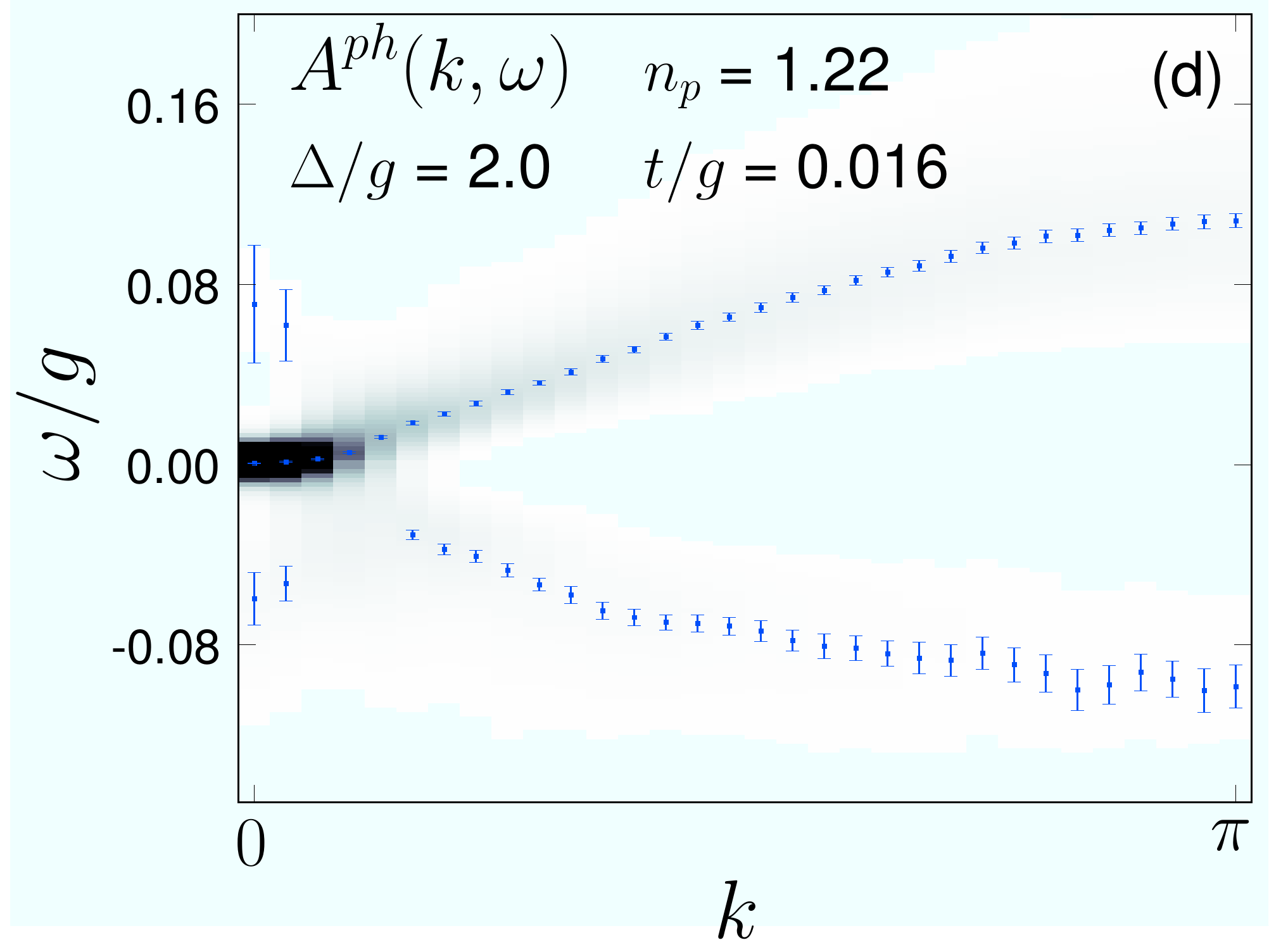}
  %}
  \caption{\label{fig:detuning_A_ph} Single-photon spectra with detuning $\Delta=\epsilon-\omega_0$
    for the excitonic case $\Delta/g= -2$, $\mu/g = -0.5$  (a,b) 
    and photonic case $\Delta/g =  2$, $\mu/g =  0.64$ (c,d),
    in the MI (a,b) and in the SF (c,d).  Here $L=64$ and $\beta g = 3L$.}
\end{figure}

\subsubsection{Detuning}

The detuning between the cavity photon mode and the atomic level splitting is
an important parameter in the polariton model which is absent in the Bose-Hubbard model. Its influence on
the physics has been discussed
before.\cite{Ai.Ho.Ta.Li.08,irish_polaritonic_2008,Zh.Sa.Ue.08} Detuning can
also be easily changed experimentally, motivating a calculation of the
excitation spectra for $\Delta\neq0$.  Our results are shown in
Fig.~\ref{fig:detuning_A_ph}. 

The extent of the different phases, namely excitonic or polaritonic MI and
photonic or polaritonic SF in the phase diagram has been analyzed for a
two-site system.\cite{irish_polaritonic_2008} The way to distinguish between
the polaritonic SF and the photonic SF is to monitor fluctuations in the
exciton occupation number (pinned in the photonic SF but fluctuating in the polaritonic SF). The conclusion has been that
for $t\approx|\Delta|$ a polaritonic SF exists only for $\Delta/g<-1$ This
would match with the conjectured photonic nature of the SF near the lobe tip
in two dimensions.\cite{Zh.Sa.Ue.08} However, it is not clear if these strict
values also hold for larger systems and the thermodynamic limit. Besides, the
work by Irish \etal\cite{irish_polaritonic_2008} is exclusively concerned
with the fixed density transition occurring at $t\approx|\Delta|$, whereas a
polariton SF may exist also for $t<|\Delta|$ if density fluctuations are
allowed (generic transition). The spectrum in Fig.~\ref{fig:detuning_A_ph}(b)
is for such a set of parameters.

Again pertaining to the fixed density case, the MI state is supposed to be of
excitonic nature for $\Delta/g<-1$, $t<|\Delta|$, and of polaritonic nature
for $|\Delta|/g<1$ and small enough $t/g$ respectively
$t/|\Delta|$.\cite{irish_polaritonic_2008} The former case is depicted in
Fig.~\ref{fig:detuning_A_ph}(a), whereas the latter corresponds to the
$\Delta=0$ results reported in Fig.~\ref{fig:PT_greens}.

For $\Delta\gg g$, photon excitations are always lower in energy, and the
effective interaction approaches zero. As a result, MI regions are very small
or nonexistent, and the photonic SF state is similar to that of the Bose-Hubbard model in
the limit of large $t/U$.\cite{Ai.Ho.Ta.Li.08}

Here we consider $\Delta/g=\pm2$ for comparison to previous calculations of
the spectra in the Mott phase.\cite{Ai.Ho.Ta.Li.08} (Note that the rotating wave
approximation formally requires $|\Delta|\ll\epsilon,\om_0$.\cite{narozhny_coherence_1981})
These correspond to
effective repulsions $U_\mathrm{eff}/g=0.096$ (for $\Delta/g=2$) respectively
$U_\mathrm{eff}/g=2.096$ (for $\Delta/g=-2$), in excellent agreement with the
width of the $n_\text{p}=1$ Mott lobes for the same
parameters.\cite{Ai.Ho.Ta.Li.08}

Our results in Fig.~\ref{fig:detuning_A_ph} show that again the spectra are
dominated by the generic features of the MI and the SF. However, the detuning
in the present case changes the ratio of the bandwidths of particle and hole
bands $W_\text{p}/W_\text{h}$ in the Mott state.\cite{Ai.Ho.Ta.Li.08} While
for $\Delta=0$, $W_\text{p}/W_\text{h} \approx 3$, we find
$W_\text{p}/W_\text{h} \approx 2$ (similar to the result for the Bose-Hubbard model) for
$\Delta/g = 2$ and $W_\text{p}/W_\text{h} \approx 7$ for $\Delta/g = -2$. The
incoherent features observed for $\Delta/g=-2$ in
Ref.~\onlinecite{Ai.Ho.Ta.Li.08} are not seen here. As mentioned before,
the energy of the upper polariton modes (not shown) increases for
$\Delta\neq0$.\cite{GrTaCoHo06}

In the SF, we find the expected gapless excitations, as well as gapped modes
indicative of a correlated superfluid.  Since for $\Delta/g=2$,
$U_\text{eff}$ is very small, the Mott gap of the dispersive bands in
Fig.~\ref{fig:detuning_A_ph}(c) is also small (0.057$g$), but it is still
larger than the temperature scale in our simulation $T/g = 0.005$.  We note
that, within our resolution, the positive energy spectrum in
Fig.~\ref{fig:detuning_A_ph}(d) looks gapless, but not clearly linear. In
this respect, the spectra for finite detuning resemble those obtained at high
temperatures. Apart from this issue and a scaling of energies (due to the
dependence of $U_\text{eff}$ on $\Delta$), the spectra obtained for
$\Delta/g=-2$ are very similar to those for $\Delta=0$, whereas those for
$\Delta/g=2$ resemble closely the results for the Bose-Hubbard model.

\subsubsection{Phase transition}
To end with, we present a scaling analysis for the generic phase transition.
As pointed out by Fisher~\etal~\cite{PhysRevB.40.546} the
scaling relation 
\begin{equation}\label{eq:fss_hyp}
  \rho_\text{s} 
  =
  L^{2-d-z} \Tilde \rho(\delta L^{1/\nu},\beta/L^z)
\end{equation}
should hold for the superfluid density across the MI-SF transition. Here
$\nu$ is the critical exponent of the correlation length which is expected to
diverge like $\xi \sim \delta^{-\nu}$, and $z$ is the dynamical critical
exponent. The generic transition in the Bose-Hubbard model has mean-field exponents
$z=2$ and $v=1/z=1/2$.\cite{PhysRevB.40.546,Ba.Sc.Zi.90,alet:024513,capogrosso-sansone:134302}
Recent field theory\cite{Ko.LH.09} and strong-coupling results\cite{Sc.Bl.09}
predict the same universality  classes for the polariton model, in conflict
with numerical results in two dimensions which suggest the absence of
multicritical points.\cite{Zh.Sa.Ue.08}

\begin{figure} % 14
  \centering
 \includegraphics[width=.95\linewidth]{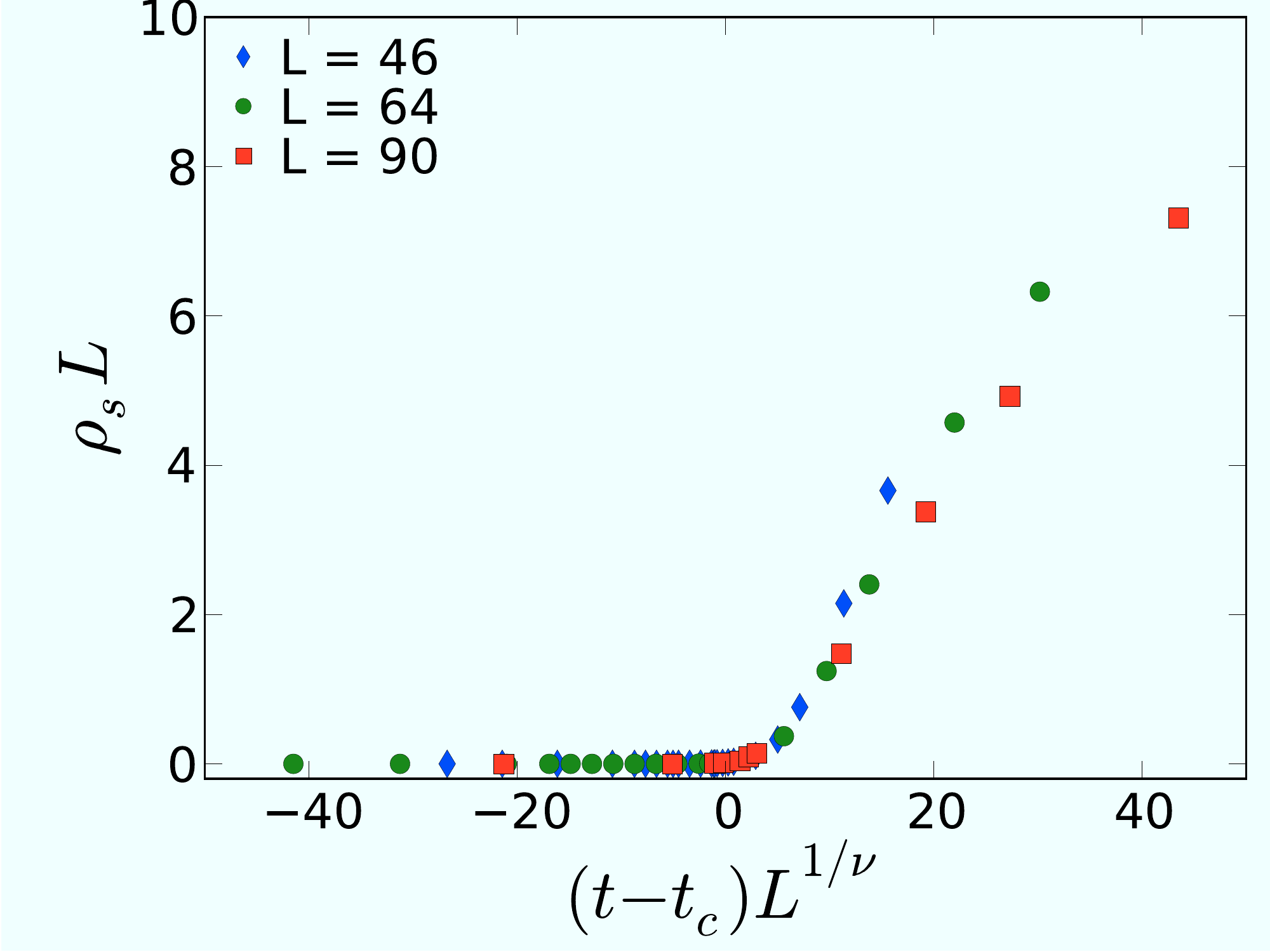}
  \caption{(color online) Finite size scaling for the generic transition in
    the polariton model at $\mu/g = 0.4$, testing the scaling
    hypothesis~Eq.~(\ref{eq:fss_hyp}).}
  \label{fig:fss_mu0_4}
\end{figure}

We test in 1D the scaling hypothesis Eq.~(\ref{eq:fss_hyp}) with $z=2$ and
the hyperscaling relation $z = 1/\nu$, along the line $\mu/g=0.4$ where the
generic transition is expected [see Fig.~\ref{fig:phasediagrams}(b)].

To this end we keep the temperature constant at $\beta=L^2/10$ and plot
$\rho_\text{s} L^{d+z-2}$ over $(t-t_\mathrm{c}) L^{1/\nu}$ to obtain the
universal function $\tilde \rho$ (Fig.~\ref{fig:fss_mu0_4}).  Defining a cost
function in the spirit of Ref.~\onlinecite{harada_XY_1998}, allows us to
evaluate the quality of the finite-size-scaling plot quantitatively.

% Testing $z=2$ and keeping $\nu = 1/z$ constant
We find the minimum of our cost function at $t_\mathrm{c} = 0.0626(1)$ and
$\nu = 0.50(2)$ when matching $\tilde \rho$ close to the phase transition
[$|(t-t_c)L^{1/\nu}|<1$ in Fig.~\ref{fig:fss_mu0_4}].  We note that with the
system sizes available, this result is not very stable.  A fit in a larger
region of $|(t-t_c)L^{1/\nu}|$ provides a better data collapse overall (but
worse close to the phase transition), with $\nu\approx 0.65$ and very
slightly smaller $t_\mathrm{c}$.

A similar scaling with $z=1$ did not succeed, so that we conclude that the
universality class of the generic transition in the polariton model is the
same as that in the Bose-Hubbard model, despite the composite nature of the
quasiparticles. This is consistent with recent field-theory and
strong-coupling results.\cite{Ko.LH.09,Sc.Bl.09}

An accurate scaling analysis for the fixed density transition through the
lobe tip has been found to require much larger system sizes and is therefore
not shown. In the 1D case considered, the shape of the lowest Mott lobe in 1D
(see Fig.~\ref{fig:phasediagrams}) suggests that the similarity to the Bose-Hubbard model
holds also in this respect, \ie a Kosterlitz-Thouless type phase transition.

\section{Conclusions}\label{sec:conclusions}

We calculated the single-boson spectral function and the
dynamic structure factor of the Bose-Hubbard model, and for a recently
proposed model of itinerant polaritons in coupled-cavity arrays.
These models undergo a quantum phase transition from a Mott insulator to a
superfluid state upon increasing the hopping integral of the bosons
respectively photons with respect to the interaction. Results in one
dimension, within and close to the Mott lobe with density one, have been
obtained.

Despite the generally different nature of the conserved particles, the models
exhibit very similar spectral properties, including gapped particle and hole
bands in the Mott insulating phase, and Bogoliubov type excitations in the
superfluid phase. Additional excitations related to the second branch of
upper polariton states exist in the single-particle spectrum of the polariton
model,\cite{Sc.Bl.09} but cancel out in the dynamic structure factor. In
general, these features have high energy and very small spectral weight, so
that for practical purposes the excitation spectra are qualitatively similar
to the Bose-Hubbard model.

Correlation effects are particularly strong in the one
dimensional case considered. Our results in the superfluid phase represent the first
unbiased nonperturbative spectra for $A(k,\om)$ in both models and for
$S(k,\om)$ in the polariton model (in both phases).  Good qualitative
agreement with recent analytical work on the two-dimensional Bose-Hubbard
model was found, and we have compared our results in the superfluid phase to
Bogoliubov theory. The limiting cases of the Mott insulator close to the atomic
limit, as well as the weakly interacting superfluid are described quite well
by analytical approximations, whereas in the phase transition region, our
nonperturbative results show considerable deviation. Emerging particle-hole symmetry on
approach of the multicritical lobe tip has been demonstrated for the polariton model.

For the polariton model, we have also explored the influence of detuning and
finite temperature on the spectral properties, and have presented a scaling
analysis to determine the universality class of the generic phase transition.
Keeping in mind experimental realizations of coupled cavity arrays,
interesting open issues for future work include the excitation spectra in the
two-dimensional case (and comparison to analytical results\cite{Sc.Bl.09}),
the behavior of the sound velocity across the phase transition (also for the
Bose-Hubbard model) and disorder.

The present work further highlights the fact that the physics of strongly
correlated bosons as described by the Bose-Hubbard model may be observed
in terms of optical models that, if realized, would have some distinct
experimental advantages and further contain new degrees of freedom due to the
mixed nature of the quasiparticles.

%%%%%%%%%%%%%%%%%%%%%%%%%%%%%%%%%%%%%%%%%%%%%%%%%%%%%%%%%%%%%%%%%%%%%
\begin{acknowledgments}
%%%%%%%%%%%%%%%%%%%%%%%%%%%%%%%%%%%%%%%%%%%%%%%%%%%%%%%%%%%%%%%%%%%%%
  
MH was supported by the FWF Schr\"odinger Fellowship No.~J2583. PP
acknowledges support from the FWF, projects P18551 and P18505.  We made use
of the ALPS library\cite{ALPS_I,ALPS_II} and the ALPS
applications.\cite{ALPS_DIRLOOP} We acknowledge fruitful discussions with F.
Assaad, M. J. Bhaseen, J. Keeling, D. Khmelnitskii and P. B. Littlewood. We are
grateful to H. Monien and D. Rossini for providing us with data for
Figure~\ref{fig:phasediagrams}.

%%%%%%%%%%%%%%%%%%%%%%%%%%%%%%%%%%%%%%%%%%%%%%%%%%%%%%%%%%%%%%%%%%%%%
\end{acknowledgments}
%%%%%%%%%%%%%%%%%%%%%%%%%%%%%%%%%%%%%%%%%%%%%%%%%%%%%%%%%%%%%%%%%%%%%

%\bibliographystyle{h-physrev}
%\bibliography{../bibliography}

\end{document}